# Deep learning directed synthesis of fluid ferroelectric materials

Charles Parton-Barr[1], Stuart R. Berrow [1], Calum J. Gibb [2], Jordan Hobbs [1], Wanhe Jiang [1], Caitlin O'Brien [2,3], Will C. Ogle [2,3], Helen F. Gleeson[1], Richard J. Mandle[1,2,*]

[1]School of Physics and Astronomy, University of Leeds, Leeds, UK, LS29JT
[2]School of Chemistry, University of Leeds, Leeds, UK, LS29JT
[3]School of Applied Mathematics, University of Leeds, Leeds, UK, LS29JT

* r.mandle@leeds.ac.uk

## Abstract

We present a data-to-molecule deep-learning workflow for the targeted design of organic fluid ferroelectrics and validate it through the synthesis and experimental characterization of machine-learning generated candidate materials. Fluid ferroelectrics, a class of liquid crystals that exhibit switchable long-range polar order, offer opportunities in ultrafast electro-optic technologies, responsive soft matter, and next-generation energy materials. Without well-validated design rules, development of new fluid ferroelectric materials remains challenging. We curate a comprehensive dataset of all known longitudinally polar liquid-crystal materials and train graph neural networks to predict ferroelectric nematic transition temperatures. A graph variational autoencoder generates de novo structures which are filtered by an ensemble of classifiers and regressors to identify candidates with predicted ferroelectric nematic behaviour and accessible transition temperatures. Integration with a computational retrosynthesis engine and a digitised chemical inventory narrows the design space to a synthesis-ready longlist. Eleven machine-learning-selected candidates were synthesised and characterized, with experimental ferroelectric nematic transitions compared against model predictions. The experimental feedback augments the original dataset and a further iteration uses graph neural networks and XGBoost models to produce a ranked set of synthesis candidates. These results demonstrate a step towards practical, closed-loop discovery of synthesizable fluid ferroelectrics and autonomous design of functional soft materials.

## Introduction

The design of molecular materials requires domain specific expertise and understanding of the complex, often multidimensional relationship between chemical structure and macroscopic properties [1]. The vastness of chemical space means that discovering new molecular materials through exploratory synthesis is inherently slow [2]. Significant experimental effort is required to iteratively expand from known structures, a process prone to the encoding of human bias. Advances in computational power and algorithmic efficiency make deep learning an ever more practical approach for extracting structure-property relationships from large molecular datasets [3]. Coupled to generative models capable of designing new molecular structures [4], this enables the data-driven exploration of chemical space and opens a new paradigm in the design of molecular materials. The end point of molecule design is physical realisation, and to best achieve this, the generative workflow output must be guided by a set of requirements [5]. The molecules need to be realistically synthesisable, property prediction must be rigorously justified, expert human feedback integrated, and analysis of predictions should elucidate structural property relations.

Approaches for generating novel molecules can be broadly classified into rule-based systems and generative machine-learning models. Rule-based approaches construct new molecules by recombining predefined structural fragments or applying encoded chemical reactions [6]. In contrast, generative models learn a latent representation of chemical space from training data, sampling this latent space to propose new chemical structures [7]. Deep generative models have been developed using recurrent neural networks [8], variational autoencoders [9], and generative adversarial networks; [10] each learning the implicit molecular syntax (e.g. from SMILES strings) and producing novel candidate structures. Recently, molecular graph neural networks have been combined with variational autoencoder architectures to generate new graph structures [11]. While these methods have been successfully and widely applied in the context of drug discovery [12], protein structure [13], and inorganic/solid-state materials [14], they are yet to significantly impact the design of soft materials [15].

Liquid crystals provide a demanding testbed for molecular design: their bulk properties emerge from molecular structure through subtle and collective interactions that are not captured by simple molecular descriptors. Liquid crystals are soft phases of matter that combine degrees of orientational and positional order with fluidity, giving rise to states of matter (mesophases) with distinct symmetry and physical properties[16-18]. Amongst these phases is the recently discovered ferroelectric nematic ($N_F$) phase (Fig. 1a) which long range polar order arising from the parallel arrangement of molecular dipoles yields bulk spontaneous polarisation comparable to solid state ferroelectrics [19]. This can be compared to a conventional nematic phase in which the nematic phase's inversion symmetry of its dipole moments is broken.

The formation of liquid crystal phases and their specific mesophase type depends on a delicate balance of molecular shape and electronic structure, which impact on collective molecular organisation. Small structural modifications can give rise to significant differences in phase behaviour, making traditional structure-property heuristics unreliable. Only a small number of structural motifs are currently known to support the $N_F$ phase, notably DIO [20], RM734 [21, 22] (Fig. 1b) and derivatives thereof [23-26]. This narrow chemical space stems from an absence of design rules for the structural-property relationship that causes the spontaneous polarity to emerge [27-30]. Identifying new structural motifs capable of stabilising polar order therefore presents a significant challenge in the molecular design of such materials.

Machine learning studies in liquid crystal science have focused on property prediction or classification rather than molecular design. Examples include models for the classification of ferroelectric nematic materials [31], predicting clearing point (isotropisation) temperatures [32, 33], phase transitions in bent-core type materials [34],  lyotropic phase behaviour [35], image-based frameworks for phase identification via optical texture analysis[36, 37], hydrodynamic properties via director field dynamics [38], and optimisation of LCD backlight technology [39]. Each of these studies illustrates the promise of data-driven methods in liquid crystal research but none address generative molecular design with explicit synthesis of new materials, which is the focus of the present work.

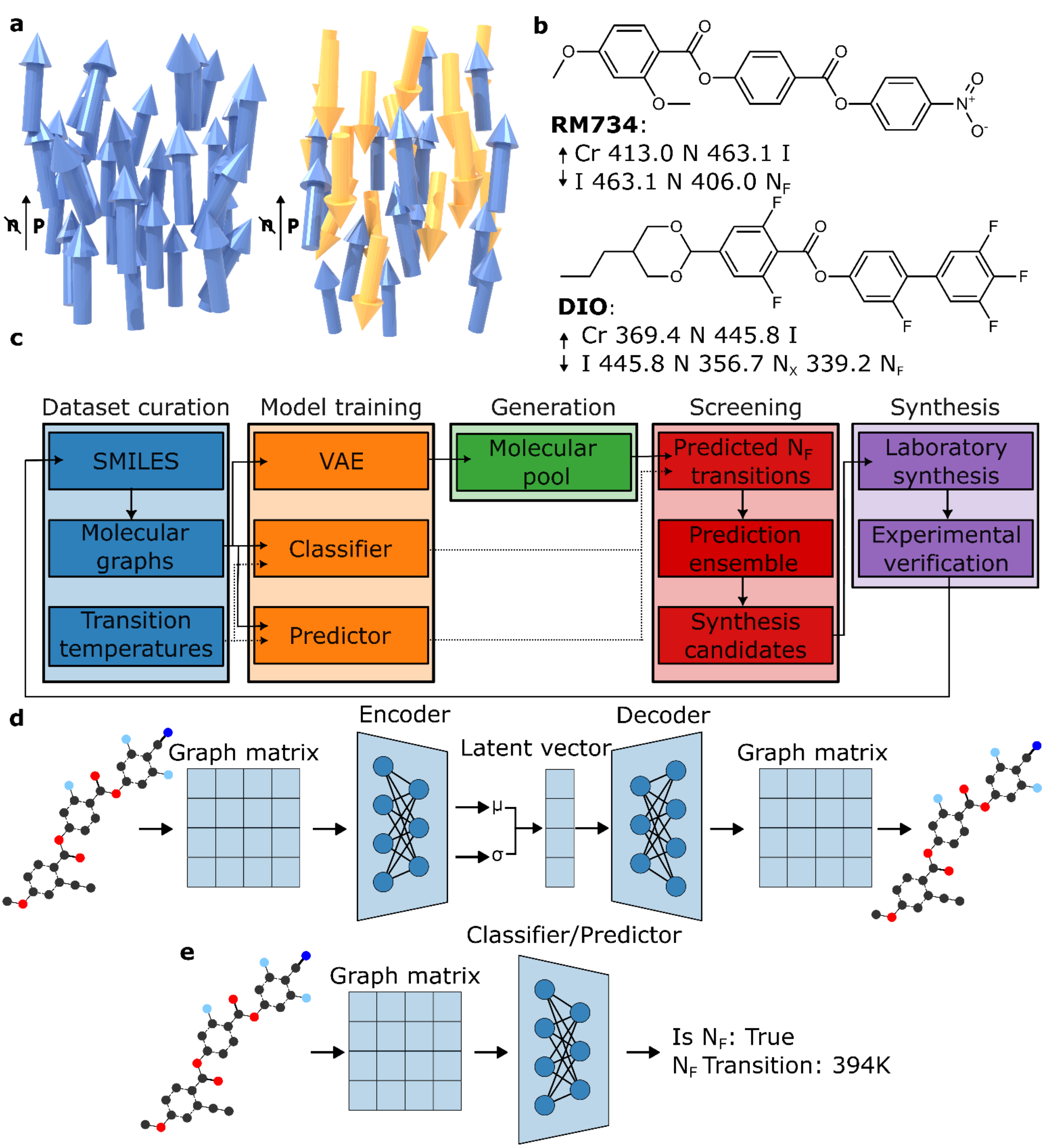

**Figure 1.** a) A schematic representation of the ferroelectric nematic ($N_F$) phase (right), in which molecular dipoles align collectively to produce macroscopic polar order and lack inversion symmetry. (b) Representative $N_F$ materials RM734 and DIO, with molecular structures, transition temperatures in K. (c) Overview of the generative–predictive workflow for data-driven discovery of $N_F$ materials, including dataset curation, model training, molecular generation, screening, and experimental validation. (d) Variational autoencoder architecture used for molecular generation. (e) Neural-network classifier and property prediction model architectures.

Here we introduce a generative-predictive framework for small-molecule design and apply it to the discovery of new ferroelectric nematic materials. This workflow (Fig. 1c) generates candidate structures and predicts $N_F$ behaviour and transition temperatures, providing a data-driven approach to exploring the chemical space associated with this state of matter. We validate the framework by synthesis and characterisation of model-generated candidates; these form an *ex-silico* testing set, its investigation of the $N_F$ chemical space providing valuable feedback for further training of machine learning of models. The complete workflow establishes an experimental benchmark for data-driven materials discovery. The complete set of synthesised candidates with extrapolated transition temperatures can be found in the ESI.

## Methods

### Data setup

We curated a dataset comprising >500 small-molecule liquid crystal materials which includes all reported $N_F$ materials, up to October 2025 and a further >200 materials from unpublished in-house measurements. Each entry consists of the molecular structure as SMILES and experimentally measured phase transition temperatures. To avoid biasing the model towards $N_F$ behaviour the dataset was balanced with structurally similar non-$N_F$ materials.

Smiles strings were converted to molecular graphs using RDKIT. In the graph representation, each molecule is represented by a node feature matrix, $V$ (atom descriptors) and an edge matrix, $E$ (bond connectivity). These graph representations were annotated with transition temperatures and used as inputs for deep-learning models (Fig. 1(d,e)). A more complete description of the featurisation methodology is given in section 5 of the ESI. For the second iteration of the workflow, the scikit-fingerprint package is used to compute fingerprint representations of molecules [40].

### Model training

We employ three deep learning architectures: (a) a graph variational autoencoder (VAE) for generation of *de novo* structures (b) a binary classifier to predict the presence or absence of $N_F$ behaviour; (c) a scalar predictor (regression model) for predicting $N_F$ transition temperatures (Fig. 1e).

The VAE is a modified MGVAE architecture [11] and learns a probabilistic latent representation of molecular structure. The encoder maps molecular graphs into a continuous latent space, and the decoder reconstructs graphs from latent vectors (i.e. the inverse process), enabling the generation of novel molecules

For classification and regression graph neural networks transform molecular graphs in scalar outputs corresponding to phase assignment or $N_F$ transition temperature. Our model library includes classifiers and regressors built on the following message-passing layers, GATConv [41], GATv2Conv [42], GINConv [43], GCNConv [44], GatedGraphConv [45], TransformerConv[46].

We additionally train XGBoost [47] classifiers and regressors on fingerprint representations. In the second iteration of the workflow, their predictions are combined with those of the GNNs, producing a cross-model consensus filtering that includes distinct model architectures.

Given the modest size of our dataset (~700 molecules) repeated k-fold cross-validation was used for model training and performance evaluation. Hyperparameter settings for the training of models can be found in the supplementary information.

**Generation**

After training the VAE model was used to generate novel molecular structures by sampling its latent space and decoding the resulting vectors into molecular graphs. Because the latent space derives from experimentally characterised liquid crystal molecules, the generated structures occupy a similar chemical manifold ("dataset-like") while also allowing exploration of training set's chemical space.

The decoded molecular graphs were converted into canonical SMILES strings; after filtering any duplicates this yields a pool of novel molecules occupying the learned $N_F$ liquid crystal chemical manifold, ~30,000 novel structures overall.

**Screening**

We construct an ensemble of classifiers by selecting the three most accurate models (Fig. 2a) from across four different graph neural network architectures. Each classifier simply returns a Boolean value indicating the presence or absence of the $N_F$ phase. An analogous ensemble of predictors (Fig. 2b) was selected in the same manner, choosing for the lowest root mean square error (RMSE) of $N_F$ transition temperature predictions.

Molecules for which the ensemble of classifiers demonstrated high consensus were subsequently evaluated using the ensemble of predictors, with the mean predicted $N_F$ transition temperature as a ranking criterion. This yields the generated pool of materials

From the screened pool we selected candidate molecules representing both high predicted likelihood of $N_F$ behaviour and low-probability negative controls. Computational retrosynthesis tools [48] allied to our digitised chemical inventory and vendor catalogues allowed us to identify longlist of synthetically accessible targets. This machine generated longlist was then reviewed and prioritised by a team of human synthetic chemists

**Synthesis**

All compounds were synthesised using standard organic chemistry techniques in conventional laboratory glassware, either open to air or under an atmosphere of dry nitrogen. All chemical reactions were ultimately subject to purification by flash chromatography over silica gel using a Combiflash NextGen 300+ system (Teledyne Isco) with an appropriate solvent gradient. Final materials were filtered through 0.2-micron PTFE filters to remove particulates. Yields refer to isolated, spectroscopically homogenous material. Chemical characterisation was made by NMR spectroscopy ($^{1}$H, $^{13}$C and $^{19}$F) and high-resolution mass spectrometry (HRMS), all materials were confirmed to have purity of >99% (peak area) by reverse phase HPLC. Full synthetic details are given in the ESI to this article. Synthesised materials were studied by polarised optical microscopy (POM) and differential scanning calorimetry (DSC). Transition temperatures for polar phases were extrapolated from mixture studies with DIO as the host (~90 wt%) and the material in question as the guest (~ 10 wt%).

## Results

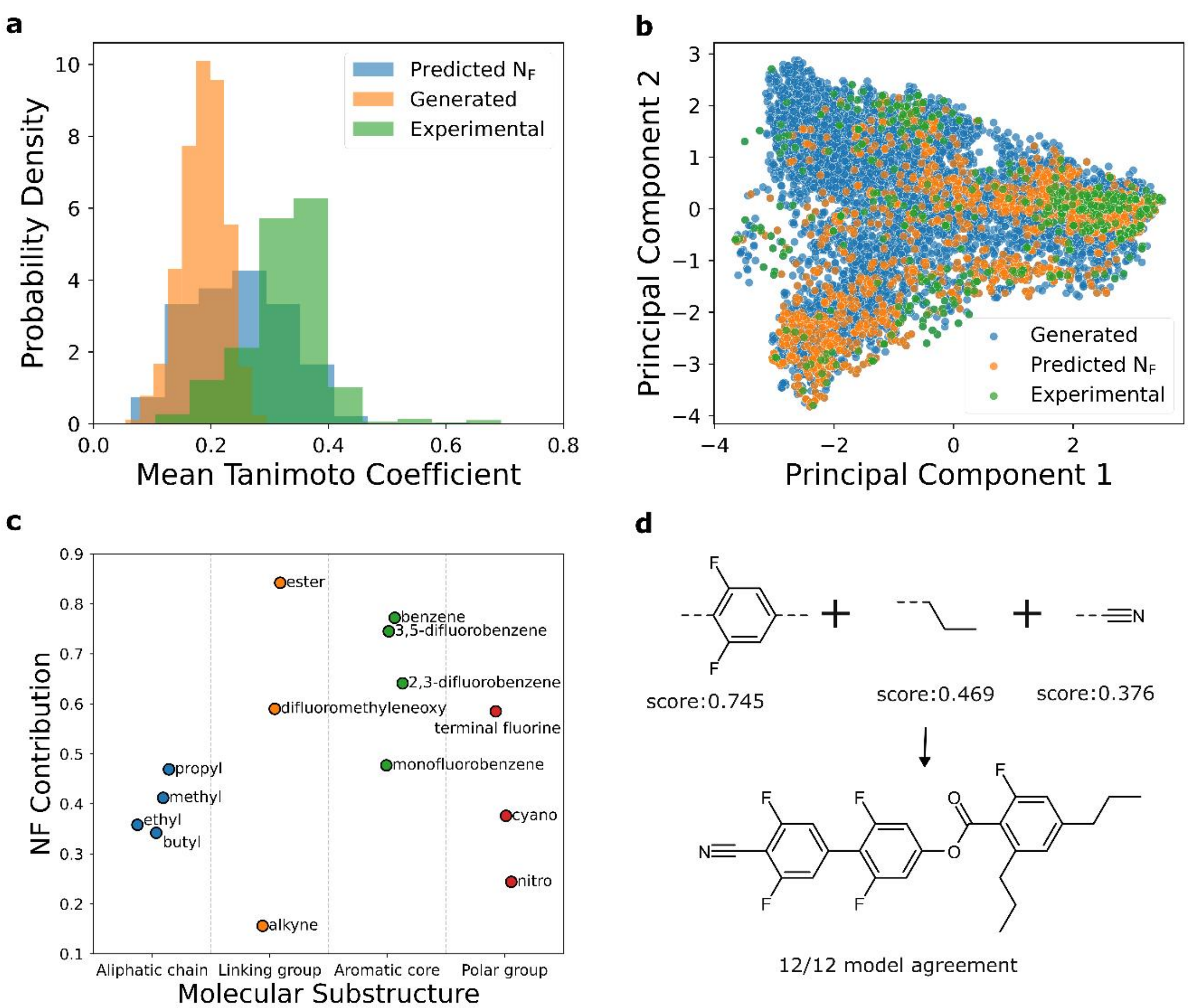

**Figure 2.** Analysing the model decisions in classifying a material as $N_F$. a) The Tanimoto coefficients comparing the molecule similarity within the three datasets. b) Visualisation of the structural similarity of the total three datasets. c) subgraph masking explanation scoring of the contribution to the $N_F$ phase for molecular substructures present in the generated molecules. d) contribution-guided design coupled with scaffold analysis of the dataset.

To evaluate how the generative model explores chemical space relative to the experimental dataset, we compared distributions of selected structural descriptors (Fig. S10). The generated molecules exhibit broader variability in size and ring count, while retaining similar overall carbon distributions. Nitrogen-containing structures are less common in the generated set, whereas fluorination increases modestly. Oxygen counts shift toward simpler structures, suggesting the model samples lower-polarity motifs while retaining ester functionality. Substructure analysis (SMARTS-based) indicates reduced frequency of specific motifs that are over-represented in the training set (e.g., multiple difluorobenzene rings), alongside emergence of new functionalities such as ketones. Collectively, these patterns indicate that the generative model learns the dominant structural trends in liquid-crystal materials while exploring adjacent chemical space rather than memorising training examples.

Our classifier and predictor ensembles are comprised of four different types of graph convolutional layers and we identify the most accurate models for each architecture via hyperparameter searching (Table S2).

To assess whether the model explores new chemical space, we compared molecular similarity distributions across the experimental dataset, the generated pool, and the predicted $N_F$ subset using Tanimoto coefficients of Morgan fingerprints (Fig. 2a). As expected, generated structures are more diverse than the experimental set, while predicted $N_F$ candidates cluster closer to experimental $N_F$ molecules, indicating that the model prioritises plausible $N_F$ like scaffolds while expanding chemical diversity beyond known motifs

Principal component analysis (PCA) of molecular fingerprints further shows that generated and predicted $N_F$ molecules populate regions adjacent to (but extending beyond) the experimentally explored $N_F$ space (Fig. 2b), confirming that the generative model proposes novel yet chemically realistic structures.

To gain insight into model decisions, we applied the substructure-masking explanation method (SME) [49] to quantify the influence of functional groups on $N_F$ predictions (Fig. 2c). Removing specific fragments and recomputing predictions yields a per-fragment contribution score, identifying structural features that promote or suppress $N_F$ behaviour. Combining high-contribution motifs provides a rational, interpretable route to designing candidate molecules (Fig. 2d), complementing chemical intuition.

**Synthesis and experimental validation of *de novo* compounds**

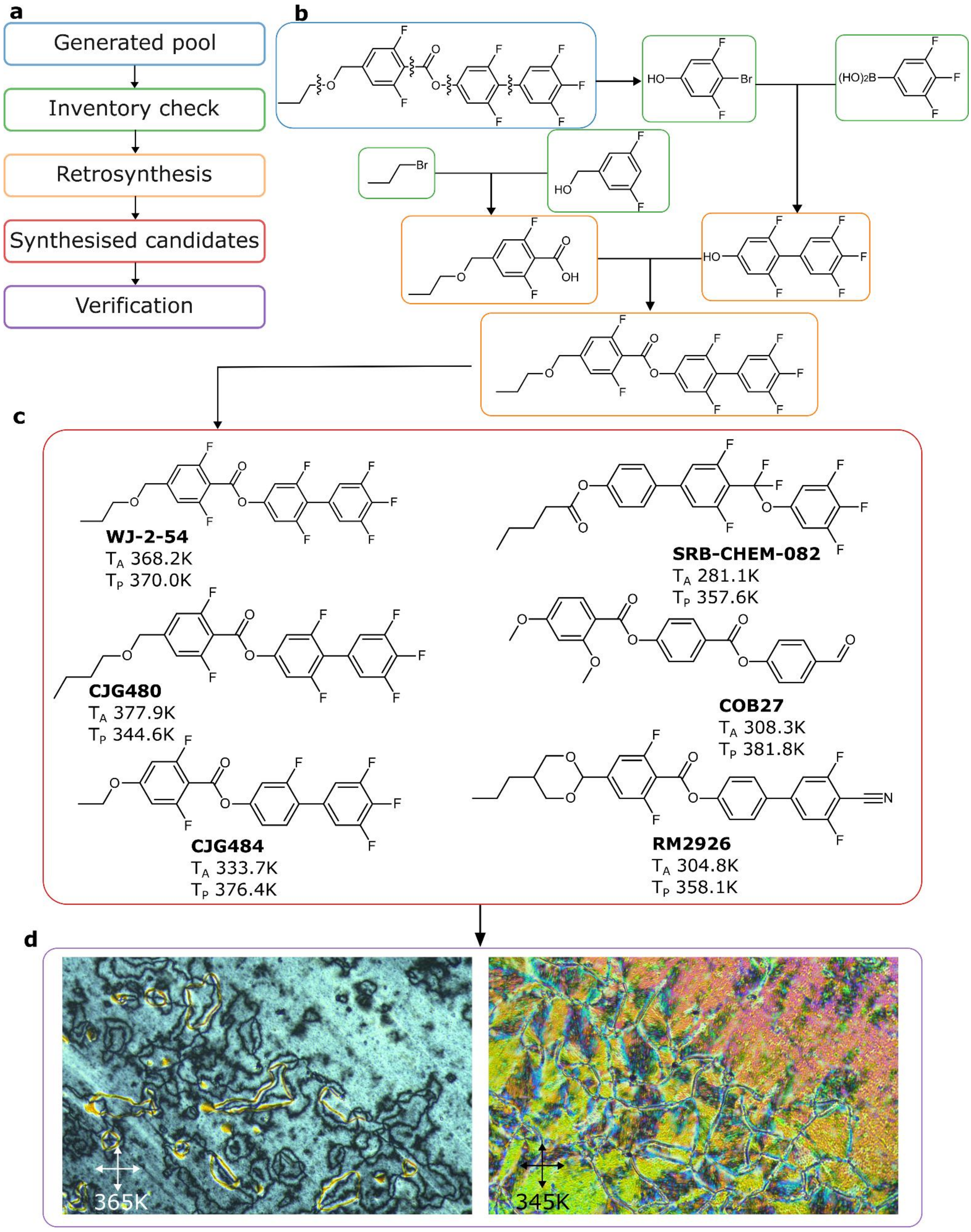

**Figure 3.** From machine learning predictions to experimental verification. A) The stages of the workflow that takes the molecules from generated to physical materials. b) Retrosynthesis coupled with a laboratory stock inventory assists in the identification of readily synthesisable molecules within the generated pool. c) Synthesised $N_F$ structures and their extrapolated ($T_A$) and predicted ($T_P$) $N_F$ transition temperatures. d) Once synthesized, POM analysis of 90wt% DIO: 10 wt% guest mixtures are used to extrapolate $N_F$ transition temperatures

To select synthetically tractable targets from the generated pool we employed AiZynthFinder [48] to identify feasible retrosynthetic routes. The requires precursors were cross referenced with our in-house digitised chemical inventory (Fig. 4b), yielding a list of materials which could be rapidly accessible. Synthesis candidates were subsequently reviewed by synthetic chemists to confirm feasibility. From this process, we synthesised 11 previously unreported materials. All materials obtained in >99% purity (peak area) as judged by C18-HPLC analysis; full synthetic details and chemical characterisation data are given in the ESI to this article. We utilise extrapolated $T_{NF}$ values to remove the explicit dependency on melting point and supercooling range. The structures of a selection of these synthesised $N_F$ candidates are given in figure 5d with their extrapolated actual transition temperatures ($T_A$; K) and predicted transition temperatures ($T_P$; K). The resulting dataset provides a direct test of the generative-predictive pipeline. The molecules can be assessed and used to augment the original experimental dataset, providing further structural data from which the models can learn from in further iterations of this process, thus continuing to guide molecular discovery. Molecules with a high chemical similarity to existing $N_F$ materials (CJG480, CJG484, WJ- 2-54) have extrapolated $N_F$ transition temperatures within ~ 40 K of the predicted values. There is a growth in error as the structures increasingly differ from experimental dataset culminating in a maximum discrepancy between predicted and actual $T_{NF}$ of 76 K (SRB-CHEM-082). Full details of extrapolated $N_F$ transitions for the complete set of synthesised molecules can be found in the supplementary information. The novel structural and transitional data is added to the $N_F$ dataset to supplement further predictions.

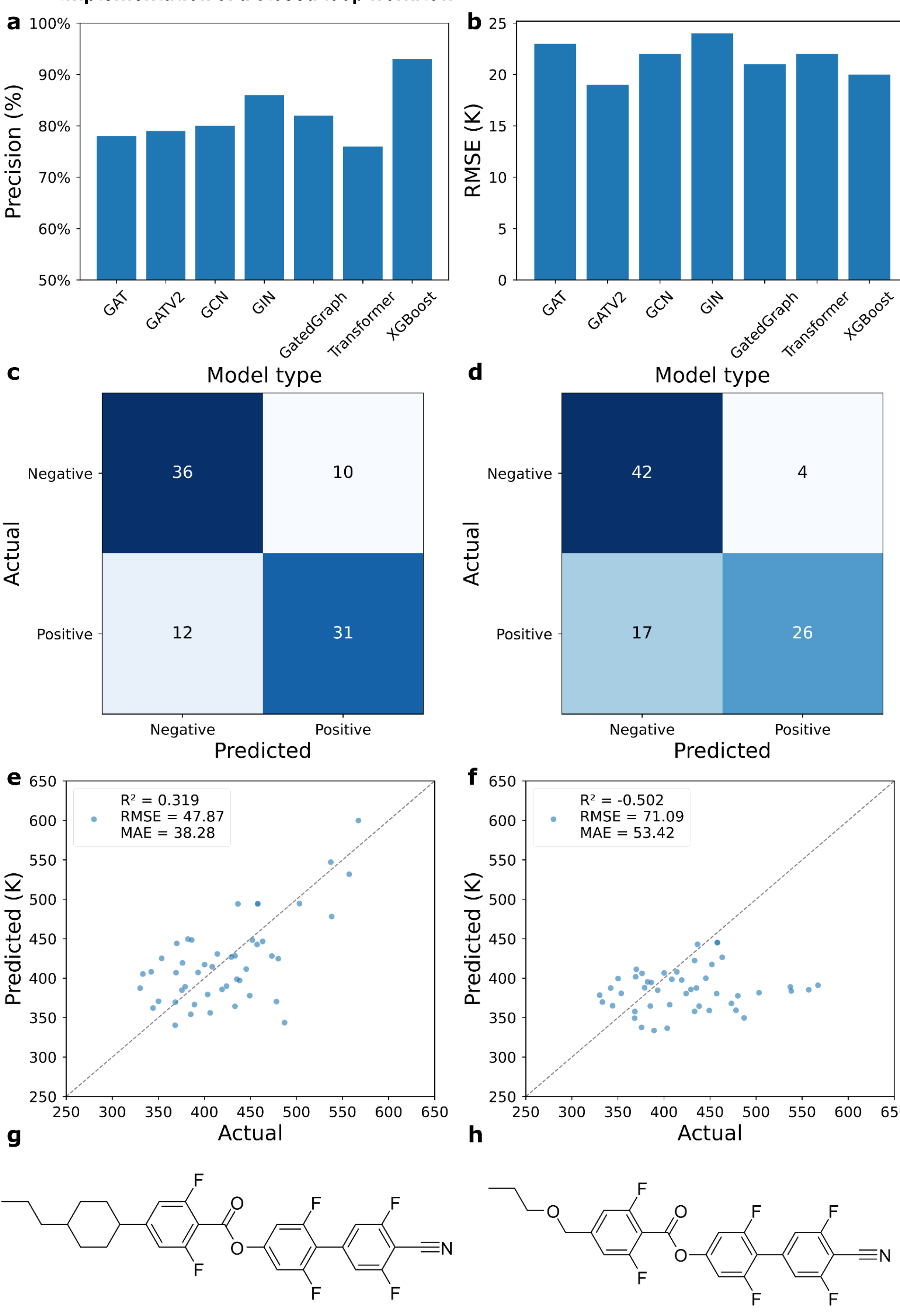

**Figure 4.** Model performance analysis and candidate structures for the second iteration through the NFML workflow. Classifier precision (a) and regressor RMSE (b) metrics for random split GNNs and XGBoost models. Confusion matrices for c) GATv2 scaffold and d) XGBoost scaffold classifiers. Predicted vs actual plots for e) GATv2 scaffold and f) XGBoost scaffold regressors. Structures of synthesis candidates annotated with predicted $N_F$ transition temperatures and the standard deviation across ensemble predictions. Candidates were selected by consensus filtering between the scaffold-split trained GATv2 and XGBoost classifiers. Transition temperatures were predicted by an ensemble of GATv2 regressors.

In the previous section, the eleven synthesis candidates functioned as an experimental probe of the GNN ensemble predictions, these revealing a performance gap between model predictions and experimentally realized transition temperatures. We elected to augment the $N_F$ dataset with additional 200 materials taken from the literature [50-56] published after our initial cutoff date (October 2025), as well as the 11 de novo materials from this work and additional in-house measurements. At this point, the dataset comprises ~ 1000 entries. We then perform a second iteration of the model training pipeline, implementing several methodological additions. Firstly, we further partitioned the data to include a held-out test set against which the in-silico performance of the GNN models was evaluated, allowing us to perform Optuna [57] hyperparameter searches tested against metrics of the validation folds. Secondly, we trained XGBoost classifiers and regressors to predict the $N_F$ transition temperature. Third, ensembles were trained for the six GNN architectures (GAT, GATv2, GCN, GIN, GatedGraph and TransformerConv) and XGBoost models with the latter being assembled from selection of molecular fingerprint representations. The fingerprints used in the XGBoost ensembles are given section 2 of the ESI. The ensemble sizes for the GNNs were increased from the first iteration (30 models for each architecture vs a 12-model selection) to reduce the variance of the predictions. Scaffold splitting of the dataset was introduced to reduce the structural data leakage between the training, validation and testing splits and from this scaffold-trained classifiers and regressors were produced for each architecture.

Figure 4a and 4b present the precision and RMSE for the GNN models trained on a random split of the dataset. Classifiers were evaluated through their precision and recall metrics. All random split models possess a precision of above 75% and a recall above 90%. The random split XGBoost models have a greater precision for classification tasks whereas the GATv2 model performs the best for RMSE. We consider the precision to be the more significant metric for assessing the viability of synthesis candidates as a false positive is a waste of resources. The false negative reduces the predicted candidate pool, but given this is many tens of thousands of molecules in size we deem this acceptable.

Figures 4c and 4d compare the confusion matrices of the scaffold split trained c) GATv2 and d) XGBoost classifiers. The XGBoost classifier has a greater precision, 86% vs 75% respectively, but predicts fewer actual positives than the GATV2 classifier. Demonstrated by the XGBoost classifier possessing a recall of 60% vs the GATv2 classifier's 72%. Figures 4e and 4f compare the predicted vs actual figures for the e) GATv2 and f) XGBoost regressors. For a $T_{NF}$ of over 500K the XGBoost regressor fails to give accurate predictions, whereas the GATv2 regressor can more accurately predict on the same dataset. Scaffold models except for GATv2 and XGBoost are discarded for possessing a recall performance of <60%. All scaffold regressors except for GATv2 were discarded from predictions ensembles due to negative $R^2$ values.

The scaffold-vs-random comparison along with the experimental results shows that random split models overstate their generalisability to unseen data. GATv2 and XGBoost are the only classifier models that show an acceptable level of performance degradation when trained on scaffold splits of the data, suggesting that these architectures are better able to capture the structural features that govern $N_F$ phase formation. These results show that the split strategy and model architecture should be considered when assessing a molecule's suitability for synthesis. Confusion matrices and predicted vs actual plots for all model architectures and splitting types can be found in section 6 of the ESI.

Two sets of prediction candidates filtered by molecules with accessible retrosynthetic routes are presented in section 7 of the ESI. Set A is constructed from the predictions of all random split model architectures and rely on a cross-architecture consensus for the validation of their predictions. B) is constructed from models trained on a scaffold split of the data with XGBoost and GATv2 classifiers and the GATv2 regressor being selected for use based on the analysis of their performance metrics performed above. The sets were constructed by evaluating the total votes each molecule receives and filtered into a final set of NF candidates filtered by strong consensus (60% ensemble confidence) to yield a tractable final pool of molecules. These molecules were then passed to regressors to predict their range of $N_F$ transition temperatures. The predicted candidates that survive this filtering are therefore consistent across model choice and molecular representation. All molecules within these retrosynthesis-filtered sets possess a Tanimoto similarity to the training dataset of above 0.6, indicating that there's a moderate structural similarity to the existing $N_F$ dataset and there is confidence in the structures falling within the applicability domain of the models. Candidate sets without retrosynthesis filtering can be found at can be found in the GitHub repository linked at the end of this manuscript. Figures 4g and h presents notable candidate molecules from set B along with their predicted $N_F$ transition temperatures and the standard deviation across ensemble predictions.

The experimental outcome of the first iteration led us to implement the stated methodological additions, and we believe produce a greater confidence in the workflow's predictions.

## Conclusion

A generative-predictive framework for data-driven design and experimental validation of small molecules is presented. By integrating molecular graph learning, ensemble prediction, retrosynthetic filtering, and laboratory synthesis, this provides a practical end-to-end approach for accelerating materials discovery.

Applied to the discovery of ferroelectric nematic liquid crystals, we use this workflow to propose synthetically accessible candidates which are then synthesised and subjected to experimental verification. Several possess extrapolated $N_F$ transitions within reasonable boundaries, demonstrating that machine-learned chemical intuition can successfully guide molecular design in a domain where subtle structural changes govern emergent properties.

The synthesised molecules were added to the training data for a second iteration through the workflow, producing two sets of synthesis candidates. Set A was selected by consensus across random-split trained classifiers and regressors. Set B was selected by consensus between scaffold-split trained GATv2 and XGBoost classifiers with $N_F$ predicted by an ensemble of GATv2 regressors

This work illustrates a generalisable strategy for closed-loop molecular innovation: learn from experimental structure–property data, generate candidates within a realistic chemical manifold, triage using model consensus and retrosynthetic constraints, and validate in the laboratory via chemical synthesis and experimental physical characterisation. We expect this approach to extend to other classes of functional soft materials and to accelerate the rational exploration of chemical space. Moreover, the integration of comparable workflows with high throughput synthesis and characterisation offers the prospect of autonomous, machine led, exploration of chemical space.

### Code and data availability

The code used in this article is available at https://github.com/123cpb/nfml. The data associated with this article is openly available from the University of Leeds Data Repository at https://doi.org/10.5518/1798

### Author Contributions

The dataset was curated by CPB, JH, CJG, and RJM. CPB and RJM developed the machine learning code used in this work. Chemical synthesis was performed by SRB, COB, CJG, WJ, WCO, RJM. Characterisation of liquid crystalline behaviour was undertaken by (SRB, CPB, CJG, JH, WJ, WCO, COB, RJM). RJM and HFG secured funding. The first draft was written by CPB and RJM, with contributions from all authors.

### Acknowledgements

RJM thanks: UKRI for funding via a Future Leaders Fellowship (Grant no MR/W006391/1; CJG, JH); Merck Electronics for funding via the LAMP2 Project (WJ, COB) and discussions around materials, The University of Leeds for funding via a University Academic Fellowship and a PhD Studentship for CPB; The SOFI2 CDT for funding PhD studentships for WCO and COB.

HFG thanks: S.R.B and H.F.G acknowledge funding from the Engineering and Physical Sciences Research Council, grant number EP/V054724/1.

Computational work was undertaken on the Aire HPC system at the University of Leeds, UK.

# Deep learning directed synthesis of fluid ferroelectric materials

Charles Parton-Barr[1], Stuart R. Berrow [1], Calum J. Gibb [2], Jordan Hobbs [1], Wanhe Jiang [1], Caitlin O'Brien [1,2,3], Will C. Ogle [1,2,3], Helen F. Gleeson[1], Richard J. Mandle[1,2,*]

[1]School of Physics and Astronomy, University of Leeds, Leeds, UK, LS29JT
[2]School of Chemistry, University of Leeds, Leeds, UK, LS29JT
[3]School of Applied Mathematics, University of Leeds, Leeds, UK, LS29JT

* r.mandle@leeds.ac.uk

**Contents:**

## 1. Model Architectures

### 1.1. Variational Autoencoder (VAE)

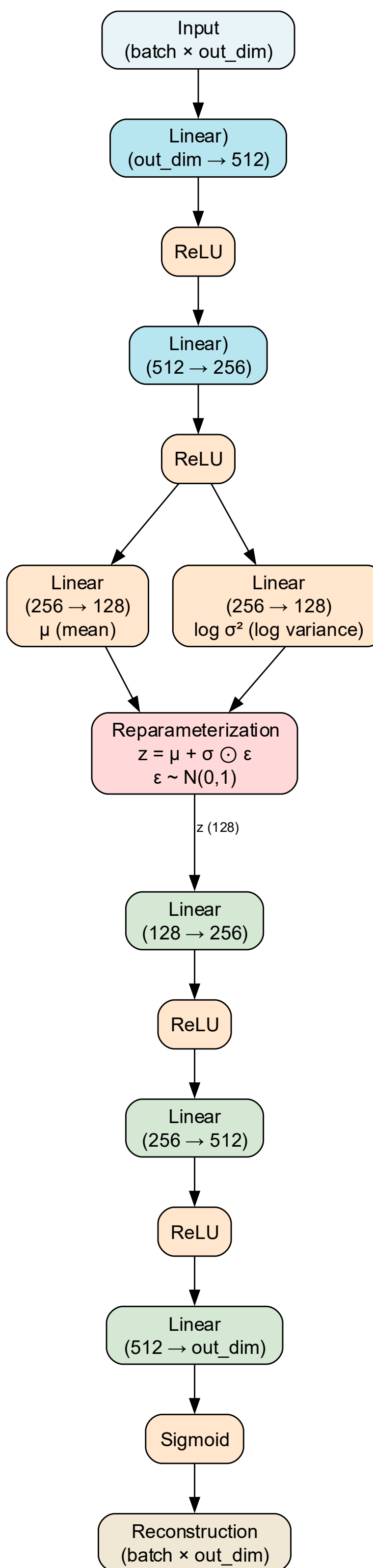

**Figure SI-1:** Schematic representation of the VAE model with example hyperparameters

### 1.2. GATClassifier

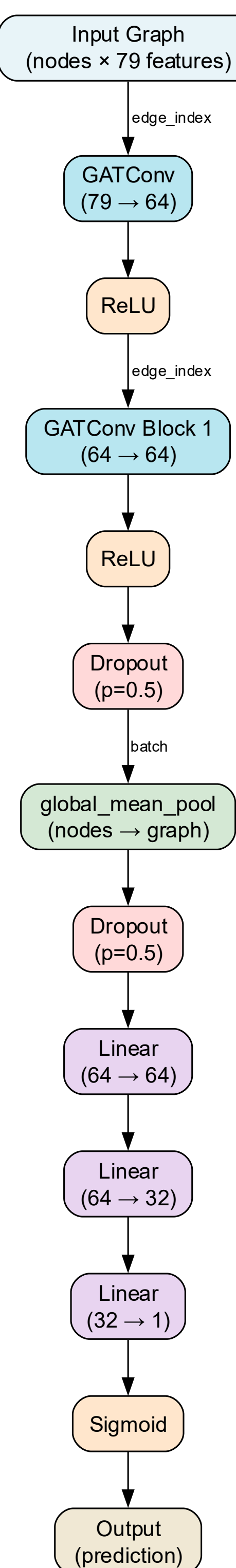

**Figure SI-2:** Schematic representation of the GATClassifier model architecture with example hyperparameters.

### 1.3. GATV2 Classifier

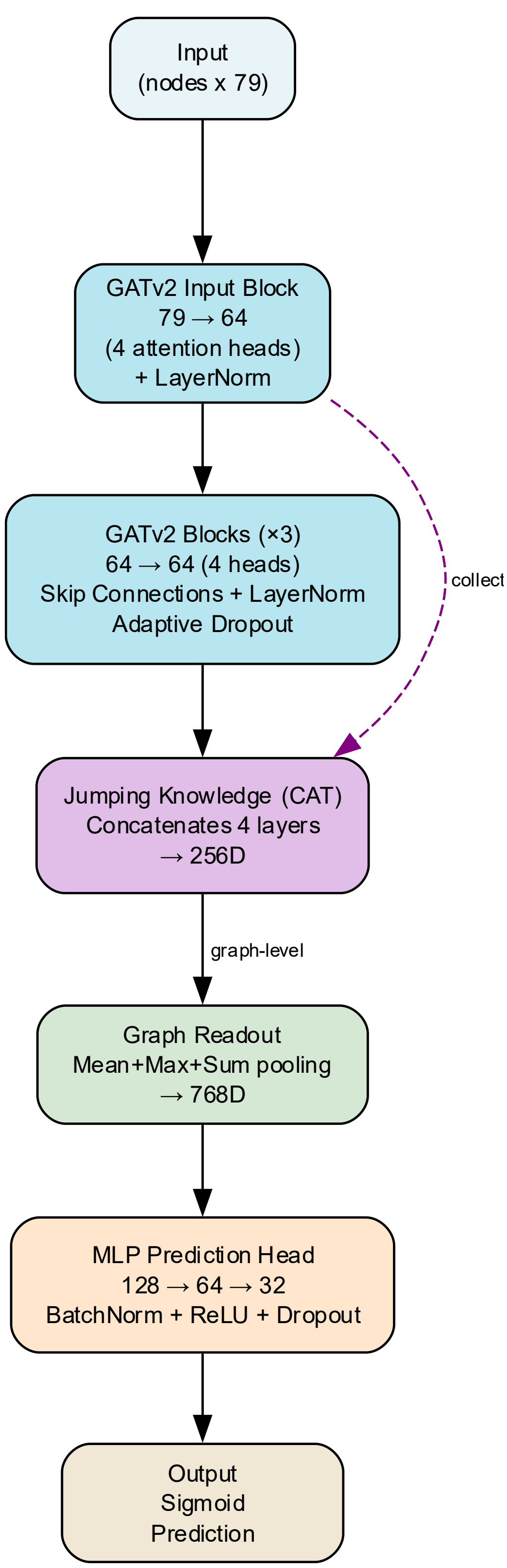

**Figure SI-3:** Schematic representation of the GATv2 Classifier model architecture with example hyperparameters. The architecture has been simplified for layout purposes.

## 1.4. GINClassifier

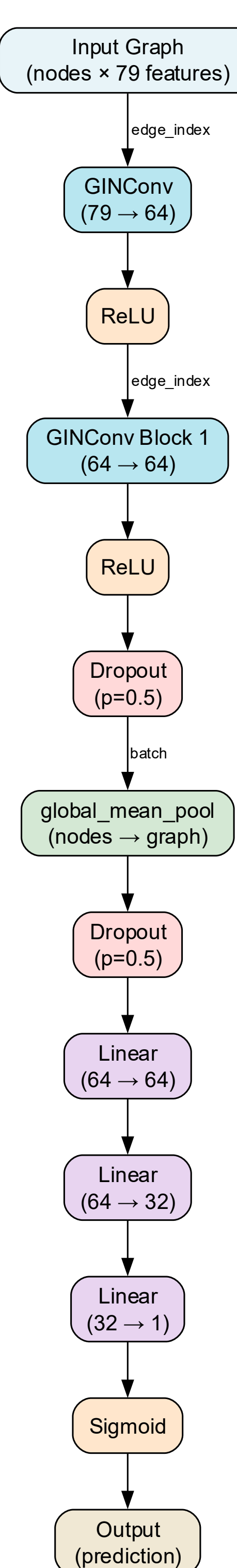

**Figure SI-4**: Schematic representation of the GINClassifier model architecture with example hyperparameters

## 1.5. GCNClassifier

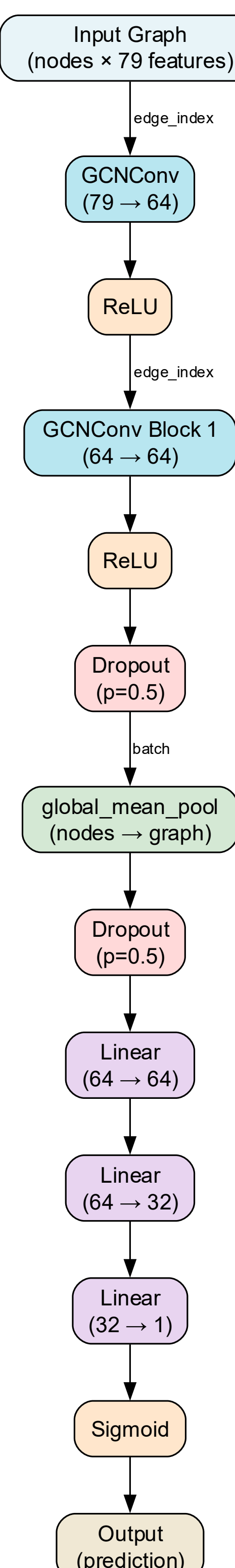

**Figure SI-5:** Schematic representation of the GCNClassifier model architecture with example hyperparameters

## 1.6. GATPredictor

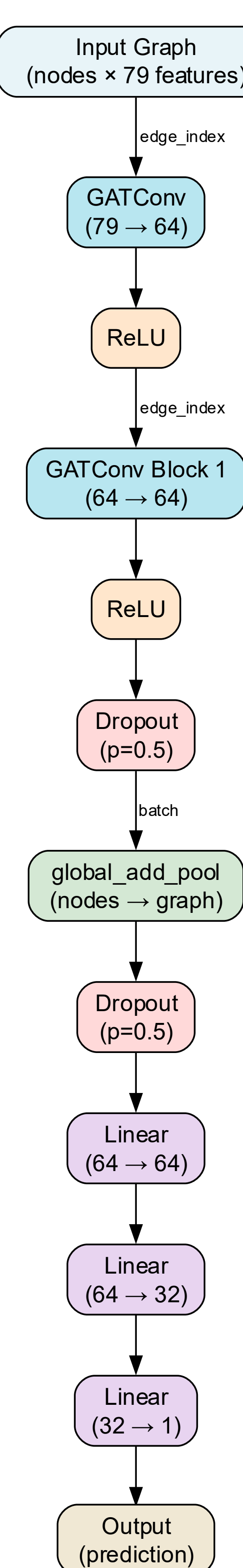

**Figure SI-6** Schematic representation of the GATPredictor model architecture with example hyperparameters

### 1.7. GATV2 Predictor

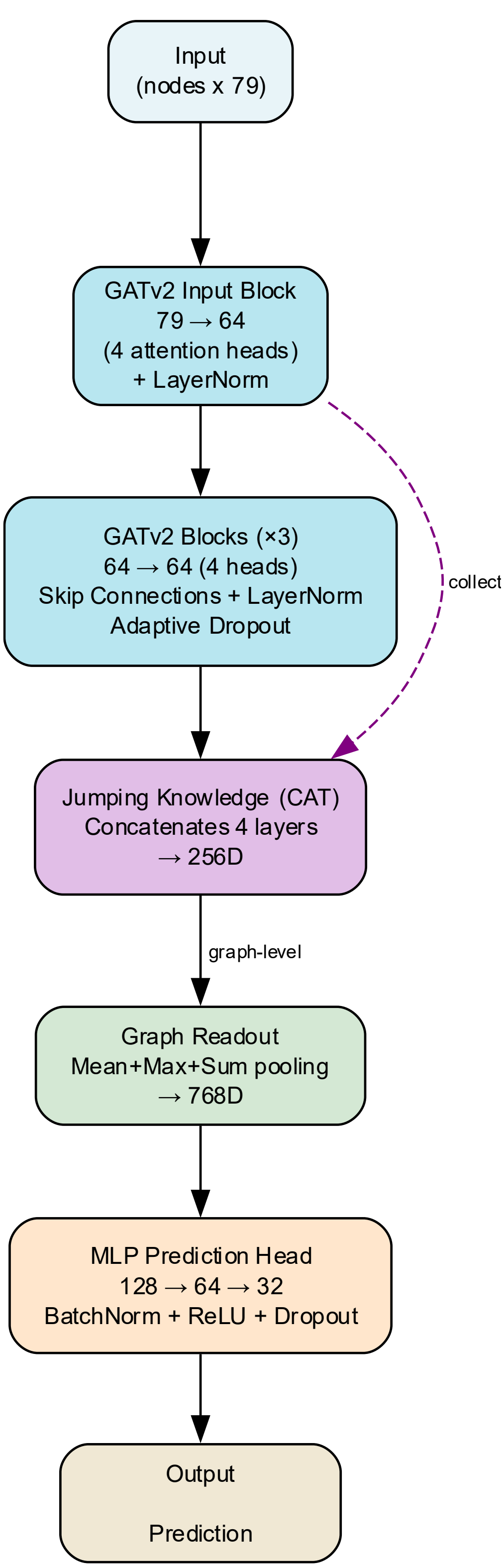

**Figure SI-7:** Schematic representation of the GATV2 predictor model architecture with example hyperparameters. The architecture has been simplified for layout purpose

## 1.8. GINPredictor

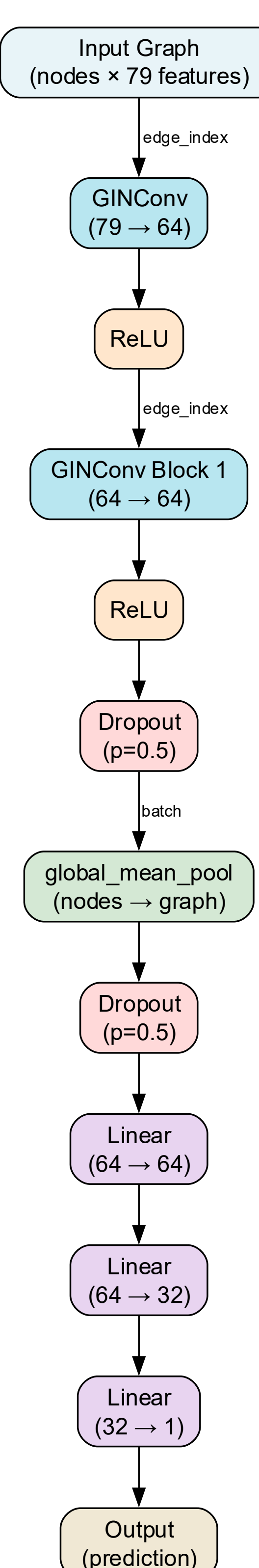

**Figure SI-8:** Schematic representation of the GINPredictor model architecture with example hyperparameters

## 1.9. GCNPredictor

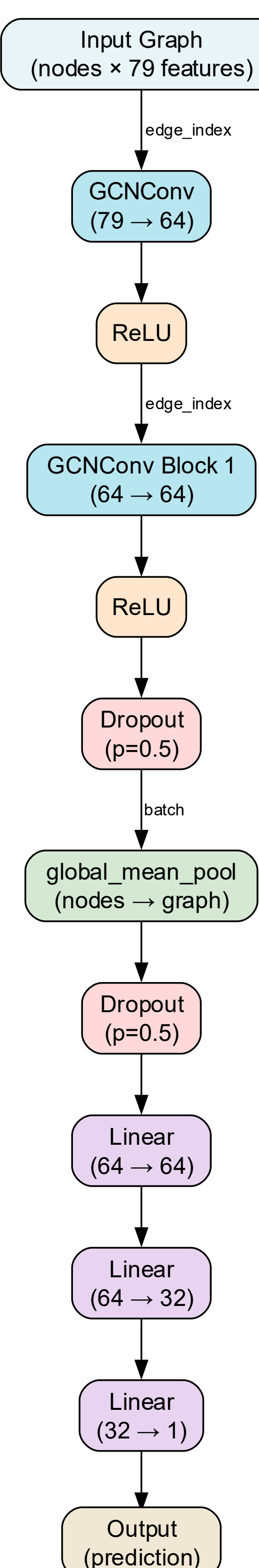

**Figure SI-9:** Schematic representation of the GCNpredictor model architecture with example hyperparameters

## 2. Model hyperparameters

#### 2.1. Iteration 1

| Model | Hyperparameters |
| --- | --- |
| 1 | Model type: Classifier<br>Convolutional blocks: 3<br>Dropout: 0.5<br>Convolutional layer hidden dimension: 256<br>Learning rate: 0.001<br>Batch size: 32<br>Epochs: 50<br>k-fold splits: 5<br>number of repeats: 3 |
| 2 | Model type: Classifier<br>Convolutional blocks: 10<br>Dropout: 0.25<br>Convolutional layer hidden dimension: 256<br>Learning rate: 0.001<br>Batch size: 16<br>Epochs: 50<br>k-fold splits: 5<br>number of repeats: 3 |
| 3 | Model type: Classifier<br>Convolutional blocks: 10<br>Dropout: 0.25<br>Convolutional layer hidden dimension: 256<br>Learning rate: 0.001<br>Batch size: 32<br>Epochs: 50<br>k-fold splits: 5<br>number of repeats: 3 |
| 4 | Model type: EnhancedClassifier<br>Convolutional blocks: 3<br>Dropout: 0.25<br>Convolutional layer hidden dimension: 256<br>Learning rate: 0.001<br>Batch size: 32<br>Epochs: 50<br>k-fold splits: 5<br>number of repeats: 3 |
| 5 | Model type: EnhancedClassifier<br>Convolutional blocks: 3<br>Dropout: 0.25<br>Convolutional layer hidden dimension: 512<br>Learning rate: 0.001 |

| | Batch size: 16<br>Epochs: 50<br>k-fold splits: 5<br>number of repeats: 3 |
|---|---|
| 6 | Model type: EnhancedClassifier<br>Convolutional blocks: 3<br>Dropout: 0.25<br>Convolutional layer hidden dimension: 512<br>Learning rate: 0.001<br>Batch size: 32<br>Epochs: 50<br>k-fold splits: 5<br>number of repeats: 3 |
| 7 | Model type: GCNClassifier<br>Convolutional blocks: 3<br>Dropout: 0.25<br>Convolutional layer hidden dimension: 512<br>Learning rate: 0.001<br>Batch size: 32<br>Epochs: 50<br>k-fold splits: 5<br>number of repeats: 3 |
| 8 | Model type: GCNClassifier<br>Convolutional blocks: 3<br>Dropout: 0.25<br>Convolutional layer hidden dimension: 256<br>Learning rate: 0.001<br>Batch size: 16<br>Epochs: 50<br>k-fold splits: 5<br>number of repeats: 3 |
| 9 | Model type: GCNClassifier<br>Convolutional blocks: 3<br>Dropout:0.25<br>Convolutional layer hidden dimension: 512<br>Learning rate:0.001<br>Batch size: 16<br>Epochs: 50<br>k-fold splits: 5<br>number of repeats: 3 |
| 10 | Model type: GINClassifier<br>Convolutional blocks: 3<br>Dropout: 0.5 |
| | Convolutional layer hidden dimension: 256<br>Learning rate: 0.001<br>Batch size: 32<br>Epochs: 25<br>k-fold splits: 5<br>number of repeats: 3 |
| 11 | Model type: GINClassifier<br>Convolutional blocks: 3<br>Dropout: 0.5<br>Convolutional layer hidden dimension: 512<br>Learning rate: 0.001<br>Batch size: 32<br>Epochs: 50<br>k-fold splits: 5<br>number of repeats: 3 |
| 12 | Model type: GINClassifier<br>Convolutional blocks: 3<br>Dropout: 0.5<br>Convolutional layer hidden dimension: 512<br>Learning rate: 0.001<br>Batch size: 32<br>Epochs:50<br>k-fold splits:3<br>number of repeats: 5 |

**Table SI-1:** Hyperparameters of the twelve classifier-type models that make up our iteration 1 voting ensemble

| Model | Hyperparameters |
|---|---|
| 1 | Model type: Predictor<br>Convolutional blocks: 3<br>Dropout: 0.25<br>Convolutional layer hidden dimension: 256<br>Learning rate: 0.001<br>Batch size:16<br>Epochs: 50<br>k-fold splits: 5<br>number of repeats: 3 |
| 2 | Model type: Predictor<br>Convolutional blocks: 3<br>Dropout: 0.25<br>Convolutional layer hidden dimension: 512<br>Learning rate: 0.001<br>Batch size: 32<br>Epochs: 50<br>k-fold splits: 5<br>number of repeats: 3 |
| 3 | Model type: Predictor<br>Convolutional blocks: 3<br>Dropout: 0.5<br>Convolutional layer hidden dimension: 512<br>Learning rate: 0.001<br>Batch size: 32<br>Epochs: 50<br>k-fold splits: 5<br>number of repeats: 3 |
| 4 | Model type: EnhancedPredictor<br>Convolutional blocks: 3<br>Dropout: 0.25<br>Convolutional layer hidden dimension: 256<br>Learning rate: 0.001<br>Batch size: 16<br>Epochs: 50<br>k-fold splits: 5<br>number of repeats: 3 |
| 5 | Model type: EnhancedPredictor<br>Convolutional blocks: 3<br>Dropout: 0.25<br>Convolutional layer hidden dimension: 256<br>Learning rate: 0.001<br>Batch size: 32 |

| | Epochs: 50<br>k-fold splits: 5<br>number of repeats: 3 |
|---|---|
| 6 | Model type: EnhancedPredictor<br>Convolutional blocks: 3<br>Dropout: 0.25<br>Convolutional layer hidden dimension: 512<br>Learning rate: 0.001<br>Batch size: 32<br>Epochs: 50<br>k-fold splits: 5<br>number of repeats: 3 |
| 7 | Model type: GCNPredictor<br>Convolutional blocks: 10<br>Dropout: 0.25<br>Convolutional layer hidden dimension: 256<br>Learning rate: 0.001<br>Batch size: 16<br>Epochs: 50<br>K-fold splits: 5<br>Number of repeats: 3 |
| 8 | Model type: GCNPredictor<br>Convolutional blocks: 3<br>Dropout: 0.25<br>Convolutional layer hidden dimension: 512<br>Learning rate: 0.001<br>Batch size: 16<br>Epochs: 50<br>k-fold splits: 5<br>Number of repeats: 3 |
| 9 | Model type: GCNPredictor<br>Convolutional blocks: 3<br>Dropout: 0.25<br>Convolutional layer hidden dimension: 512<br>Learning rate: 0.001<br>Batch size: 16<br>Epochs: 50<br>k-fold splits: 5<br>number of repeats: 3 |
| 10 | Model type: GINPredictor<br>Convolutional blocks: 3<br>Dropout: 0.5<br>Convolutional layer hidden dimension: 512 |

| | Learning rate: 0.001<br>Batch size: 32<br>Epochs: 50<br>k-fold splits: 5<br>number of repeats: 3 |
|---|---|
| 11 | Model type: GINPredictor<br>Convolutional blocks: 3<br>Dropout: 0.25<br>Convolutional layer hidden dimension: 256<br>Learning rate: 0.001<br>Batch size: 16<br>Epochs: 50<br>k-fold splits: 5<br>number of repeats: 3 |
| 12 | Model type: GINPredictor<br>Convolutional blocks: 3<br>Dropout: 0.25<br>Convolutional layer hidden dimension: 256<br>Learning rate:0.001<br>Batch size:32<br>Epochs: 50<br>k-fold splits: 5<br>number of repeats: 3 |

**Table SI-2**: Hyperparameters of the twelve predictor-type models that make up our mean ensemble

**2.2. Iteration 2**

| GATv2_classifier_random | Model type: GATv2Classifier<br>Convolutional blocks: 4<br>Split type: random<br>Dropout: 0.06453371686984193<br>Convolutional layer hidden dimension: 256<br>Learning rate: 0.0002963783947778693<br>Batch size:128<br>Epochs: 134<br>k-fold splits: 5<br>number of repeats: 3 |
|---|---|
| GATv2_classifier_scaffold | Model type: GATv2Classifier<br>Convolutional blocks: 7<br>Split type: scaffold<br>Dropout: ": 0.08572640535698162<br>Convolutional layer hidden dimension: 512<br>Learning rate: 0.00014598173992939992<br>Batch size:32 |

| | Epochs: 77<br>k-fold splits: 5<br>number of repeats: 1 |
|---|---|
| GATv2_regression_random | Model type: GATv2Predictor<br>Convolutional blocks: 3<br>Split type: random<br>Dropout: 0.050648010737235766<br>Convolutional layer hidden dimension: 256<br>Learning rate: 0.001326755059733017<br>Batch size:128<br>Epochs: 164<br>k-fold splits: 5<br>number of repeats: 3 |
| GATv2_regression_scaffold | Model type: GATv2Classifier<br>Convolutional blocks: 3<br>Split type: scaffold<br>Dropout: 0.07262193184426281<br>Convolutional layer hidden dimension: 2048<br>Learning rate: 0.0005627555836364237<br>Batch size:128<br>Epochs: 1345<br>k-fold splits: 5<br>number of repeats: 1 |
| GAT_classification_random | Model type: GATClassifier<br>Convolutional blocks: 4<br>Split type: random<br>Dropout: 0.36560519439636197<br>Convolutional layer hidden dimension: 512<br>Learning rate: 0.0019720784929833294<br>Batch size:128<br>Epochs: 178<br>k-fold splits: 5<br>number of repeats: 3 |
| GAT_regression_random | Model type: GATClassifier<br>Convolutional blocks: 2<br>Split type: random<br>Dropout: 0.12851254578153898<br>Convolutional layer hidden dimension: 128<br>Learning rate: 0.0027288673346892948<br>Batch size:64<br>Epochs: 134<br>k-fold splits: 5<br>number of repeats: 3 |

| GG_classification_random | Model type: GatedGraphClassifier<br>Convolutional blocks: 2<br>Split type: random<br>Dropout: 0.3719301907573434<br>Convolutional layer hidden dimension: 256<br>Learning rate: 0.00011685814088177345<br>Batch size:16<br>Epochs: 297<br>k-fold splits: 5<br>number of repeats: 3 |
|---|---|
| GG_predictor_random | Model type: GatedGraphPredictor<br>Convolutional blocks: 2<br>Split type: random<br>Dropout: 0.29667032373742863<br>Convolutional layer hidden dimension: 256<br>Learning rate: 0.00046414581776362673<br>Batch size:32<br>Epochs: 211<br>k-fold splits: 5<br>number of repeats: 3 |
| GCN_classification_random | Model type: GCNClassifier<br>Convolutional blocks: 3<br>Split type: random<br>Dropout: 0.36856148280182427<br>Convolutional layer hidden dimension: 256<br>Learning rate: 0.0020097134871543154<br>Batch size:16<br>Epochs: 222<br>k-fold splits: 5<br>number of repeats: 3 |
| GCN_predictor_random | Model type: GCNPredictor<br>Convolutional blocks: 2<br>Split type: random<br>Dropout: 0.21407488272172256<br>Convolutional layer hidden dimension: 2048<br>Learning rate: 0.0015880323955553737<br>Batch size:32<br>Epochs: 278<br>k-fold splits: 5<br>number of repeats: 3 |
| GIN_classification_random | Model type: GINClassifier<br>Convolutional blocks: 2<br>Split type: random |
| | Dropout: 0.33620310886845617<br>Convolutional layer hidden dimension: 2048<br>Learning rate: 0.0002963783947778693<br>Batch size:16<br>Epochs: 193<br>k-fold splits: 5<br>number of repeats: 3 |
| GIN_predictor_random | Model type: GINPredictor<br>Convolutional blocks: 2<br>Split type: random<br>Dropout: 0.18981454545142928<br>Convolutional layer hidden dimension: 2048<br>Learning rate: 0.00013120667432220057<br>Batch size:32<br>Epochs: 254<br>k-fold splits: 5<br>number of repeats: 3 |
| Transformer_classifier_random | Model type: TransformerClassifier<br>Convolutional blocks: 2<br>Split type: random<br>Dropout: 0.07820827682277663<br>Convolutional layer hidden dimension: 1024<br>Learning rate: 0.0003861495044875059<br>Batch size:16<br>Epochs: 162<br>k-fold splits: 5<br>number of repeats: 3 |
| Transformer_regressor_random | Model type: TransformerPredictor<br>Convolutional blocks: 2<br>Split type: random<br>Dropout: 0.10148251352352801<br>Convolutional layer hidden dimension: 256<br>Learning rate: 0.0008401731273741588<br>Batch size:32<br>Epochs: 235<br>k-fold splits: 5<br>number of repeats: 3 |
| XGBoost_classifier_random | Model type: GATv2Classifier<br>Split type: random<br>{<br>"mordred": {<br>"fp_type": "mordred",<br>"n_estimators": 1500,<br>"max_depth": 10, |

```
    "learning_rate": 0.031486170398999376,
    "subsample": 0.5582330181548657,
    "colsample_bytree": 0.693347452454977,
    "min_child_weight": 1,
    "gamma": 3.5976471689023346,
    "reg_alpha": 0.0014175360348497963,
    "reg_lambda": 0.014462580415499563
  },
  "map": {
    "fp_type": "map",
    "n_estimators": 1300,
    "max_depth": 7,
    "learning_rate": 0.08485430190997417,
    "subsample": 0.7778245492737748,
    "colsample_bytree": 0.8746696788598094,
    "min_child_weight": 2,
    "gamma": 1.1685285928180547,
    "reg_alpha": 0.07732776208662087,
    "reg_lambda": 0.003824234652671362
  },
  "atom_pair": {
    "fp_type": "atom_pair",
    "n_estimators": 1300,
    "max_depth": 12,
    "learning_rate": 0.039633884790200496,
    "subsample": 0.6658084625171566,
    "colsample_bytree": 0.9155220309089538,
    "min_child_weight": 1,
    "gamma": 1.514055060779356,
    "reg_alpha": 2.111593329069601e-06,
    "reg_lambda": 5.91851873119353e-07
  },
  "layered": {
    "fp_type": "layered",
    "n_estimators": 300,
    "max_depth": 8,
    "learning_rate": 0.08432952262963654,
    "subsample": 0.849471335172441,
    "colsample_bytree": 0.6407220063216659,
    "min_child_weight": 1,
    "gamma": 0.3144416035463985,
    "reg_alpha": 4.677451066046656e-06,
    "reg_lambda": 2.782542042341084e-06
  },
  "rdkit": {
    "fp_type": "rdkit",
    "n_estimators": 1500,
```

| | |
|---|---|
| | "max_depth": 9,<br>"learning_rate": 0.05560026540900335,<br>"subsample": 0.7073338793838971,<br>"colsample_bytree": 0.8230826480029119,<br>"min_child_weight": 1,<br>"gamma": 0.02520464715498605,<br>"reg_alpha": 4.835183307728258e-07,<br>"reg_lambda": 5.161725451432342e-07<br>}<br>} |
| XGBoost_classifier_scaffold | Model type: XGBoost<br>Split type: Scaffold<br>{<br>"atom_pair": {<br>"fp_type": "atom_pair",<br>"n_estimators": 500,<br>"max_depth": 12,<br>"learning_rate": 0.2252676609591457,<br>"subsample": 0.8914245301735767,<br>"colsample_bytree": 0.9227362849274916,<br>"min_child_weight": 5,<br>"gamma": 0.17155659300385515,<br>"reg_alpha": 4.012475167509174e-06,<br>"reg_lambda": 7.061523689928079e-07<br>},<br>"estate": {<br>"fp_type": "estate",<br>"n_estimators": 500,<br>"max_depth": 8,<br>"learning_rate": 0.2581435651469605,<br>"subsample": 0.5121468703862474,<br>"colsample_bytree": 0.7545536011908952,<br>"min_child_weight": 10,<br>"gamma": 2.928675493341906,<br>"reg_alpha": 0.0008302818959886782,<br>"reg_lambda": 0.00025296745746954314<br>},<br>"autocorr": {<br>"fp_type": "autocorr",<br>"n_estimators": 500,<br>"max_depth": 3,<br>"learning_rate": 0.16192202099242461,<br>"subsample": 0.5878883162992878,<br>"colsample_bytree": 0.8012978303987651,<br>"min_child_weight": 7,<br>"gamma": 0.2899015059505158,<br>"reg_alpha": 0.006131395879664587, |

<table>
<tr>
<td></td>
<td>"reg_lambda": 0.0033148885862714582<br>},<br>"mordred_3d": {<br>"fp_type": "mordred_3d",<br>"n_estimators": 1500,<br>"max_depth": 8,<br>"learning_rate": 0.26267206761930645,<br>"subsample": 0.5202521009243568,<br>"colsample_bytree": 0.5658158564399262,<br>"min_child_weight": 7,<br>"gamma": 3.9741467802626844,<br>"reg_alpha": 0.0021846357624013147,<br>"reg_lambda": 1.220676651751883e-06<br>},<br>"layered": {<br>"fp_type": "layered",<br>"n_estimators": 1300,<br>"max_depth": 11,<br>"learning_rate": 0.24074448556600725,<br>"subsample": 0.6408954049802805,<br>"colsample_bytree": 0.7071581300464582,<br>"min_child_weight": 7,<br>"gamma": 1.5480074057483273,<br>"reg_alpha": 7.153871966850925e-05,<br>"reg_lambda": 0.002319740413099216<br>}<br>}</td>
</tr>
<tr>
<td>XGBoost_regressor_random</td>
<td>Model type: xgboost<br>Split type: scaffold<br>{<br>"atom_pair": {<br>"fp_type": "atom_pair",<br>"n_estimators": 1400,<br>"max_depth": 7,<br>"learning_rate": 0.01139029138018265,<br>"subsample": 0.6187762647689145,<br>"colsample_bytree": 0.5163423366099305,<br>"min_child_weight": 7,<br>"gamma": 0.0047902841587389566,<br>"reg_alpha": 3.448864897914892e-08,<br>"reg_lambda": 1.1684157039695513<br>},<br>"map": {<br>"fp_type": "map",<br>"n_estimators": 2000,<br>"max_depth": 4,<br>"learning_rate": 0.029154010688616866,</td>
</tr>
</table>

```
    "subsample": 0.5064826179376595,
    "colsample_bytree": 0.9284484509260027,
    "min_child_weight": 2,
    "gamma": 0.010154210459249465,
    "reg_alpha": 0.16391342146504914,
    "reg_lambda": 0.0707713891409971
  },
  "mordred": {
    "fp_type": "mordred",
    "n_estimators": 700,
    "max_depth": 3,
    "learning_rate": 0.022829069171836226,
    "subsample": 0.853411781767488,
    "colsample_bytree": 0.6992139228895411,
    "min_child_weight": 2,
    "gamma": 0.09274179346019618,
    "reg_alpha": 0.022430192324124215,
    "reg_lambda": 0.011924065914937148
  },
  "mordred_3d": {
    "fp_type": "mordred_3d",
    "n_estimators": 1100,
    "max_depth": 3,
    "learning_rate": 0.02496670085732168,
    "subsample": 0.5044885725127056,
    "colsample_bytree": 0.9731535218925407,
    "min_child_weight": 5,
    "gamma": 4.96855661647963e-05,
    "reg_alpha": 0.046245713871923326,
    "reg_lambda": 6.5655449961715275
  },
  "autocorr": {
    "fp_type": "autocorr",
    "n_estimators": 1600,
    "max_depth": 4,
    "learning_rate": 0.09147304911466263,
    "subsample": 0.5154845690832526,
    "colsample_bytree": 0.9715306684890168,
    "min_child_weight": 10,
    "gamma": 0.002048768404310189,
    "reg_alpha": 0.4150146349409657,
    "reg_lambda": 2.3384875158828744
  }
}
```

**Table SI-1:** Hyperparameters of the model ensebles that contribute to iteration 2 predictions

## 3. Substructure distribution histograms

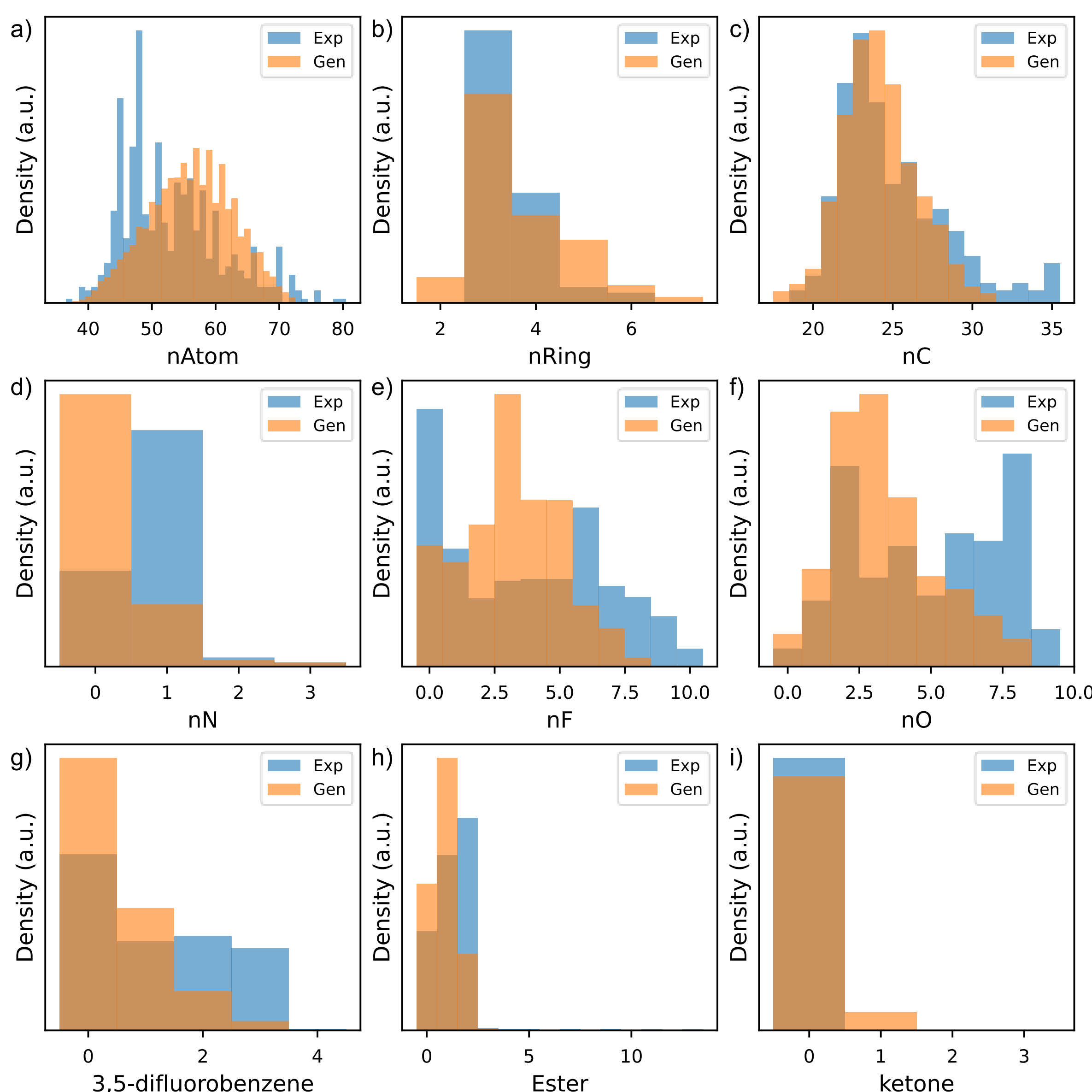

**Figure SI-10:** Comparing the per molecule probability distributions of descriptor and substructure counts for the experimental and generated dataset. A) number of atoms. B) number of rings. C) number of carbon atoms. D) number of nitrogen atoms. E) number of fluorine atoms. F) number of oxygen atoms. g) number of 3,5-difluorobenzene substructures. h) number of ester substructures. i) number of ketone substructures.

## 4. Mixture transition textures

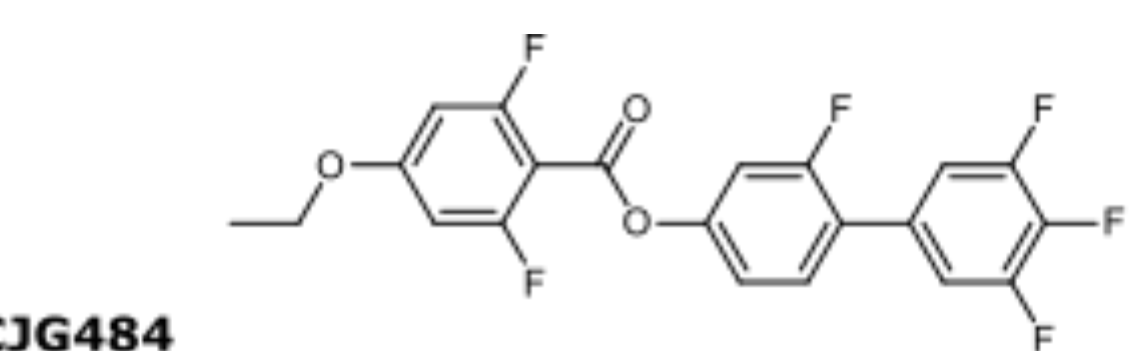

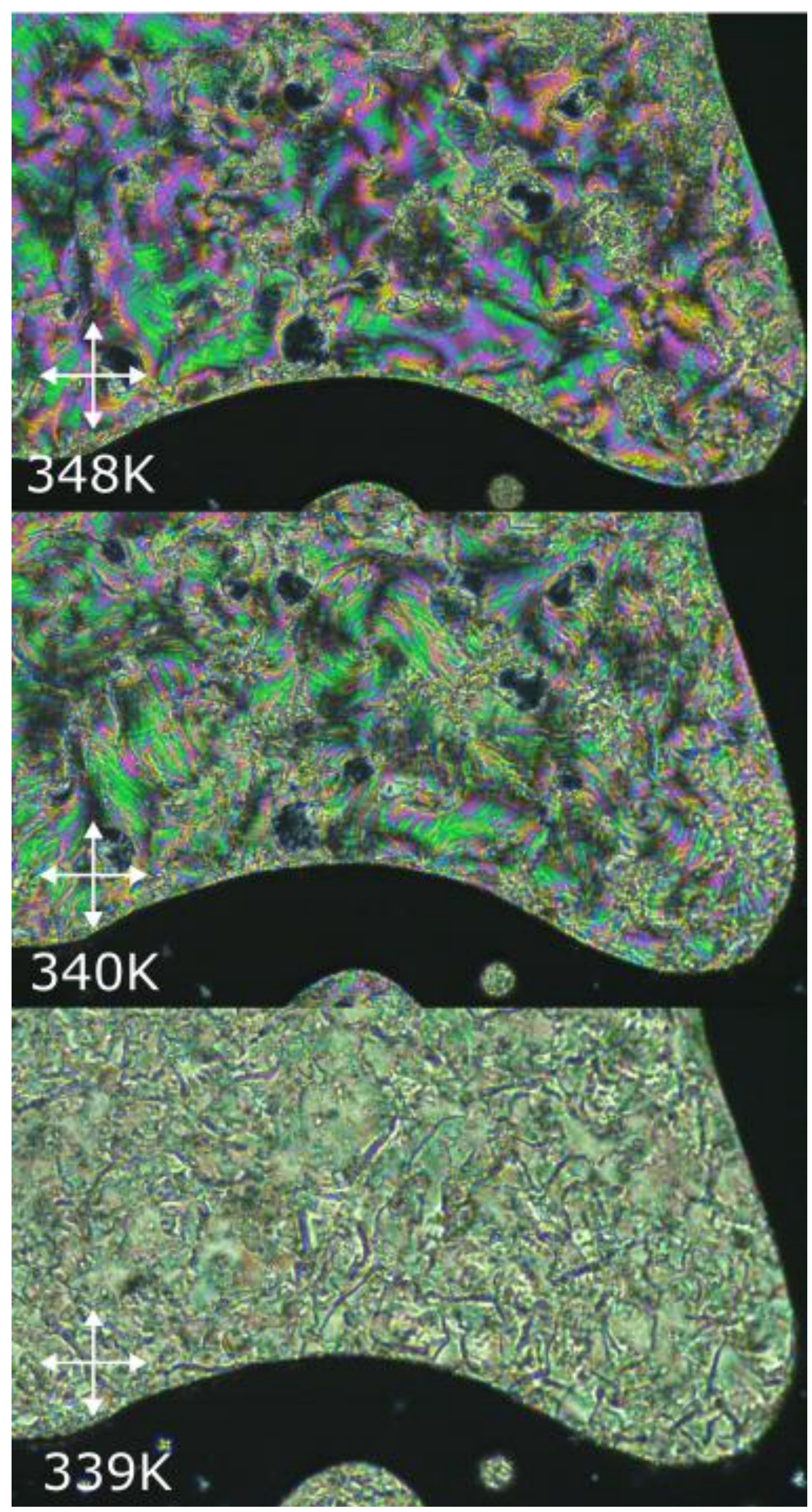

**Figure SI-11:** POM textures of a 90% wt DIO : 10% wt CJG484 mixture undergoing a transition to the $N_F$ phase. The extrapolated transition temperature of CJG484 is calculated at 333K

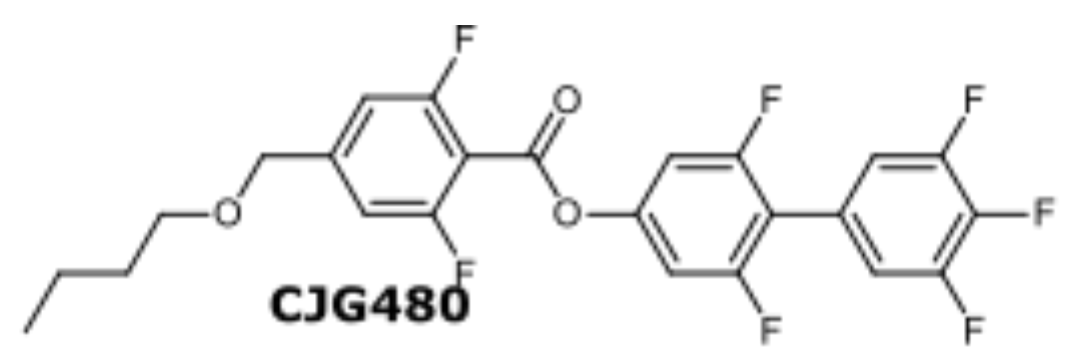

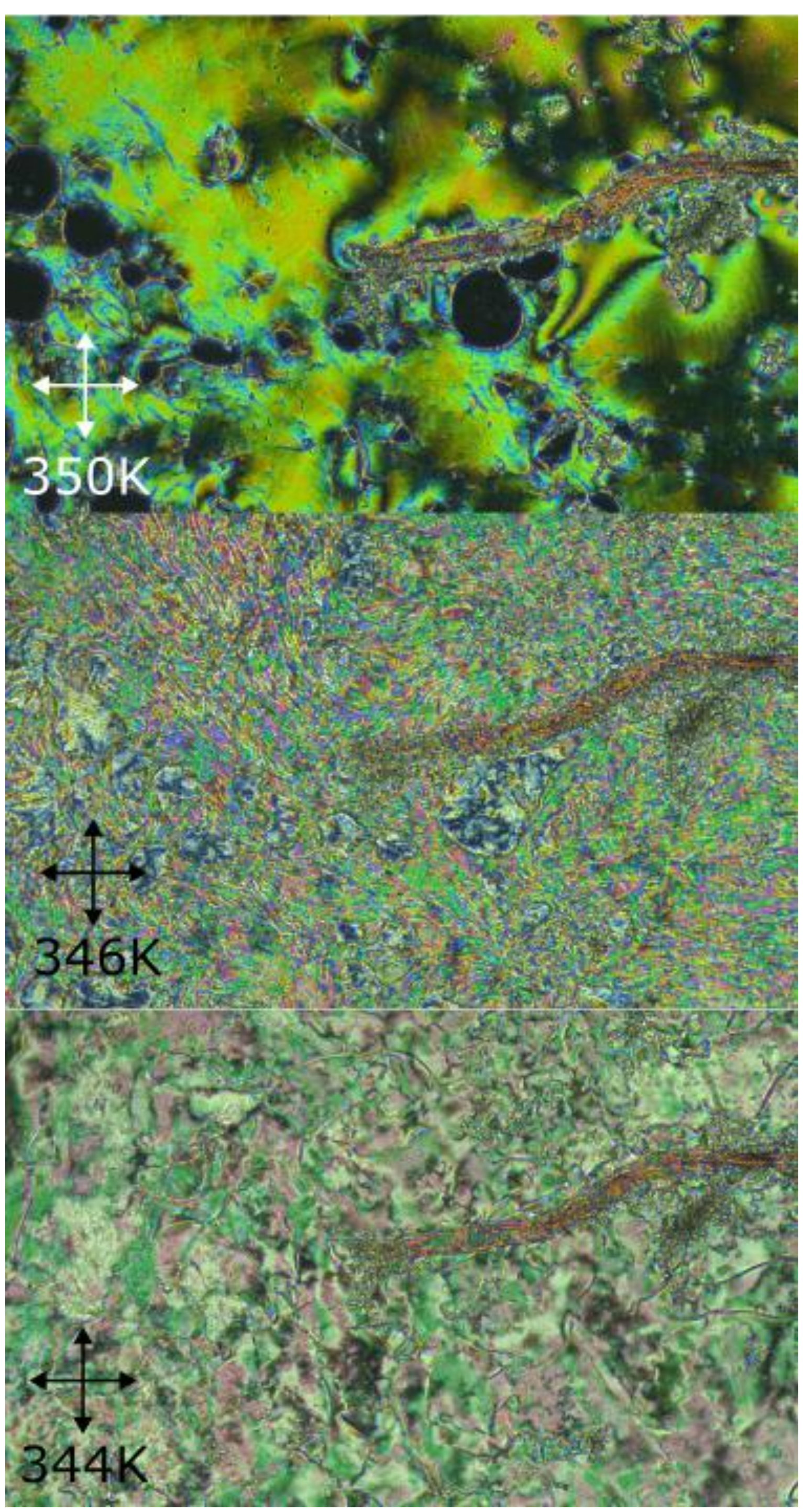

**Figure SI-12**: POM textures of a 90% wt DIO : 10% wt CJG480 mixture undergoing a transition to the $N_F$ phase. The extrapolated transition temperature of CJG480 is calculated at 377K

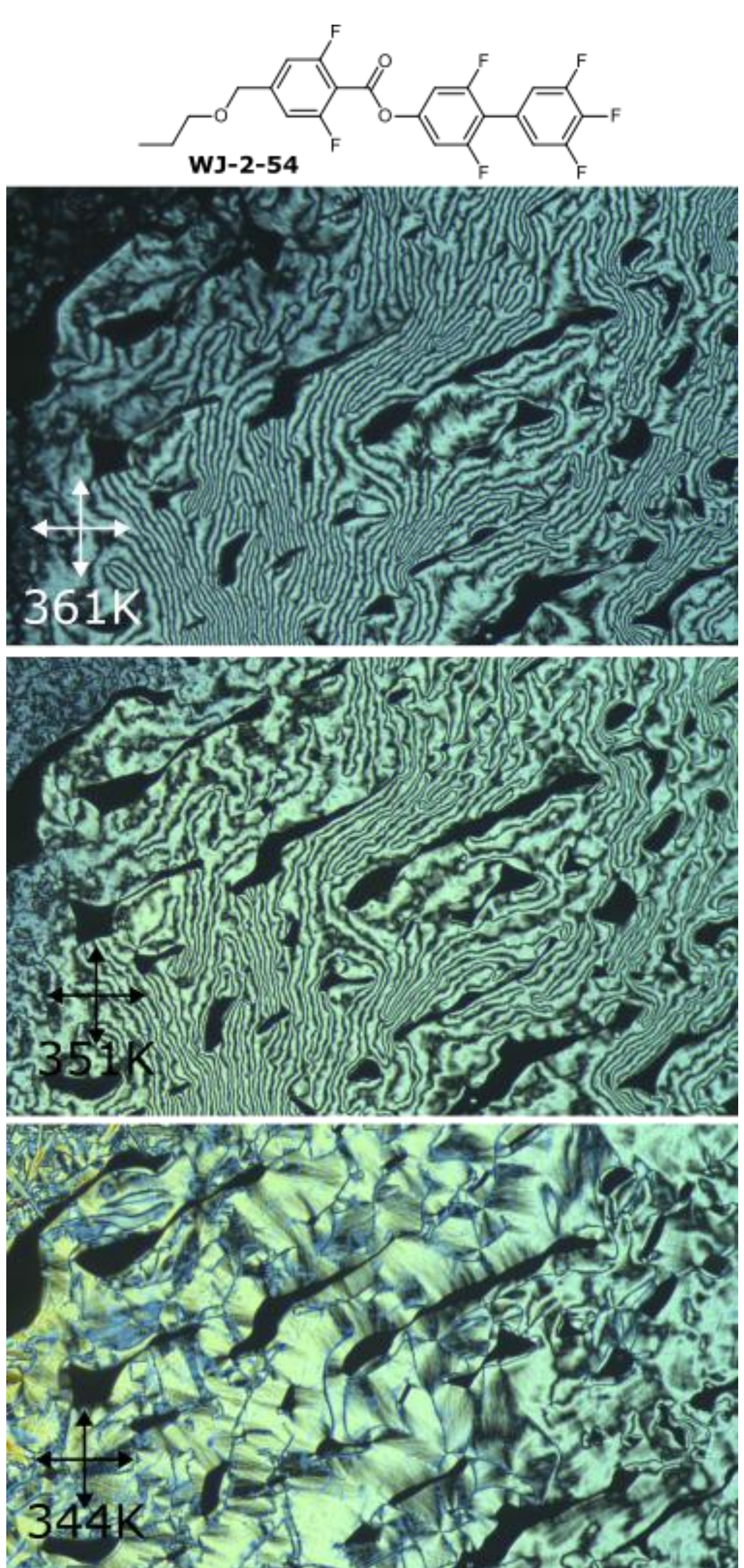

**Figure SI-13**: POM textures of a 90% wt DIO : 10% wt WJ-2-54 mixture undergoing a transition to the $N_F$ phase. The extrapolated transition temperature of WJ-2-54 is calculated at 370K

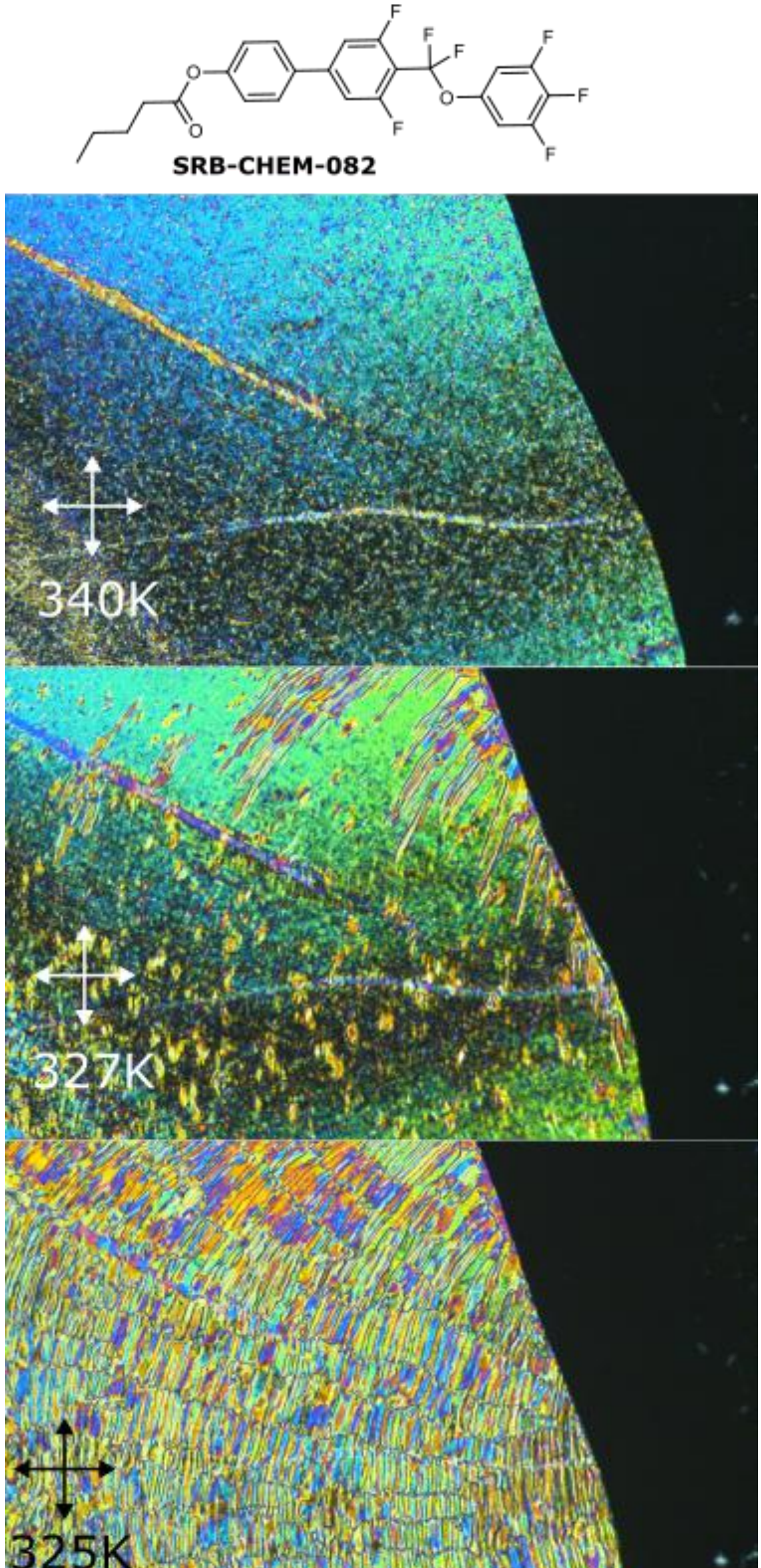

**Figure SI-14**: POM textures of a 90% wt DIO : 10% wt SRB-CHEM-082 mixture undergoing a transition to the $N_F$ phase. The extrapolated transition temperature of SRB-CHEM-082 is calculated at 280K

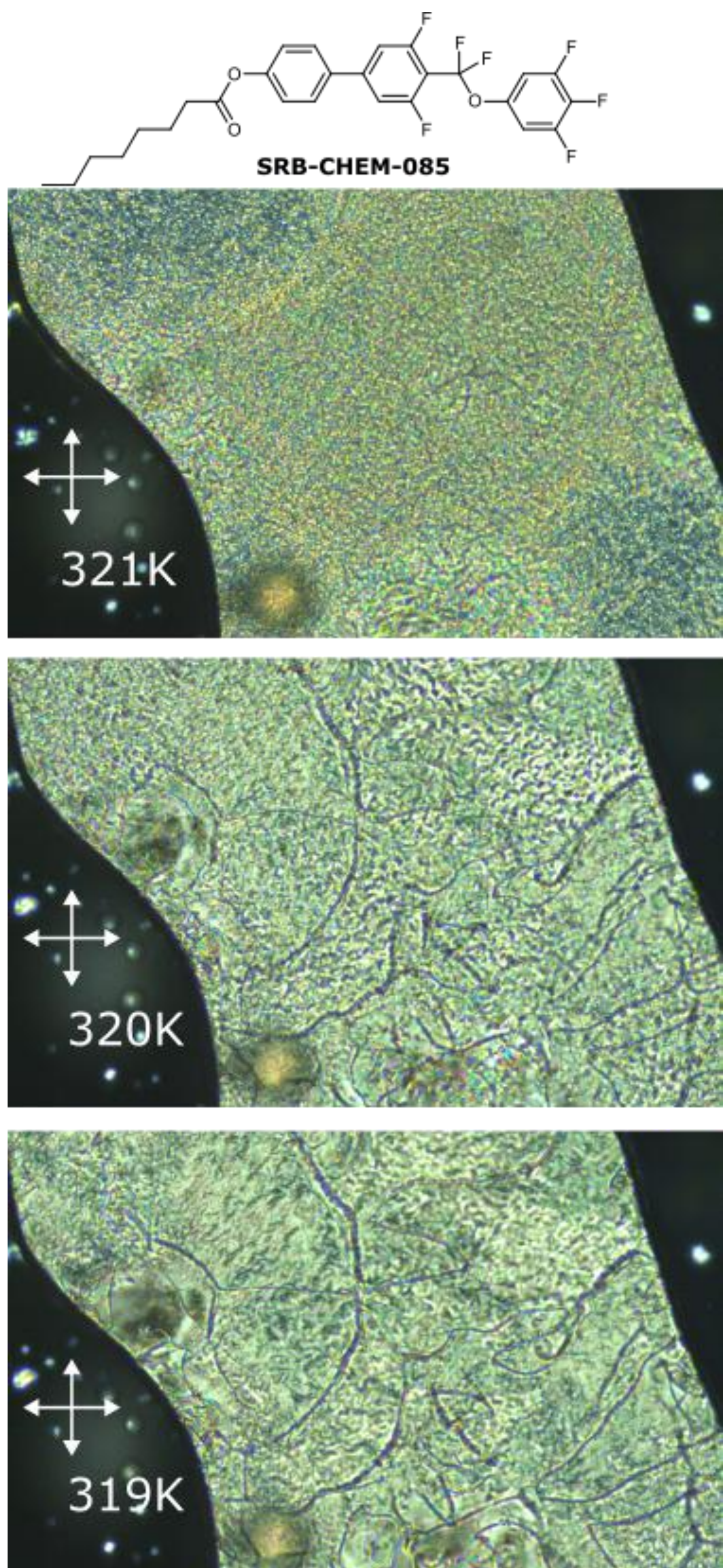

**Figure SI-15**: POM textures of a 90% wt DIO : 10% wt SRB-CHEM-085 mixture undergoing a transition to the $N_F$ phase. The extrapolated transition temperature of SRB-CHEM-085 is calculated at 200K

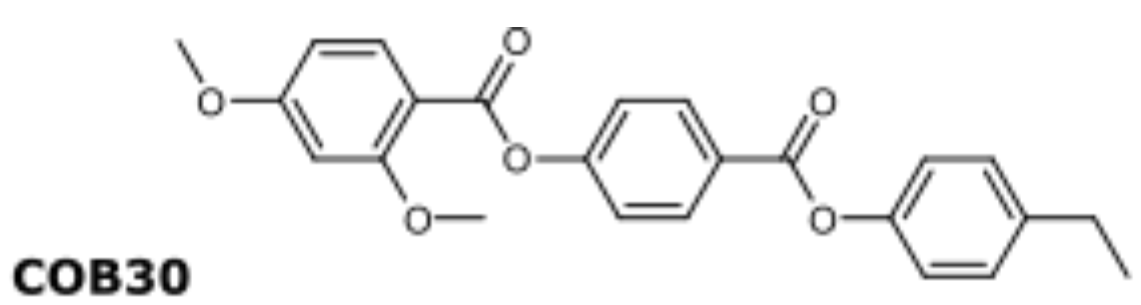

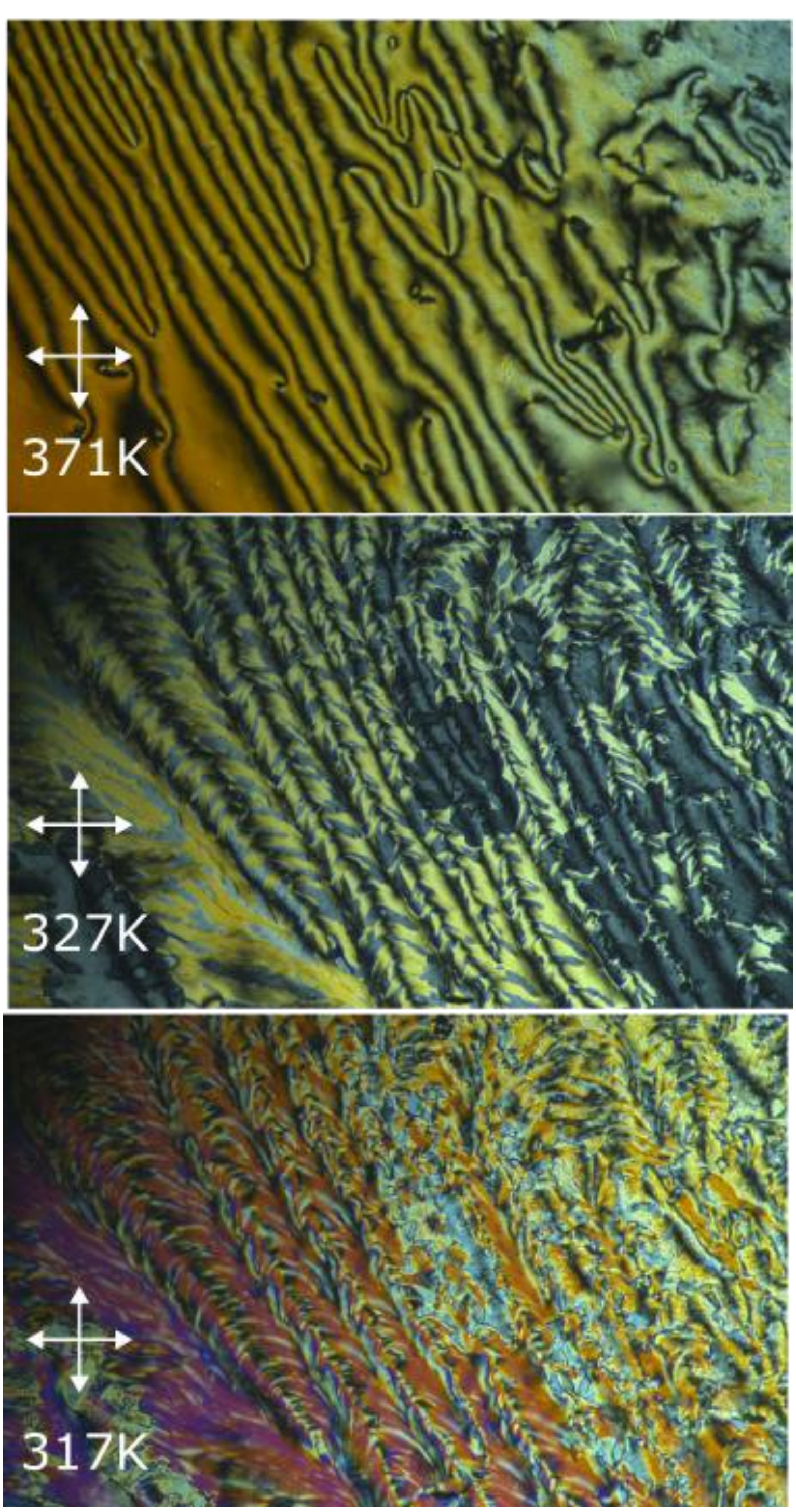

**Figure SI-16:** POM textures of a 90% wt DIO : 10% wt COB30 mixture undergoing a transition to the $N_F$ phase. The extrapolated transition temperature of COB30 is calculated at 180K

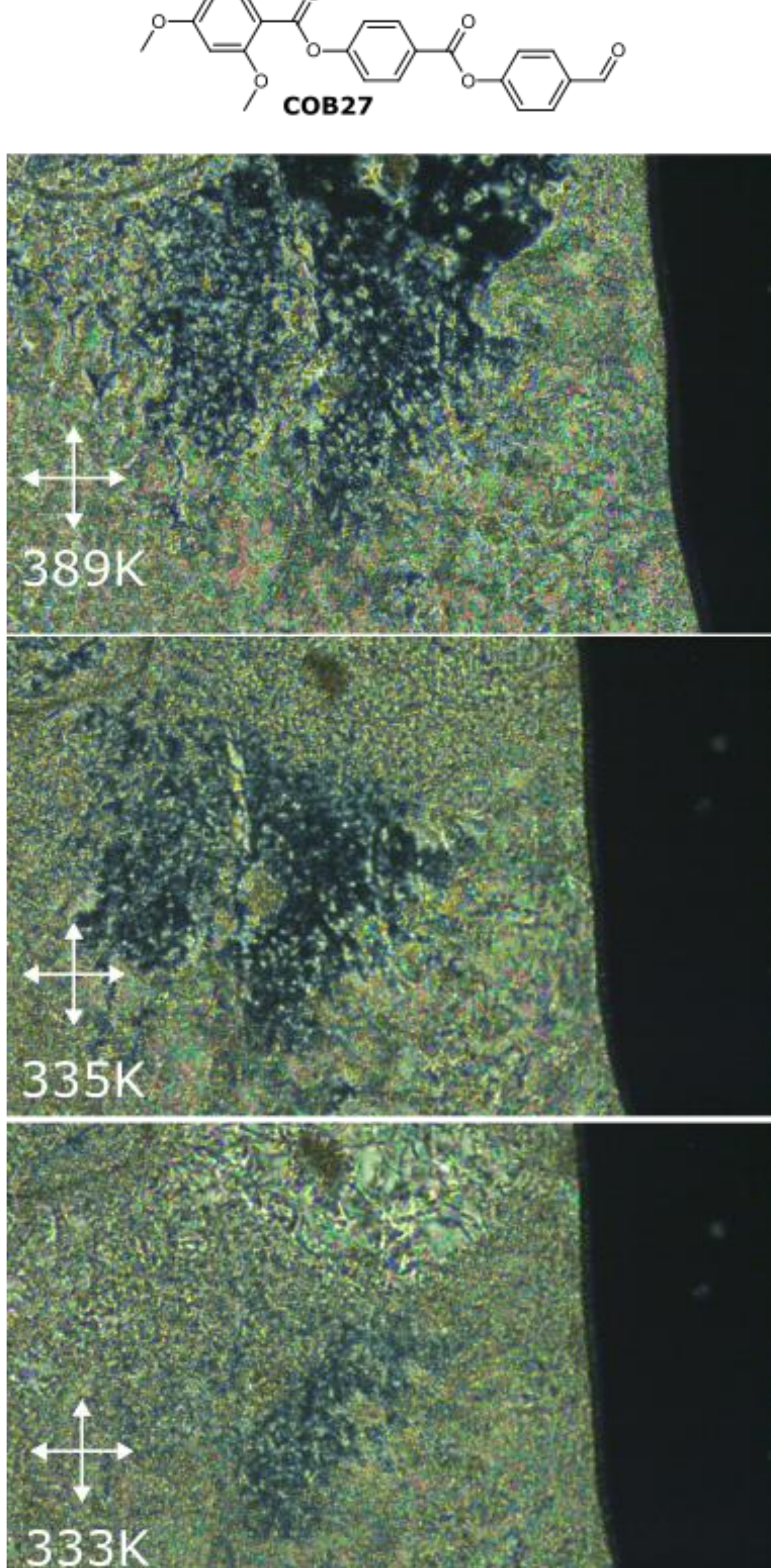

**Figure SI-17:** POM textures of a 90% wt DIO : 10% wt COB27 mixture undergoing a transition to the $N_F$ phase. The extrapolated transition temperature of COB27 is calculated at 308K

Three distinct featurisation approaches used in this work. Graph representations are constructed from molecular SMILES strings using RDKit and python code to determine their node and edge matrices. Nodes represent the heavy atoms of a molecule and the edge matrix its covalent bonds. Hydrogen atoms are considered implicitly and encoded as a node feature. The node feature vectors are built from one hot encodings of atom types from a list of permitted atoms (liquid crystals are organic molecules but additional atoms were included for generalisation purposes), the degree of heavy atom neighbours, the atoms formal charge, its hybridisation and the implicit hydrogen count. Ring membership and aromaticity are denoted by a binary value. The atomic mass, Van der Waals radius, and covalent radius are normalised by scaling constants. The constructed feature vector has a dimension of 79 components. The bond feature vector is constructed from a one hot encoding of the bond type and its stereochemistry, and binary values for its conjugation and ring membership. The connectivity is calculated from the RDKit adjacency matrix and converted to the COO format

Fingerprints were computed using the scikit-fingerprints library [1]. Through the course of hyperparameter searching all fingerprints within the library were available to be considered.

The featurisation of the VAE input is taken from the code published in the MGCVAE paper [2]. The nodes of each molecule is represented by a hot encoded tensor of shape ($N_{max}$ , $A$) where $N_{max}$ is the maximum number of molecules, a constant defined in the training script, and $A$ is atom type encoded from coders that create a feature vector from every atom type in the training dataset. Each row indicates the element of atom I with rows beyond the molecule size assigned zero values to indicate no atoms. Bonds are encoded as a one hot encoded tensor of shape ($N_{max}$, $N_{max} \cdot |B|$) where $B$ is the number of bond types in the dataset. The final tensor for each molecule concatenates the atom, bond and a length tensor of shape ($N_{max}$, )

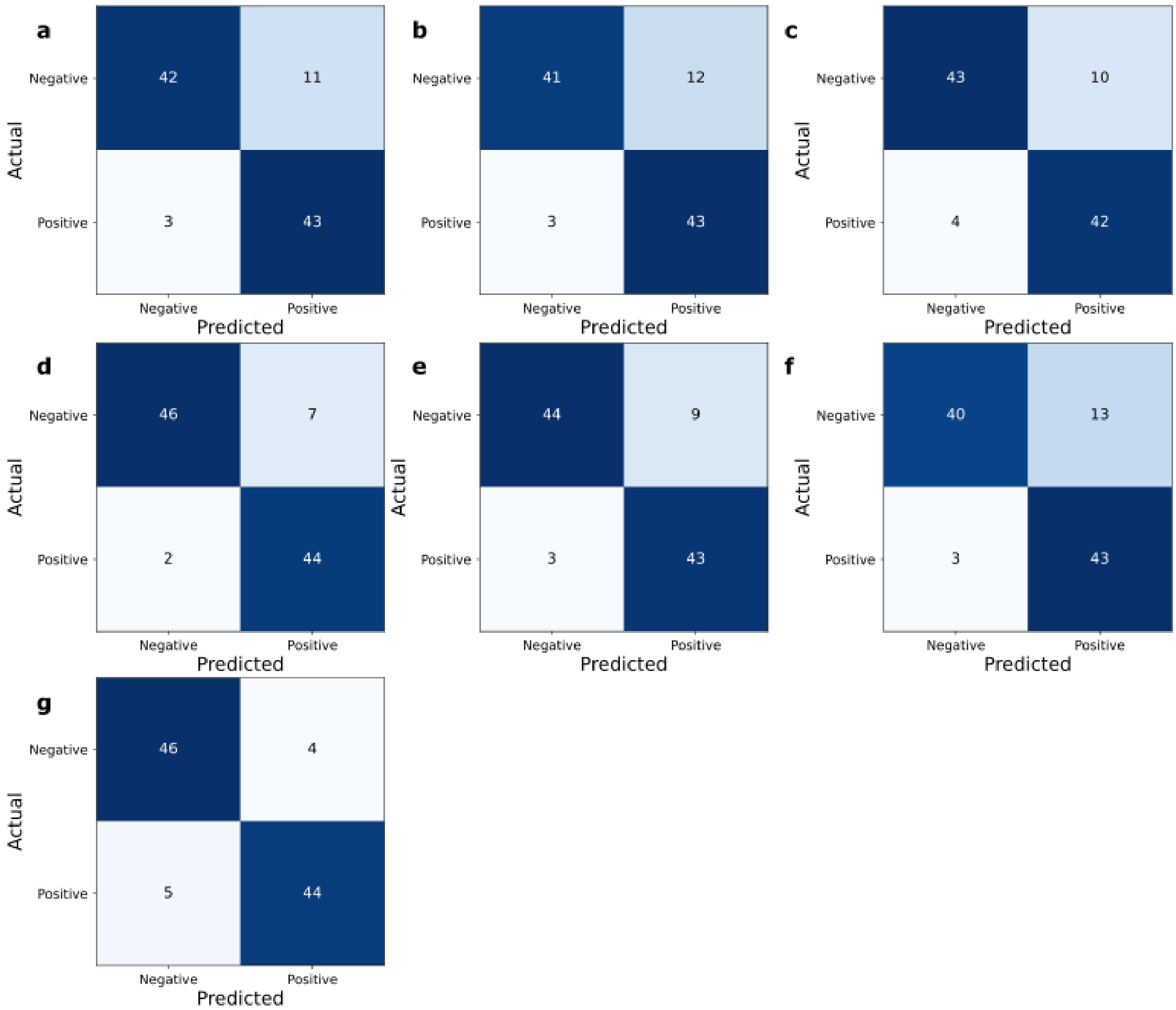

**Figure SI-17:** Confusion matrices of the random split trained a) GAT2, b) GAT, c) GCN, d) GIN, e) GG, f) TransformerConv, g) XGBoost classifiers

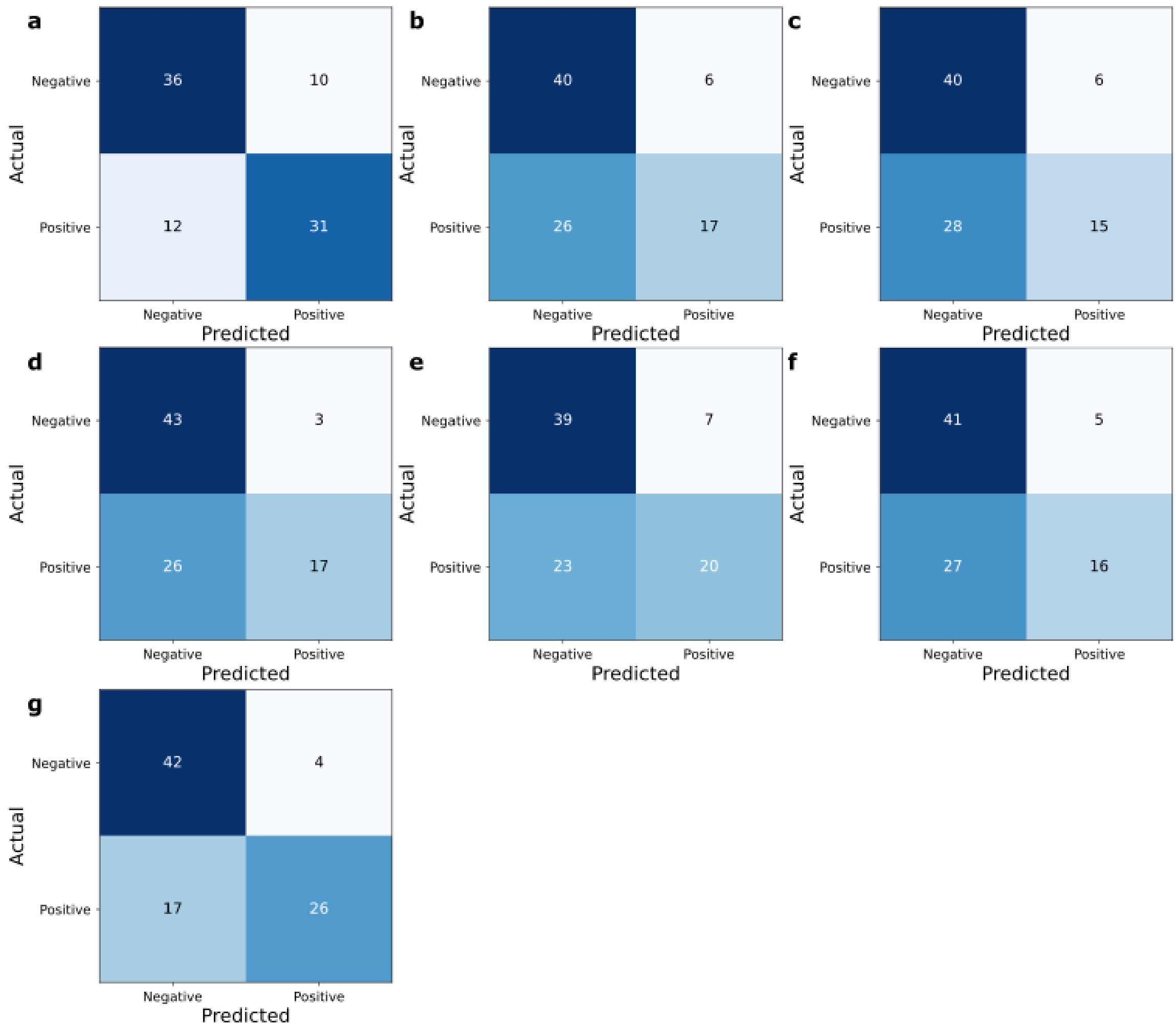

**Figure SI-18:** Confusion matrices of the scaffold split trained a) GAT2, b) GAT, c) GCN, d) GIN, e) GG, f) TransformerConv, g) XGBoost classifiers

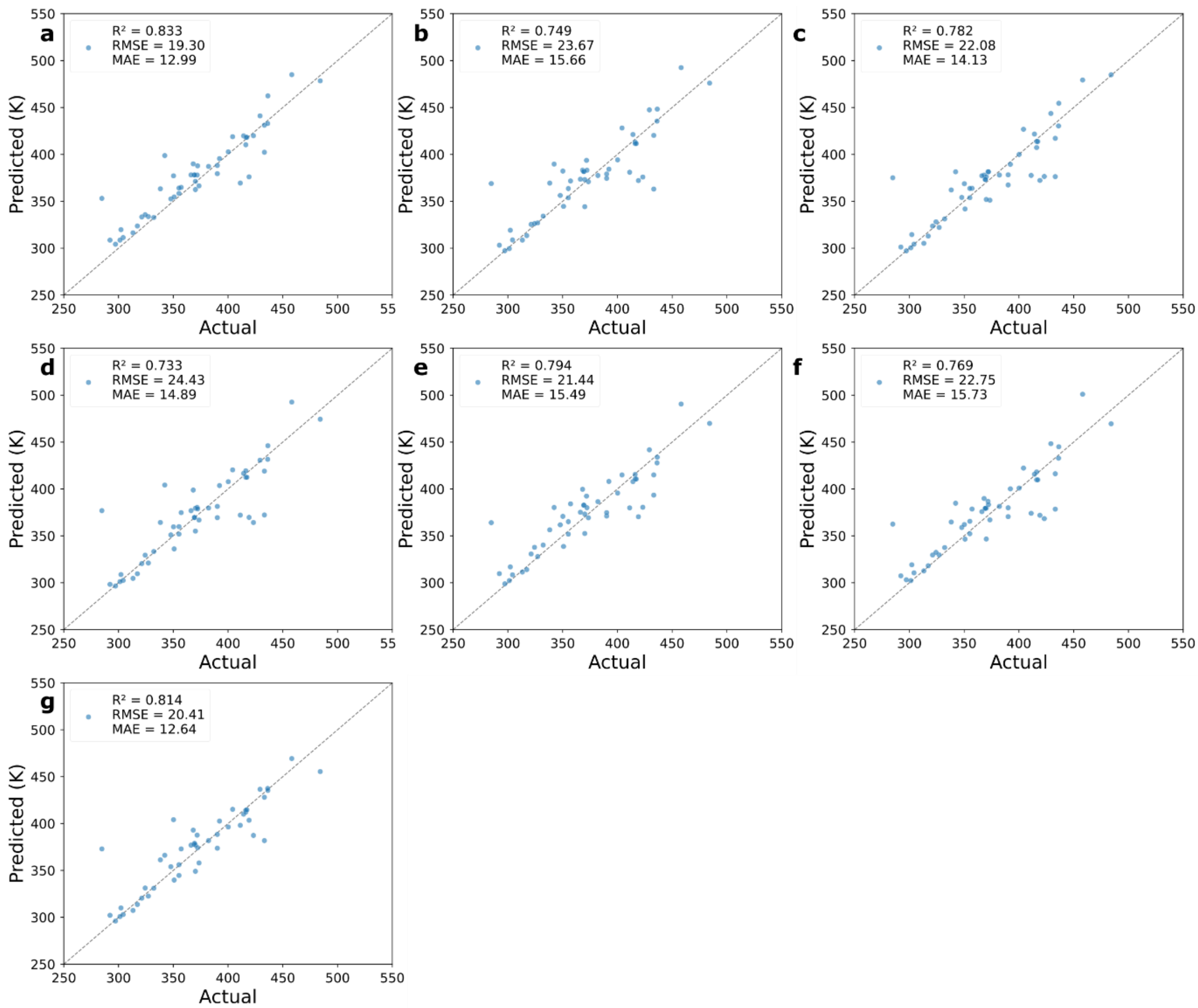

**Figure SI-19:** Predicted vs actual plots of the random split trained a) GAT2, b) GAT, c) GCN, d) GIN, e) GG, f) TransformerConv, g) XGBoost regressors

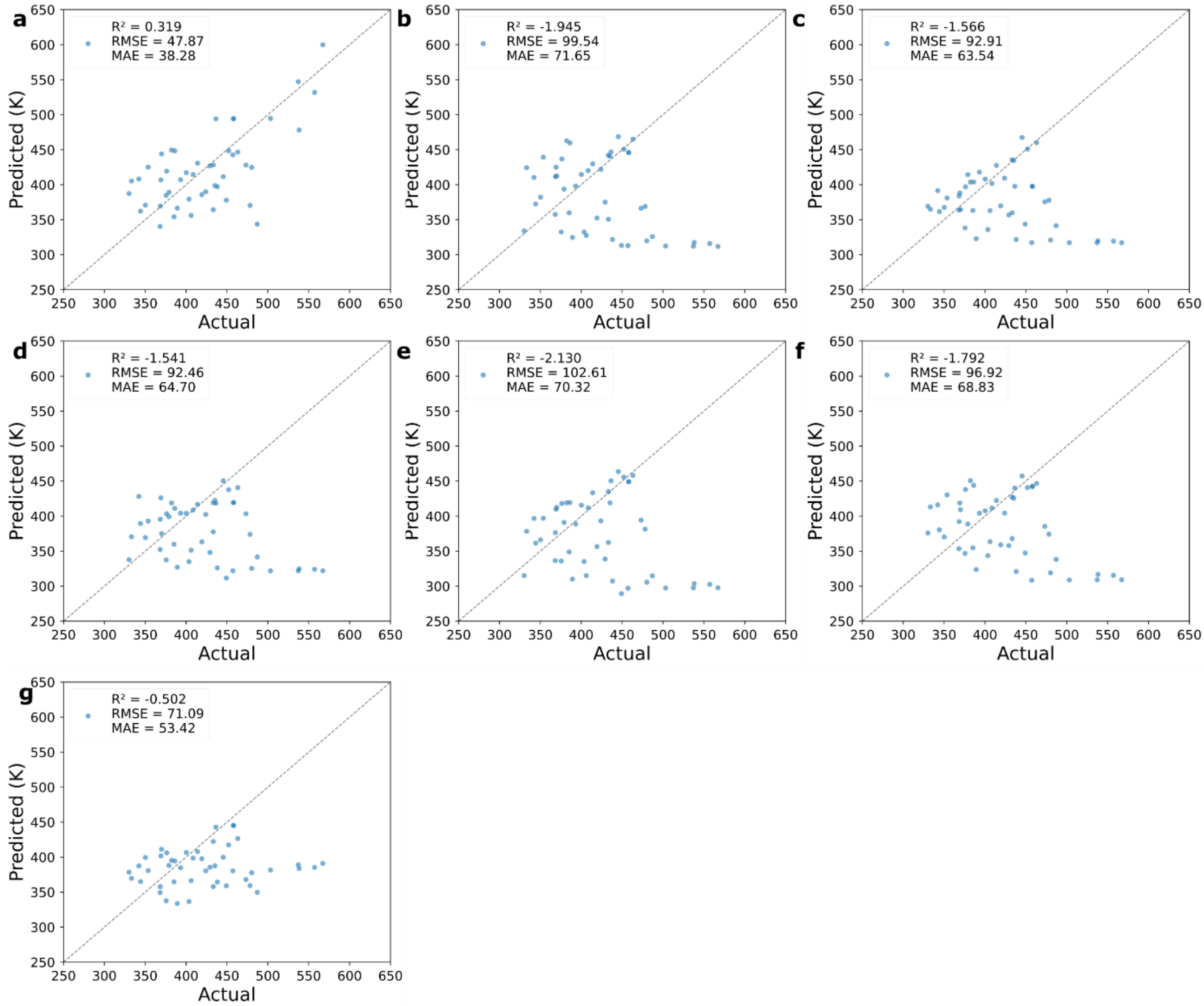

**Figure SI-20:** Predicted vs actual plots of the scaffold split trained a) GAT2, b) GAT, c) GCN,

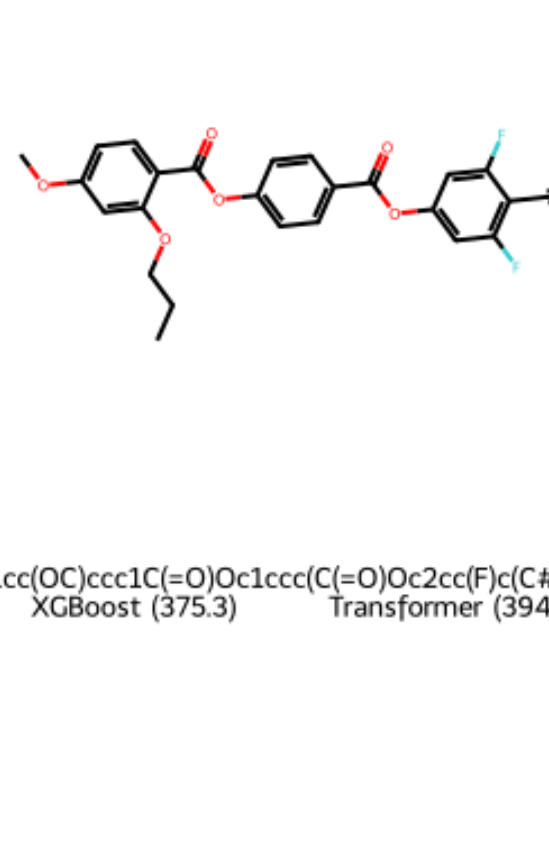

CCCOc1cc(OC)ccc1C(=O)Oc1ccc(C(=O)Oc2cc(F)c(C#N)c(F)c2)cc1
XGBoost (375.3) Transformer (394.1)

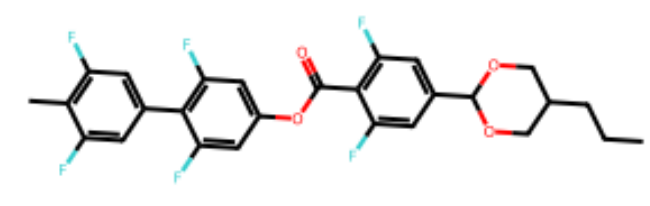

CCCC1COC(c2cc(F)c(C(=O)Oc3cc(F)c(-c4cc(F)c(C)c(F)c4)c(F)c3)c(F)c2)OC1
GAT (382.6) GG (416.2)

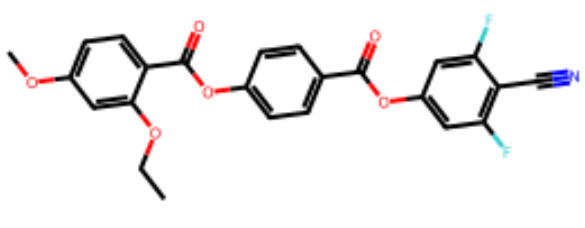

CCOc1cc(OC)ccc1C(=O)Oc1ccc(C(=O)Oc2cc(F)c(C#N)c(F)c2)cc1
XGBoost (389.5) GIN (407.1)

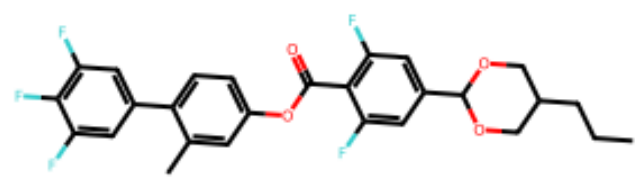

CCCC1COC(c2cc(F)c(C(=O)Oc3ccc(-c4cc(F)c(F)c(F)c4)c(C)c3)c(F)c2)OC1
GG (343.3) GIN (361.8)

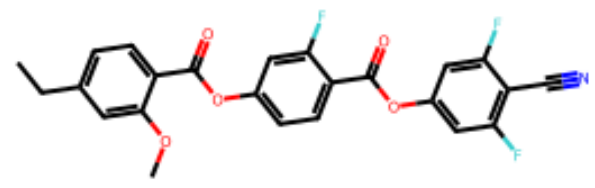

CCc1ccc(C(=O)Oc2ccc(C(=O)Oc3cc(F)c(C#N)c(F)c3)c(F)c2)c(OC)c1
GIN (377.1) GAT2 (410.3)

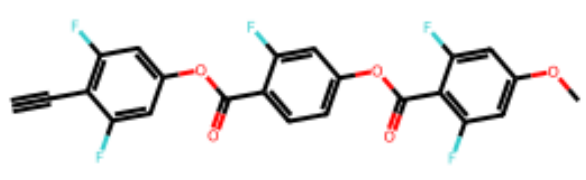

C#Cc1c(F)cc(OC(=O)c2ccc(OC(=O)c3c(F)cc(OC)cc3F)cc2F)cc1F
GAT (424.6) GG (469.2)

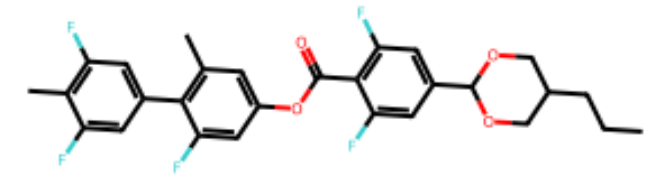

CCCC1COC(c2cc(F)c(C(=O)Oc3cc(C)c(-c4cc(F)c(C)c(F)c4)c(F)c3)c(F)c2)OC1
Transformer (363.3) GAT2 (386.8)

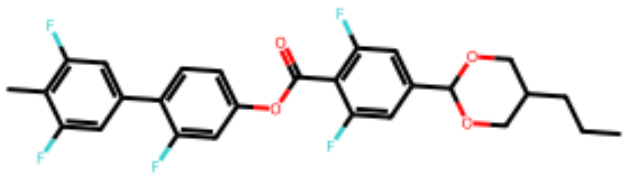

CCCC1COC(c2cc(F)c(C(=O)Oc3ccc(-c4cc(F)c(C)c(F)c4)c(F)c3)c(F)c2)OC1
GCN (347.0) GG (375.3)

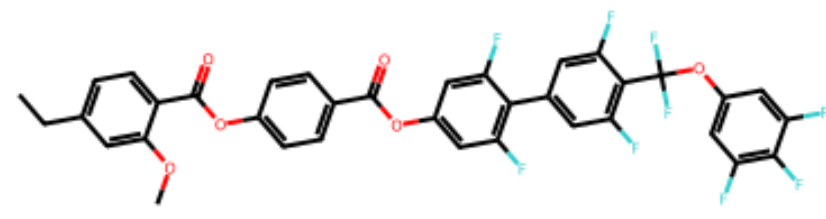

CCc1ccc(C(=O)Oc2ccc(C(=O)Oc3cc(F)c(-c4cc(F)c(C(F)(F)Oc5cc(F)c(F)c(F)c5)c(F)c4)c(F)c3)cc2)c(OC)c1
GIN (436.9) GAT2 (471.3)

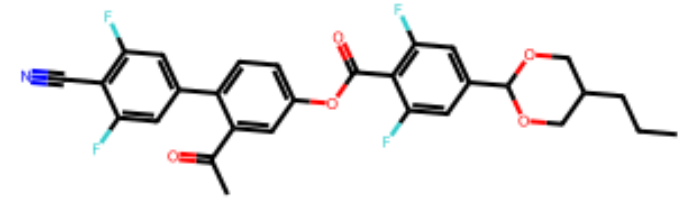

CCCC1COC(c2cc(F)c(C(=O)Oc3ccc(-c4cc(F)c(C#N)c(F)c4)c(C(C)=O)c3)c(F)c2)OC1
GIN (356.8) GG (378.0)

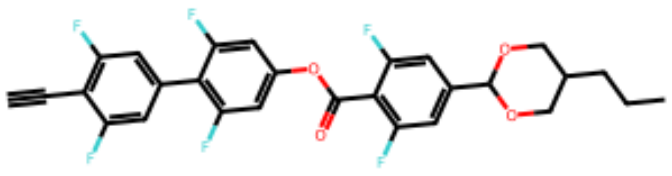

C#Cc1c(F)cc(-c2c(F)cc(OC(=O)c3c(F)cc(C4OCC(CCC)CO4)cc3F)cc2F)cc1F
GAT (376.4) XGBoost (422.1)

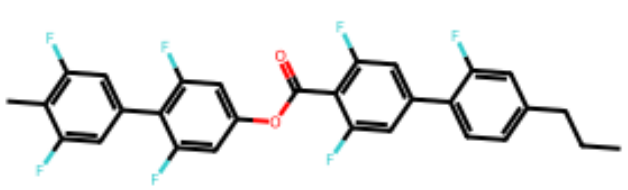

CCCc1ccc(-c2cc(F)c(C(=O)Oc3cc(F)c(-c4cc(F)c(C)c(F)c4)c(F)c3)c(F)c2)c(F)c1
GCN (390.8) XGBoost (414.5)

CCCC1COC(c2cc(F)c(C(=O)Oc3ccc(-c4cc(F)c(C#N)c(F)c4)c(C)c3)c(F)c2)OC1
GG (357.1) GIN (368.9)

**Figure SI-21:** The candidate pool of molecules validated by a consensus across the seven random split trained molecule and a predicted retrosynthetic route. The min-max temperature range along with model responsible for the prediction is also shown

CCCOc1cc(OC)ccc1C(=O)Oc1ccc(C(=O)Oc2cc(F)c(C#N)c(F)c2)cc1
371.5 392.1

CCCC1COC(c2cc(F)c(C(=O)Oc3cc(F)c(-c4cc(F)c(C)c(F)c4)c(F)c3)c(F)c2)OC1
381.0 411.5

CCOc1cc(OC)ccc1C(=O)Oc1ccc(C(=O)Oc2cc(F)c(C#N)c(F)c2)cc1
392.2 410.2

CCc1ccc(C(=O)Oc2ccc(C(=O)Oc3cc(F)c(C#N)c(F)c3)c(F)c2)c(OC)c1
394.7 414.6

CCCC1COC(c2cc(F)c(C(=O)Oc3cc(O)c(-c4cc(F)c(F)c(F)c4)c(F)c3)c(F)c2)OC1
341.1 380.7

C#Cc1c(F)cc(OC(=O)c2ccc(OC(=O)c3c(F)cc(OC)cc3F)cc2F)cc1F
443.2 469.0

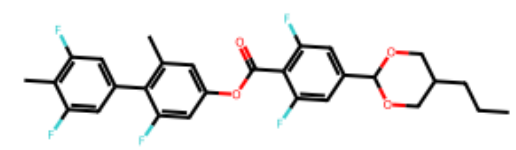
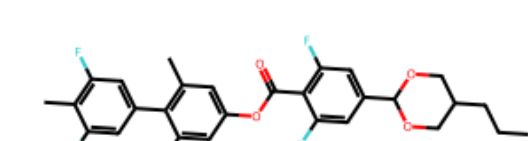
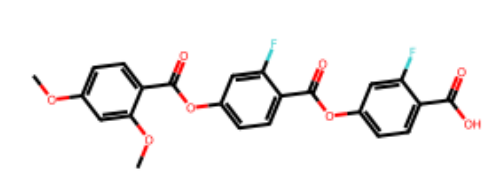
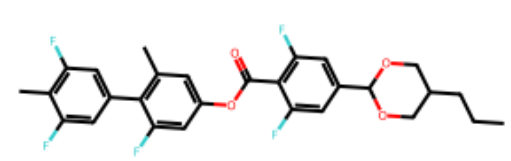
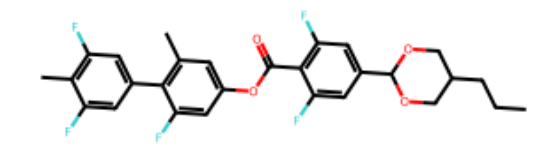
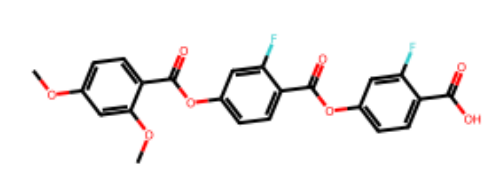
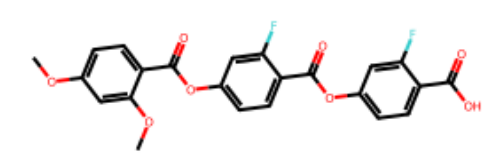
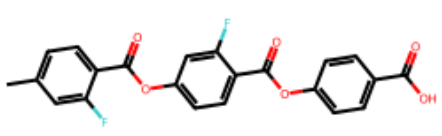
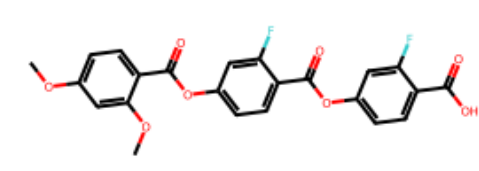
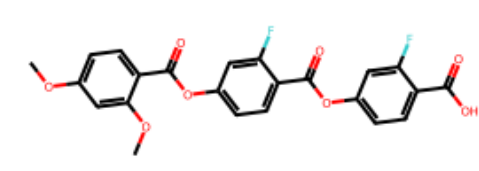
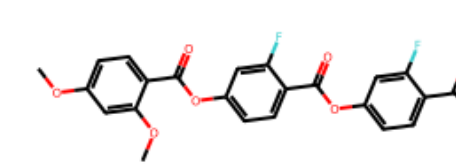
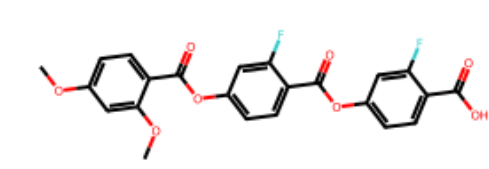
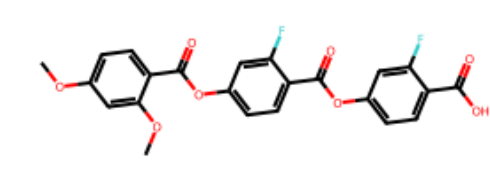
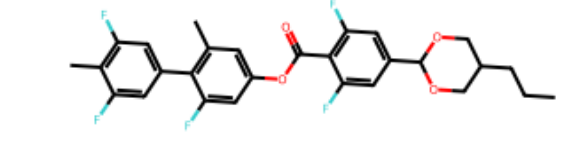
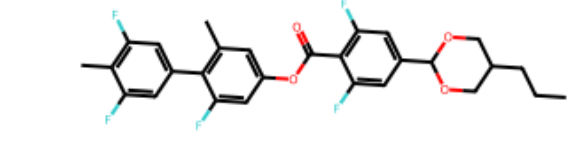
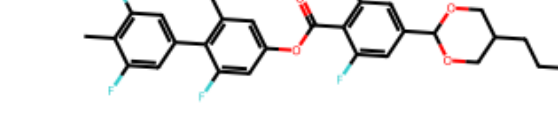
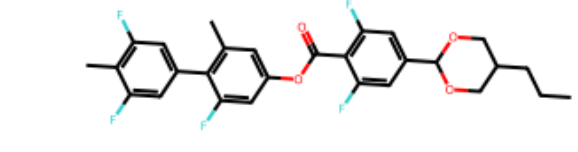
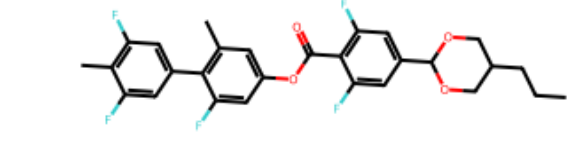
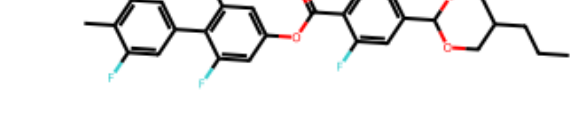
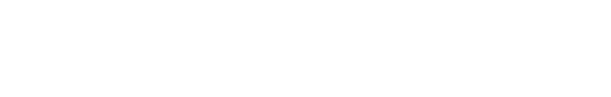
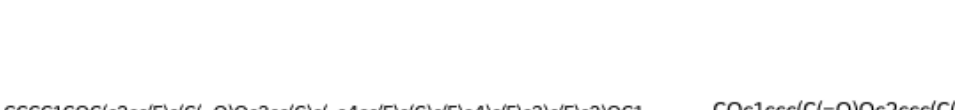
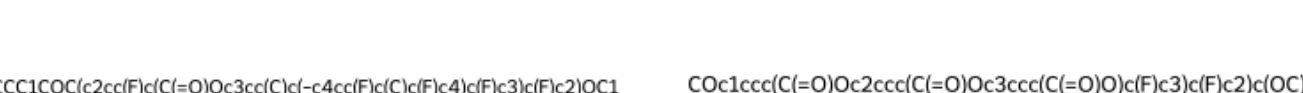
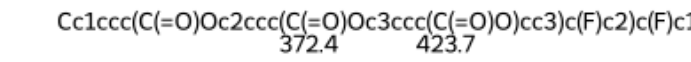

CCCC1COC(c2cc(F)c(C(=O)Oc3cc(C)c(-c4cc(F)c(C)c(F)c4)c(F)c3)c(F)c2)OC1
351.6 381.6

COc1ccc(C(=O)Oc2ccc(C(=O)Oc3ccc(C(=O)O)c(F)c3)c(F)c2)c(OC)c1
380.6 417.2

Cc1ccc(C(=O)Oc2ccc(C(=O)Oc3ccc(C(=O)O)cc3)c(F)c2)c(F)c1
372.4 423.7

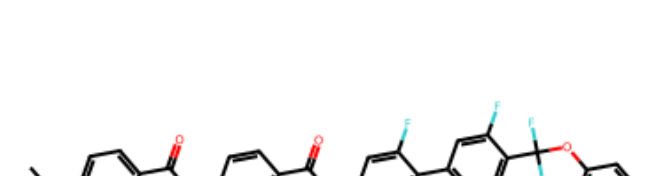
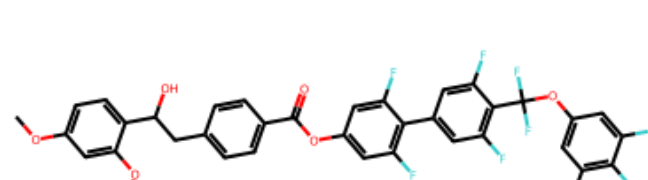
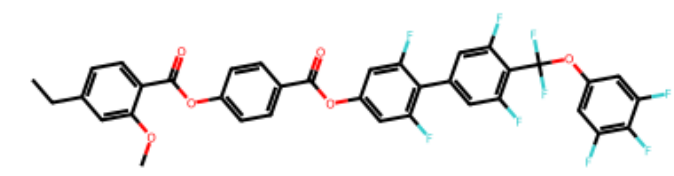
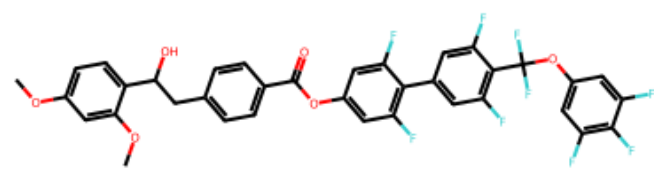
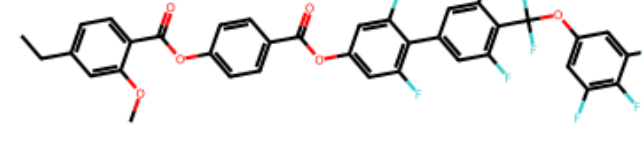
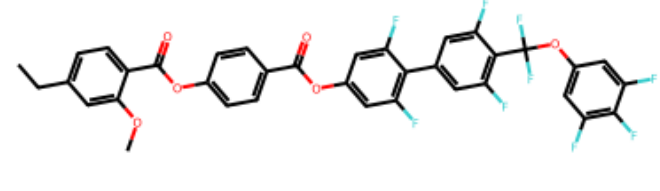
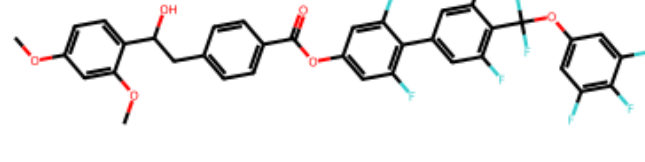
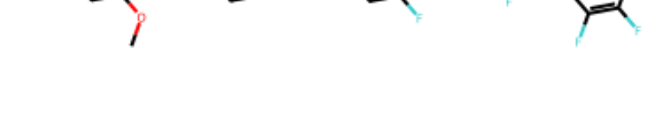

CCc1ccc(C(=O)Oc2ccc(C(=O)Oc3cc(F)c(-c4cc(F)c(C(F)(F)Oc5cc(F)c(F)c(F)c5)c(F)c4)c(F)c3)cc2)c(OC)c1
448.6 479.2

CCCC1COC(c2cc(F)c(C(=O)Oc3cc(F)cc(Oc4cc(F)c(F)c(F)c4)c3)c(F)c2)OC1
365.8 392.4

COc1ccc(C(O)Cc2ccc(C(=O)Oc3cc(F)c(-c4cc(F)c(C(F)(F)Oc5cc(F)c(F)c(F)c5)c(F)c4)c(F)c3)cc2)c(OC)c1
421.4 450.1

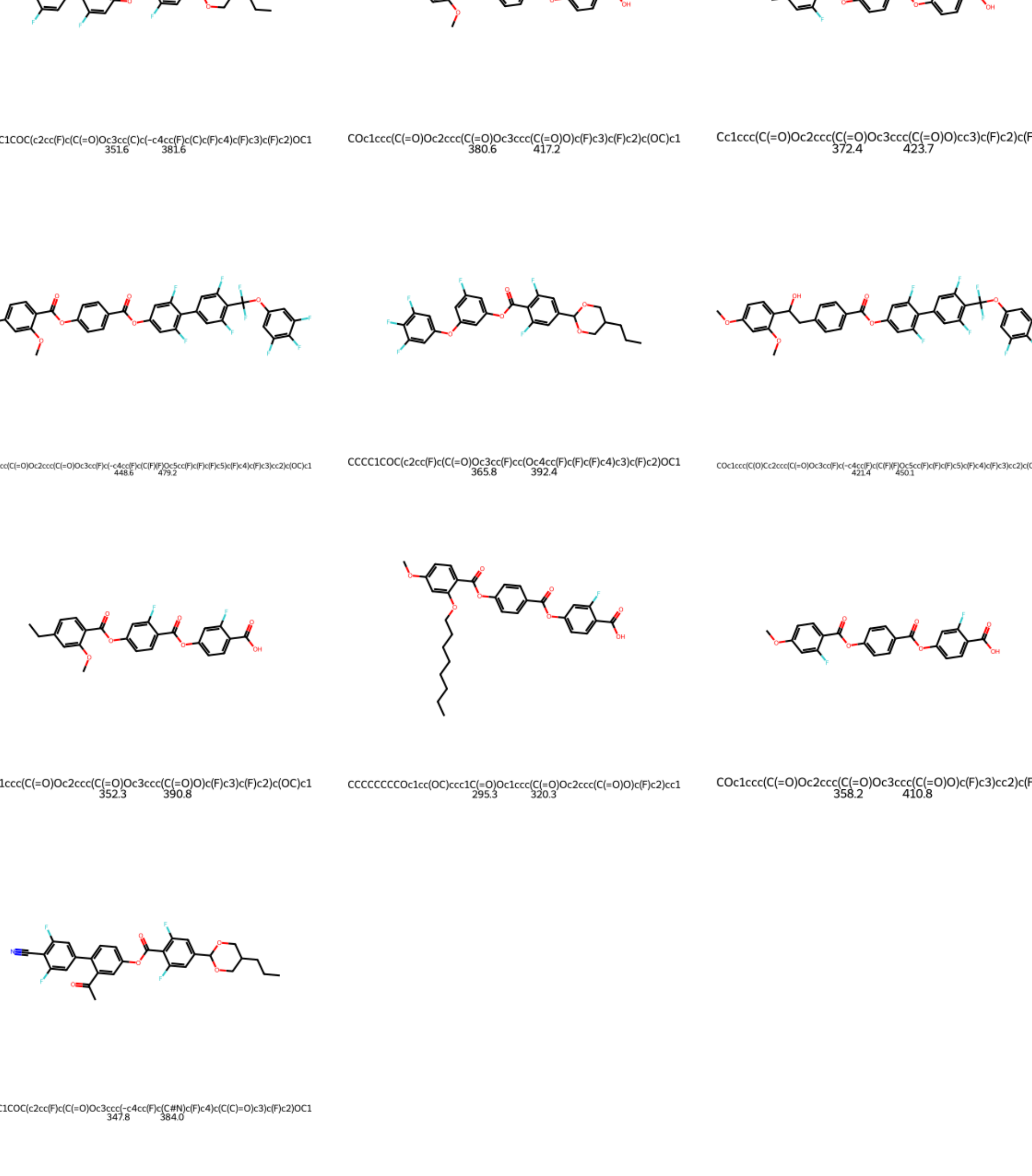
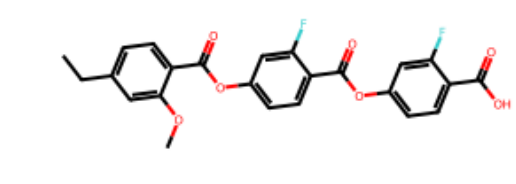
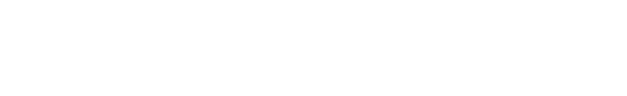

CCc1ccc(C(=O)Oc2ccc(C(=O)Oc3ccc(C(=O)O)c(F)c3)c(F)c2)c(OC)c1
352.3 390.8

CCCCCCCCOc1cc(OC)ccc1C(=O)Oc1ccc(C(=O)Oc2ccc(C(=O)O)c(F)c2)cc1
295.3 320.3

COc1ccc(C(=O)Oc2ccc(C(=O)Oc3ccc(C(=O)O)c(F)c3)cc2)c(F)c1
358.2 410.8

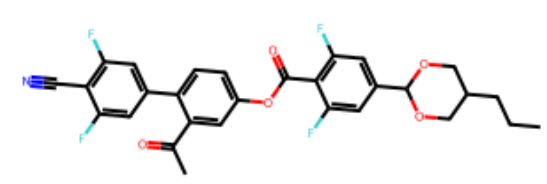
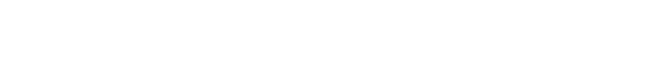

CCCC1COC(c2cc(F)c(C(=O)Oc3ccc(-c4cc(F)c(C#N)c(F)c4)c(C(C)=O)c3)c(F)c2)OC1
347.8 384.0

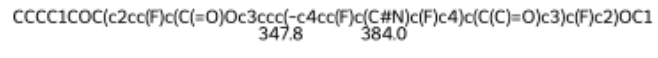

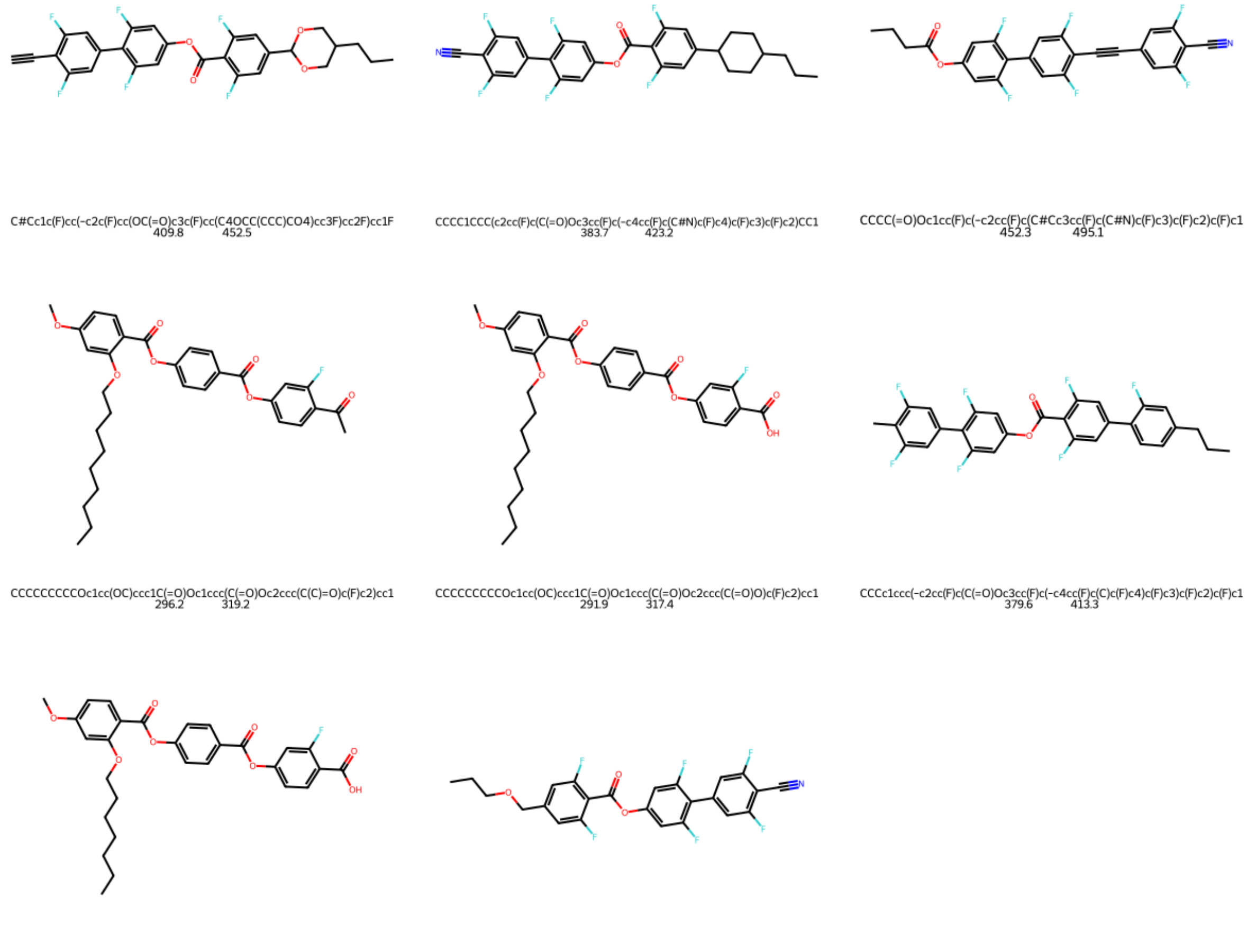

**Figure SI-22:** The candidate set produced by scaffold split GATv2 and XGBoost classifiers with 60% ensemble agreement and a valid retrosynthetic route. The range of their predicted NF transition (K) is given as the standard deviation of the ensemble predictions from their mean prediction.

### 8.1. Chemical Synthesis

Chemicals were purchased from commercial suppliers (Fluorochem, Apollo Scientific, Merck, ChemScene, Ambeed) and used as received. Solvents were purchased from Merck and used without further purification. Anhydrous solvents were obtained by percolation through activate alumina in a standard solvent purification system. Reactions were performed in standard laboratory glassware at ambient temperature and atmosphere (unless noted otherwise) and were monitored by TLC with an appropriate eluent and visualised with 254 nm light. Chromatographic purification was performed using a Combiflash NextGen 300+ System (Teledyne Isco) with a silica gel stationary phase and a binary solvent gradient as the mobile phase, with detection made in the 200-800 nm range. Chromatographed materials were subjected to re-crystallisation from an appropriate solvent system.

### 8.2. Chemical Characterisation Methods

The structures of intermediates and final products were determined using $^{1}H$, $^{13}C\{^{1}H\}$, and $^{19}F$ NMR spectroscopy. NMR spectroscopy was performed using either a Bruker Avance III HDNMR spectrometer operating at 400 MHz, 100.5 MHz or 376.4 MHz ($^{1}H$, $^{13}C\{^{1}H\}$ and $^{19}F$, respectively) or a Bruker AV4 NEO 11.75T spectrometer operating at 500 MHz, 125.5 MHz or 470.5 MHz ($^{1}H$, $^{13}C\{^{1}H\}$ and $^{19}F$, respectively).High resolution mass spectrometry data (HRMS) was collected using a Bruker MaXis Impact spectrometer with a positive ESI source (VIP-HESI) or a Bruker Impact QqTOF instrument acquired with positive DIP APCI source. In both, the sample was introduced *via* direct infusion as solution in acetonitrile. HPLC analysis was performed using an Agilent 1290 Infinity II system fitted with a poroshell 120 ec-C18 column running a $H_2O$:MeCN gradient.

### 8.3. Mesophase Characterisation

Phase transition temperatures and associated enthalpies of transition were determined by differential scanning calorimetry (DSC) using a TA instruments Q2000 heat flux calorimeter with a liquid nitrogen cooling system for temperature control. Samples were measured with 10 ℃ $min^{-1}$ heating and cooling rates. The transition temperatures and enthalpy values reported are averages obtained for duplicate runs. Phase transition temperatures were measured on cooling cycles for consistency between monotropic and enantiotropic phase transitions, while crystal melts were obtained on heating. Phase identification by polarised optical microscopy (POM) was performed using a Leica DM 2700 P polarised optical microscope equipped with a Linkam TMS 92 heating stage. Samples were studied sandwiched between two untreated glass coverslips.

### 8.4. General Methods:

**General Method A – Steglich esterification:**

A round bottom flask or reaction vial were charged with carboxylic acid (1.2 mol eqv), phenol (1 mol eqv), EDC.HCl (1.5 mol eqv) and DMAP (2 mol%). The solids were suspended in DCM and stirred vigorously until the reaction was complete (1-48h) as judged by TLC. The reaction mass was dry loaded onto celite and purified by flash chromatography over silica with a suitable solvent gradient. The chromatographed material was recrystalised from a suitable solvent to afford the target material.

**General Method B – Carboxylation:**

An oven dried flask was cooled under an atmosphere of dry nitrogen, charged with the compound to be lithiated (1 mol eqv), anhydrous THF (~ 0.2 M), and cooled to -78 °C with an external $CO_2$/acetone bath. A solution of n-butyl lithium (1.6 M in hexane, 1.2 mol eqv.) was added dropwise. Once the addition was complete, the solution was allowed to stir for 0.5h. Separately, a stream of $CO_2$ gas was dried by bubbling through neat $H_2SO_4$ and was introduced to the reaction with vigorous stirring for 20 minutes. The cooling bath was then removed, and the reaction mixture allowed to warm to ambient temperature over the course of 2h. The solution was acidified with 2M HCl. The organic layer was separated and retained, the aqueous was washed with ethyl acetate and discarded. The combined organics were dried over $MgSO_4$, and concentrated *in vacuo*. The resulting material was used without further purification.

**General Method C – Suzuki-Miyaura Cross Coupling:**

A biphasic mixture of THF and aqueous 2M sodium carbonate (1:1) was degassed by sparging with dry nitrogen whilst vigorously stirring. To this was added aryl halide (1 mol eqv), boronate species (> 1 mol eqv), and finally Pd-XPHOS-G3 (~ 1 mol%). The reaction was heated under reflux under an atmosphere of dry nitrogen, and monitored by TLC. Once complete, the reaction was cooled to ambient temperature. The organic layer was separated and retained, the aqueous layer was washed with EtOAc (x2) and discarded. The combined organics were washed with saturated aqueous sodium chloride, and dried over $MgSO_4$. The volatiles were removed in vacuo; the crude material was purified by flash chromatography and/or recrystalisation as appropriate.

## 5.5. Synthetic Schemes

(i) NaH, THF, PrBr, $N_2$, 0 °C
(ii) a) $^{n}$BuLi, THF, -78 °C
b) $CO_2$, THF, -78 °C
c) HCl, 0 °C
iii) EDC.HCl, DMAP, DCM

## Scheme 1

i) $B_2(pin_2)$, $PCl_2$(dppf), KOAc, THF, 65°C, $N_2$
ii) Pd(XPhos), $Na_2CO_3$, THF, $H_2O$. 85 °C, $N_2$
iii) DCC, DMAP, DCM

## Scheme 2

(i) EtI, $K_2CO_3$, Acetone, 65 °C
(ii) a) $^{n}$BuLi, THF, -78 °C
b) $CO_2$, THF, -78 °C
c) HCl, 0 °C
iii) EDC.HCl, DMAP, DCM

## Scheme 3

(i) NaH, THF, PrBr, $N_2$, 0 °C
(ii) a) $^nBuLi$, THF, -78 °C
b) $CO_2$, THF, -78 °C
c) HCl, 0 °C
iii) EDC.HCl, DMAP, DCM

## Scheme 4

(i) Propyl-1,3-propanediol, *p*-TsOH, Toluene, 110 °C
(ii) a) $^nBuLi$, THF, -78 °C
b) $CO_2$, THF, -78 °C
c) HCl, 0 °C
iii) EDC.HCl, DMAP, DCM

## Scheme 5

(i) Pd XPhos ($G_3$), 2M $K_2CO_{3(aq)}$,
THF, $N_{2(g)}$, 66 °C.

## Scheme 6

(i) PdXPhos $G_3$, $K_2CO_3$ (aq),
THF, reflux, $N_2$, 16 hrs
(ii) EDC.HCl, DMAP, DCM

## Scheme 7

i. EDC.HCl, DMAP, DCM

**Scheme 8**

## 6. Chemical Characterisation

### 6.1. Scheme 1

**IDX7a / WJ-2-41 / 3,5-difluorobenzyl propyl ether**

An oven dried flask was cooled under an atmosphere of dry nitrogen, and subsequently charged with THF (280 ml) and sodium hydride (1.72 g, 43 mmol, 1.5 mol eqv.). The suspension was stirred with cooling *via* an ice bath. 3,5-Difluorobenzyl alcohol (4.13 g, 28.7 mmol, 1 mol eqv.) was added dropwise, and the reaction stirred for one hour. 1-Bromopropane (24.4 g, 200 mmol, 7 mol eqv) was added dropwise. The reaction was monitored by TLC until completion. Once complete, the reaction was quenched by slow addition of water. The reaction was diluted with ethyl acetate, the organic layer was extracted and set aside. The aqueous layer was washed with ethyl acetate and discarded. The combined organics were dried over $MgSO_4$ and concentrated *in vacuo*. Chromatography of the crude material with a gradient of DCM/hexanes afforded the title compound as a colourless liquid.

| | |
|---|---|
| Yield: | 3.14 g, 59% |
| $R_F$: | 0.7 (DCM: hexane=1:1) |
| $^{1}$H NMR (400 MHz): | 6.87(m, 2H,Ar-**H**), 6.70 (tt, *J* = 9.0 Hz, *J* = 2.4 Hz, 1H,Ar-**H**), 4.47 (s, 2H,O-**$CH_2$**-Ar), 3.45 (t, *J* = 6.7 Hz, 2H,O-**$CH_2$**-$CH_2$-$CH_3$), 1.65 (m, 2H,O-$CH_2$-**$CH_2$**-$CH_3$), 0.96 (t, *J* = 7.4 Hz, 3H,O-$CH_2$-$CH_2$-**$CH_3$**). |
| $^{13}$C{$^{1}$H} NMR (101 MHz): | 163.23 (dd, *J* = 248.3, *J* = 12.6 Hz), 143.16 (t, *J* = 8.8 Hz), 109.89 (dd, J=6.7,18.6 Hz), 102.76 (t, *J* = 25.2 Hz), 72.66, 71.68, 23.06, 10.72. |

$^{19}$F NMR (376 MHz): -110.15 (t, *J* = 8.0 Hz Ar-F)

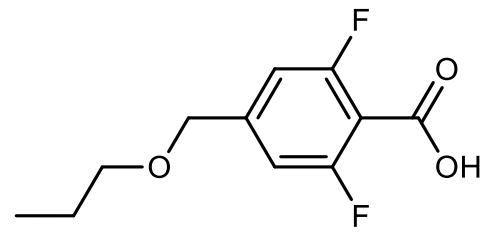

**IDX7b / WJ-2-52 / 2,6-difluoro-2-(propan-1-yloxy)benzoic acid**

Quantities used: 3,5-difluorobenzyl propyl ether (1.65 g, 8.9 mmol, 1 mol eqv), anhydrous THF (100 ml), n-butyl lithium (1.6 M in hexane, 10.6 mmol, 1.2 mol eqv.). General procedure B was followed. The resulting material was used without further purification.

Yield: 1.92 g, 94 % (pale yellow solid)

$R_F$: 0.0 (DCM)

$^{1}$H NMR (400 MHz): 11.45 (s, 1H,-COO**H**), 6.94 (m, 2H,Ar-**H**), 4.51 (s, 2H, O-$C\mathbf{H}_2$-Ar), 3.46 (t, *J* = 6.6 Hz, 2H,$CH_3$-$CH_2$-$C\mathbf{H}_2$-O), 1.64 (m, 2H,$CH_3$-$C\mathbf{H}_2$-$CH_2$-O), 0.92 (t, *J* = 7.4 Hz, 3H,$C\mathbf{H}_3$-$CH_2$-$CH_2$-O).

$^{13}$C{$^{1}$H} NMR (101 MHz): 166.20, 161.40 (dd, *J* = 259.2, *J* = 5.9 Hz), 146.74 (t, *J* = 9.7 Hz), 110.39 (dd, *J* = 23.5, *J* =3.0 Hz), 108.37 (t, *J* = 16.4 Hz), 72.79, 70.87, 22.74, 10.42.

$^{19}$F NMR (376 MHz): -108.26 (d, *J* = 9.9 Hz).

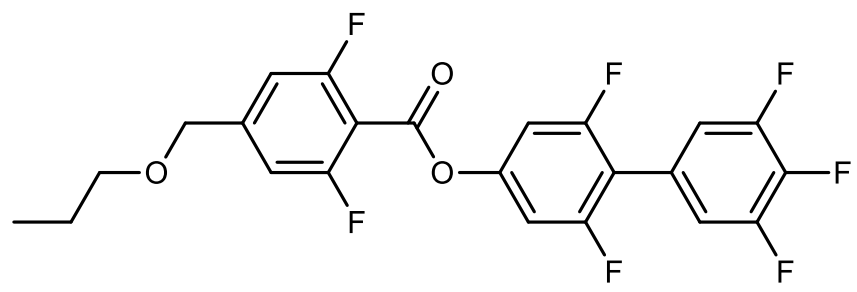

**IDX7/ WJ-2-54 / (2,3',4',5',6-pentafluoro-[1,1'-biphenyl]-4-yl)methyl 2,6-difluoro-4-(propoxymethyl)benzoate**

Quantities used: 2,6-difluoro-2-(propan-1-yloxy)benzoic acid (141 mg, 0.614 mmol, 1.2 mol eqv.), 4-(3,4,5-trifluorophenyl)-3,5-difluorophenol (133 mg, 0.511 mmol, 1 mol eqv.), EDC.HCl (147 mg, 0.77 mmol, 1.5 mol eqv.), DMAP (18.7 mg, 0.153 mmol, 30 mol%), DCM (5 ml). The reaction was performed according to general procedure A. The crude material was purified by flash chromatography over silica gel with a gradient of hexane/DCM, followed by recrystallization from n-hexane.

Yield: 120 mg (50%)

$R_F$: 0.8 (DCM/hexane, 1:1)

$^{1}$H NMR (400 MHz): 7.12 (m, $J$ = 7.6 Hz, 2H,Ar-**H**), 7.02 (m, 4H,Ar-**H**), 4.55 (s, 2H, O-C**H**$_2$-Ar), 3.49 (t, $J$ = 6.6 Hz, 2H,$CH_3$-$CH_2$-C**H**$_2$-O-), 1.68 (m, 2H,$CH_3$-C**H**$_2$-$CH_2$-O-), 0.98 (t, $J$ = 7.4 Hz, 3H,C**H**$_3$-$CH_2$-$CH_2$-O-).

$^{13}$C{$^{1}$H} NMR (101 MHz): 162.70 (dd, $J$ = 259.5, $J$ = 6.4 Hz),159.06, 158.75 (dd, $J$ = 250.4, $J$ = 8.1 Hz), 151.18 (ddd, $J$ = 250.1, $J$ = 10.3, $J$ = 4.3 Hz), 150.92 (t, $J$ = 14.4 Hz), 147.96 (t, $J$ = 9.7 Hz), 139.91 (dt, $J$ = 253.5, $J$ = 14.8 Hz), 124.36 (td, $J$ = 8.6, $J$ = 5.7 Hz), 115.27 – 114.55 (m), 113.95 (t, $J$ = 18.1 Hz), 110.49 (dd, $J$ = 22.8, 3.3 Hz), 107.74 (t, $J$ = 16.5 Hz), 106.64 (m), 73.01, 71.06, 23.01, 10.67.

$^{19}$F NMR (471 MHz): -108.36 (d, $J$ = 9.8 Hz), -112.20 (d, $J$ = 8.3 Hz), -134.31 (dd, $J$ = 20.7, 8.1 Hz), -159.93 (tt, $J$ = 20.7, $J$ = 6.5 Hz).

HRMS (APCI+): 473.0979 (calcd. For $C_{23}H_{16}F_7O_3$: 473.0982, err: 0.6 ppm; M+H)

Assay (HPLC, C18): 99.39% (254 nm , 5.176 min)

## 6.2. Scheme Two

### IDX2538 / SRB-CHEM-082 / 4-hydroxy-(3’,5’-difluoro-4’-(difluoro(3,4,5-trifluorophenoxy)methyl)-1,1’-biphenyl)yl pentanoate

Quantities used: 4-Hydroxy-(3’,5’-difluoro-4’-(difluoro(3,4,5-trifluorophenoxy)methyl)-1,1’-biphenyl) (0.402 g, 1 mmol), [3]valeric acid (0.122 g, 1.2 mmol), DCC (0.25 g, 1.2 mmol) and DMAP (12 mg, 0.1 mmol), DCM (10 ml). General procedure A was followed, with the exception that DCC was used in place of EDC.HCl. The crude material was purified via flash column chromatography over silica, using a gradient of hexane to dichloromethane as eluent. The resulting solid was then recrystallized from 2-propanol, to yield the product as colourless crystals.

Yield: 0.36 g, (73 %)

$R_F$: 0.59 (DCM/Hexane, 1:1)

$^{1}$H NMR (400 MHz): 7.59 (dt, *J* = 8.6, *J* = 2.8 Hz, 2H, Ar-**H**), 7.28 – 7.17 (m, 4H, Ar-**H**)*, 7.07 – 6.95 (m, 2H, Ar-**H**)*, 2.63 (t, *J* = 7.5 Hz, 2H, R-C**H**$_2$-COOAr), 1.79 (m, 2H, R-C**H**$_2$-CH$_2$-CH$_3$), 1.50 (m, *J* = 7.6 Hz, 2H, R-CH$_2$-C**H**$_2$-CH$_3$), 1.02 (t, *J* = 7.4 Hz, 3H, CH$_2$-C**H**$_3$). (*Overlapping Peaks)

$^{13}$C{$^{1}$H} NMR (101 MHz): 172.13, 160.28 (dd, *J* = 257.5, *J* = 10.1 Hz), 151.73, 151.02 (ddd, *J* = 251.2, *J* = 10.5, *J* = 5.1 Hz), 146.12 (t, *J* = 10.5 Hz), 144.92 – 144.41 (m), 138.24 (dt, *J* = 251.3, *J* = 15.5 Hz), 128.09, 122.50, 120.26 (t, *J* = 265.4 Hz), 117.61, 110.97 (dd, *J* = 24.1, *J* = 3.0 Hz), 108.82 – 108.01 (m), 107.45 (dd, *J* = 16.2, *J* = 6.5 Hz), 34.11, 26.96, 22.26, 13.72.

$^{19}$F NMR (471 MHz): -61.66 (t, *J* = 26.2 Hz, 2F, O-C**F**$_2$-Ar), -110.17 (td, *J* = 26.2, *J* = 11.1 Hz, 2F, Ar-**F**), -132.48 (dd, *J* = 20.8, *J* = 8.6 Hz, 2F, Ar-**F**), -163.17 (tt, *J* = 21.0, *J* = 6.0 Hz, 1F, Ar-**F**).

HRMS (ESI+): 487.1142 ([M+H]$^+$): calcd. For $C_{24}H_{18}F_{7}0_{3}$: 487.1139, err = -0.6 ppm)

Assay (HPLC, C18): 99.78% (254 nm, 5.774 min)

m.p.: 63.1 °C

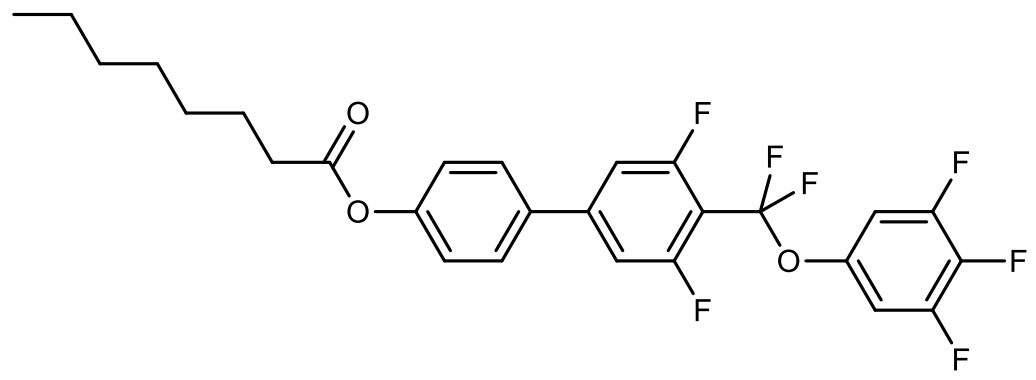

**IDX1646 / SRB-CHEM-085 / 4-hydroxy-(3’,5’-difluoro-4’-(difluoro(3,4,5-trifluorophenoxy)methyl)-1,1’-biphenyl)yl octanoate**

Quantities used: 4-Hydroxy-(3’,5’-difluoro-4’-(difluoro(3,4,5-trifluorophenoxy)methyl)-1,1’-biphenyl) (0.402 g, 1 mmol), ), [3] octanoic acid (0.173 g, 1.2 mmol), DCC (0.25 g, 1.2 mmol), DMAP (12 mg, 0.1 mmol), DCM (10 ml). General procedure A was followed, with the exception that DCC was used in place of EDC.HCl. The crude material was purified via flash column chromatography over silica, using a gradient of hexane to dichloromethane as eluent. The resulting solid was then recrystallized from 2-propanol, to yield the product as colourless crystals.

Yield: 0.30 g, (56 %)

$R_F$: 0.65 (DCM/Hexane, 1:1)

$^{1}$H NMR (400 MHz): 7.50 (dt, *J* = 8.7, *J* = 2.8 Hz, 2H, Ar-**H**), 7.17 – 7.08 (m, 4H, Ar-**H**)*, 6.96 – 6.87 (m, 2H, Ar-**H**)*, 2.52 (t, *J* = 7.5 Hz, 2H, RC**H**$_2$-COOAr), 1.70 (m, , 2H, R-C**H**$_2$-R), 1.42 – 1.17 (m, 8H, R-C**H**$_2$-R)*, 0.83 (t, *J* = 6.8 Hz, 3H, R-C**H**$_3$). (*Overlapping Peaks)

$^{13}$C{$^{1}$H} NMR (101 MHz): 172.14, 160.26 (dd, *J* = 257.7, *J* = 5.2 Hz), 151.74, 151.03 (ddd, *J* = 254.8, *J* = 10.3, *J* = 5.2 Hz), 146.12 (t, *J* = 10.6 Hz), 144.91 – 144.39 (m), 138.46 (dt, *J* = 250.5, 14.7 Hz), 134.96, 128.10, 122.50, 120.26 (t, *J* = 265.2 Hz), 117.62, 110.98 (dd, *J* = 24.1, *J* = 3.0 Hz), 108.71-108.0 (m), 107.45 (dd, *J* = 16.3, 6.4 Hz), 34.40, 31.67, 29.08, 28.93, 24.91, 22.61, 14.07.

$^{19}$F NMR (471 MHz): -61.66 (t, *J* = 26.2 Hz, 2F, O-C**F**$_2$-Ar), -110.16 (td, *J* = 26.2, *J* = 11.2 Hz, 2F, Ar-**F**), -132.47 (dd, *J* = 20.6, *J* = 8.7 Hz, 2F, Ar-**F**), -163.16 (tt, *J* = 20.9, 6.2 Hz, 1F, Ar-**F**).

HRMS (ESI+): 529.1615 ([M+H]$^+$): calcd. For $C_{27}H_{24}F_7O_3$: 529.1608, err = -1.3 ppm)

Assay (HPLC, C18): 99.16% (254 nm, 6.291 min)

m.p.: 64.8 °C

## 6.3. Scheme Three

**IDX323 / CJG 412 / 2-bromo-5-ethoxy-1,3-diflurobenzene**

4 bromo, 3,5 diflurophenol (4.16 g 20 mmol, 1 mol eqv.) was dissolved in 100 mL of acetone. Potassium carbonate (4.14 g, 30 mmol, 1.5 mol eqv.) was added, and the reaction stirred for 30 mins before iodo ethane (3.43 g, 1.77 mL, 22 mmol, 1.1 mol eqv.) was added in one portion and the reaction was heated under reflux until complete - as judged by the consumption of the phenolic reagent by TLC (circa 24 h). The reaction was then filtered, the filtrate washed with acetone, the volatiles removed *in vacuo*, before being purification by flash chromatography using a gradient of DCM:hexanes. The chromatographed product was then recrystallised from EtOH to afford the title compound as a white solid.

Yield: (White solid) 3.98 g, 84 %

$R_F$ (Hexanes): 0.10

$^{1}$H NMR (400 MHz): 6.48 (ddd, *J* = 13.2 Hz, *J* = 10.0 Hz, *J* = 8.2 Hz, 2H, Ar-**H**), 3.96 (d, *J* = 7.0 Hz, 2H, O-C$\mathbf{H_2}$-$CH_3$), 1.40 (t, *J* = 7.0 Hz, 3H, $CH_2$-C$\mathbf{H_3}$).

$^{13}$C{$^{1}$H} NMR (101 MHz): 160.24 (dd, *J* = 246.5 Hz, *J* = 7.1 Hz), 159.60 (t, *J* = 12.9 Hz), 99.08 (dd, *J* = 26.4 Hz, *J* = 2.9 Hz), 88.05 (t, *J* = 25.0 Hz), 64.24, 14.39.

$^{19}$F NMR (376 MHz): -105.23 (d, *J* = 9.5 Hz, 2F. Ar-**F**).

**IDX??? / CJG 483 / 4-ethoxy-2,6-diflurobenzoic acid**

Quantities used: 2-bromo-5-ethoxy-1,3-diflurobenzene (3.2 g, 13.5 mmol, 1 mol eqv.), anhydrous THF (100 mL), n-butyl lithium (1.6 M in hexane, 12.65 mL, 20.25 mmol, 1.5 mol eqv.) General procedure B was followed. The crude material was re-crystallised from EtOH to afford the title compound as a crystalline white solid.

Yield: 2.21 g, 81 %

$R_F$ (DCM): 0.00

$^{1}$H NMR (400 MHz, DMSO): 13.40 (s, 1H, Ar-COO**H**), 7.00 – 6.68 (m, 2H, Ar-**H**), 4.10 (q, *J* = 7.0 Hz, 2H, $\text{O-C}\mathbf{H_2}\text{-CH}_3$), 1.33 (t, *J* = 7.0 Hz, 3H, $\text{CH}_2\text{-C}\mathbf{H_3}$).

$^{13}$C{$^{1}$H} NMR (101 MHz, DMSO): 162.64 – 162.19 (m)*, 161.63 (dd, *J* = 251.8, *J* = 10.3 Hz), 104.10 (t, *J* = 18.4 Hz), 100.36 – 98.98 (m), 65.08, 14.67. *Overlapping signals.

$^{19}$F NMR (376 MHz, DMSO): -109.46 (d, *J* = 11.7 Hz, 2F, Ar-**F**).

**IDX 232 / CJG 484 / 2,3',4',5'-tetrafluoro-[1,1'-biphenyl]-4-yl 4-ethoxy-2,6-difluorobenzoate:** Quantities used: 4-ethoxy-2,6-diflurobenzoic acid (130 mg, 0.6 mmol, 1.2 mol eqv.), 4-(3,4,5-trifluorophenyl)-3-fluorophenol (121 mg, 0.50 mmol, 1 mol eqv.), EDC.HCl (143 mg, 0.75 mmol, 1.5 mol eqv.), DMAP (5 mg, 0.05 mmol, 30 mol%), DCM (5 ml). The reaction was performed according to general procedure A. The crude material was purified by flash chromatography over silica gel with a gradient of hexane/DCM, followed by recrystallization from EtOH to afford a white solid.

Yield: 307 mg, 72 %

$R_F$ (DCM): 0.90

$^{1}$H NMR (400 MHz): 7.42 (t, *J* = 8.7 Hz, 1H, Ar-**H**), 7.21 – 7.10 (m, 4H, Ar-**H**)*, 6.54 (d, *J* = 10.5 Hz, 2H, Ar-**H**), 4.09 (q, *J* = 7.0 Hz, 2H, O-C**H$_2$**-CH$_3$), 1.46 (t, *J* = 7.0 Hz, 3H, CH$_2$-C**H$_3$**). *Overlapping signals.

$^{13}$C{$^{1}$H} NMR (101 MHz): 164.05 – 163.55 (m)*, 163.00 (dd, *J* = 258.3 Hz, 8.4 Hz), 159.42 (t, *J* = 2.9 Hz), 159.36 (d, *J* = 251.3 Hz), 151.20 (ddd, *J* = 250.4 Hz, 10.3 Hz, 4.5 Hz), 151.18 (d, *J* = 10.9 Hz), 140.76, 138.24 (t, *J* = 15.3 Hz), 135.51 (t, *J* = 10.1 Hz), 131.03 – 130.82 (m), 130.56 (d, *J* = 4.0 Hz), 124.09 (d, *J* = 11.4 Hz), 118.26 (d, *J* = 3.8 Hz), 113.49 – 112.93 (m), 110.85, 110.59, 101.48 (t, *J* = 15.5 Hz), 99.28 (dd, *J* = 26.1, 3.1 Hz), 64.83, 14.39. *Overlapping signals.

$^{19}$F NMR (376 MHz): -106.16 (d, *J* = 11.6 Hz, 2F, Ar-**F**), -114.60 (t, *J* = 9.7 Hz, 1F, Ar-**F**), -134.24 (dd, *J* = 20.6, 8.7 Hz, 2F, Ar-**F**), -161.29 (tt, *J* = 20.7, 6.6 Hz, 1F, Ar-**F**).

## 6.4. Scheme 4

**CJG 411 / 3,5-difluorobenzyl butyl ether**

An oven-dried flask was cooled under an atmosphere of dry nitrogen, and subsequently charged with anhydrous THF (100 ml) and sodium hydride (60%, 3.00 g, 75 mmol, 1.5 mol eqv.). The suspension was stirred with cooling *via* an ice bath. 3,5-Difluorobenzyl alcohol (7.20 g, 50.0 mmol, 1 mol eqv.) was added dropwise, and the reaction was left to stir for one hour. 1-Bromobutane (8.2 g, 60 mmol, 1.1 mol eqv) was added dropwise. The reaction was monitored by TLC until completion. Once complete, the reaction was quenched by slow addition of water. The reaction was diluted with EtOAc, and the organic layer was extracted 2x 20 mL EtOAc. The combined organics were dried over $MgSO_4$ and concentrated *in vacuo*. Chromatography of the crude material with a gradient of DCM/hexanes afforded the title compound as a colourless oil.

Yield: 6.02 g, 62 %

$R_F$ (DCM): 0.40

$^{1}$H NMR (400 MHz): 6.93 – 6.84 (m, 2H, Ar-**H**), 6.71 (tt, *J* = 9.0 Hz, *J* = 2.5 Hz, 1H, Ar-**H**), 4.48 (s, 2H, Ar-C$\mathbf{H_2}$-O), 3.51 (t, *J* = 6.5 Hz, 2H, O-C$\mathbf{H_2}$-$CH_2$), 1.64 (m, 2H, $CH_2$-C$\mathbf{H_2}$-$CH_2$), 1.45 (m, 2H, $CH_2$-$CH_2$-C$\mathbf{H_3}$), 0.96 (t, *J* = 7.4 Hz, 3H, $CH_2$-C$\mathbf{H_3}$).

$^{13}$C{$^{1}$H} NMR (101 MHz): 163.10 (dd, *J* = 248.3 Hz, *J* = 12.6 Hz), 143.15 (t, *J* = 8.8 Hz), 109.95 – 109.42 (m), 102.52 (t, *J* = 25.4 Hz), 71.52, 70.61, 31.78, 19.35, 13.81.

$^{19}$F NMR (376 MHz): -110.17 (t, *J* = 8.0 Hz, F, Ar-**F**).

**CJG 448 / 2,6-difluoro-2-(butyl-1-yloxy)benzoic acid**

Quantities used: 3,5-Difluorobenzyl butyl ether (1.30 g, 6 mmol, 1 mol eqv), anhydrous THF (50 mL) n-butyl lithium (1.6 M in hexane, 5.63 mL 9 mmol, 1.5 mol eqv.) was added dropwise. General procedure B was followed. The resulting material was used without further purification.

| | |
|---|---|
| Yield: | 1.30 g, 89 % |
| $R_F$ (DCM): | 0.00 |
| $^{1}$H NMR (400 MHz) | 7.12 (d, *J* = 8.9 Hz, 2H, Ar-**H**), 4.50 (s, 2H, Ar-C$\mathbf{H_2}$-O), 3.45 (t, *J* = 6.5 Hz, 2H, O-C$\mathbf{H_2}$-$CH_2$), 1.54 (m, 2H, $CH_2$-C$\mathbf{H_2}$-$CH_2$), 1.35 (m, 2H, $CH_2$-C$\mathbf{H_2}$-$CH_3$), 0.89 (t, *J* = 7.4 Hz, 3H, $CH_2$-C$\mathbf{H_3}$). |

$^{19}$F NMR (376 MHz): -111.97 (d, *J* = 9.7 Hz, 2F, Ar-**F**).

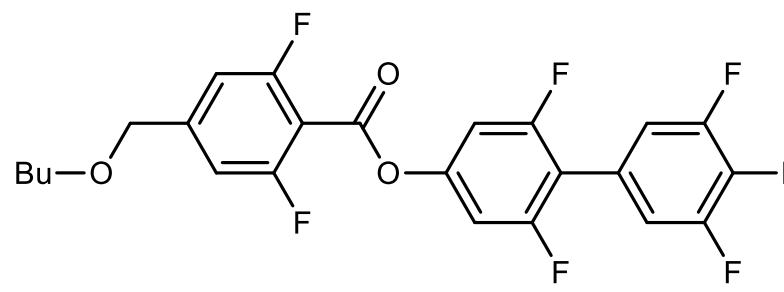

**IDX 59 / CJG 480 / 2,3',4',5',6-pentafluoro-[1,1'-biphenyl]-4-yl 4-(butoxymethyl)-2,6-difluorobenzoate**

Quantities used: 2,6-difluoro-2-(Butyl-1-yloxy)benzoic acid (383 mg, 1.65 mmol, 1.2 mol eqv.), 4-(3,4,5-trifluorophenyl)-3,5-difluorophenol (390 mg, 1.50 mmol, 1 mol eqv.), EDC.HCl (501 mg, 2.63 mmol, 1.5 mol eqv.), DMAP (18 mg, 0.15 mmol, 30 mol%), DCM (5 ml). The reaction was performed according to general procedure A. The crude material was purified by flash chromatography over silica gel with a gradient of hexane/DCM, followed by recrystallization from MeCN to give a white solid.

Yield: 510 mg, 70 %

$R_F$ (DCM): 0.85

$^1$H NMR (400 MHz) 7.17 – 7.08 (m, 2H, Ar-**H**), 7.07 – 6.98 (m, 4H, Ar-**H**)*, 4.54 (s, 2H, Ar-C$\mathbf{H_2}$-O), 3.53 (t, *J* = 6.5 Hz, 2H, O-C$\mathbf{H_2}$-CH$_2$), 1.64 (p, *J* = 6.7 Hz, 2H, CH$_2$-C$\mathbf{H_2}$-CH$_2$), 1.44 (h, *J* = 7.5 Hz, 2H, CH$_2$-C$\mathbf{H_2}$-CH$_3$), 0.96 (t, *J* = 7.4 Hz, 3H, CH$_2$-C$\mathbf{H_3}$).*Overlapping Signals

$^{13}$C{$^1$H} NMR (101 MHz) 161.29 (dd, *J* = 259.5 Hz, *J* = 5.8 Hz), 159.98, 159.58 (dd, *J* = 250.4 Hz, *J* = 8.5 Hz), 151.25 (ddd, *J* = 250.2 Hz, *J* = 10.2 Hz, *J* = 4.0 Hz), 150.77 (t, *J* = 14.3 Hz), 147.80 (t, *J* = 9.7 Hz), 139.44 (dt, *J* = 254.5 Hz, *J* = 14.6 Hz), 124.44 – 124.01 (m), 113.87 (t, *J* = 17.5 Hz), 110.39 (dd, *J* = 22.7 Hz, 3.2 Hz), 107.65 (t, *J* = 16.5 Hz), 106.90 – 106.24 (m), 71.06, 71.01, 31.73, 19.35, 13.89.

$^{19}$F NMR (376 MHz) -108.37 (d, *J* = 10.2 Hz, 2F, Ar-**F**), -112.22 (d, *J* = 8.8 Hz, 2F, Ar-**F**), -134.33 (dd, *J* = 20.6, *J* = 8.3 Hz, 2F, Ar-**F**), -159.96 (tt, *J* = 20.9, *J* = 6.5 Hz, 1F, Ar-**F**).

## 6.5. Scheme 5

### CJG 348 / 2-(3-fluorophenyl)-5-propyl-1,3-dioxane

An oven dried flask, fitted with a dean-stark apparatus, was cooled under an atmosphere of dry nitrogen and charged with 3-fluorobenzaldehyde (13.87 g, 110 mmol, 1.1 mol eqv.), propyl-1,3 propanediol (12 g, 100 mmol, 1 mol eqv.), para-toluenesulfonic acid (3.8 g 20 mmol, 0.2 mol eqv.) and 150 mL of anhydrous toluene. The reaction mixture was heated to reflux, with any evolved water being removed at regular intervals, until no further water was captured within the dean-stark apparatus (circa 48 h). The reaction was cooled to room temperature and concentred *in vacuo* before being purified by flash chromatography using a gradient of EtOAc:hexanes. The chromatographed material was then recrystallised from MeOH to afford the title compound as a white solid.

Yield: 12.1 g, 54 %

Spectral data was in keeping with literature values.[4]

**CJG 358 / 2-fluoro-4-(5-propyl-1,3-dioxan-2-yl)benzoic acid**

Quantitues used: 2-(3-fluorophenyl)-5-propyl-1,3-dioxane (12.0 g, 54 mmol, 1 mol eqv), anhydrous THF (50 mL), n-butyl lithium (1.6 M in hexane, 41.0 mL 65 mmol, 1.5 mol eqv.). General procedure B was followed. The resulting material was purified by recrystallisation from hexane: toluene (5:1) to afford the title compound as an off white solid.

Yield: (Off-white solid)13.3 g, 92 %

$R_F$ (DCM): 0.00

$^{1}$H NMR (400 MHz, DMSO): 13.47 (s, 1H, Ar-COO**H**), 7.53 – 7.39 (m, 2H, Ar-**H**)*, 7.33 – 7.25 (m, 1H, Ar-**H**), 5.66 (s, 1H), 4.13 (dd, *J* = 11.6, 4.6 Hz, 2H, O-C$\mathbf{H_{ax}}$(H$_{eq}$)-CH), 3.53 – 3.42 (m, 2H, -C$\mathbf{H_{eq}}$(H$_{ax}$)-CH), 2.02 – 1.88 (m, 1H, (CH$_2$)$_2$-C**H**-CH$_2$), 1.27 (h, *J* = 6.8 Hz, 2H, CH$_2$-C$\mathbf{H_2}$-CH$_3$), 1.04 (q, *J* = 7.5 Hz, 2H, CH$_2$-C$\mathbf{H_2}$-CH$_3$), 0.86 (t, *J* = 7.3 Hz, 3H, CH$_2$-C$\mathbf{H_3}$). *Overlapping Signals.

$^{13}$C{$^{1}$H} NMR (101 MHz, DMSO): 166.05, 158.89 (d, *J* = 247.0 Hz), 138.26 (d, *J* = 3.0 Hz), 131.42 (d, *J* = 8.8 Hz), 122.24, 122.06, 121.96 (d, *J* = 3.0 Hz), 116.44 (d, *J* = 21.6 Hz), 97.99 (d, *J* = 2.8 Hz), 72.19, 34.01, 30.09, 19.44, 14.58.

$^{19}$F NMR (376 MHz, DMSO): -116.17 (dd, *J* = 9.9, 5.6 Hz, 1F, Ar-**F**).

Spectral data was in keeping with literature values. [5]

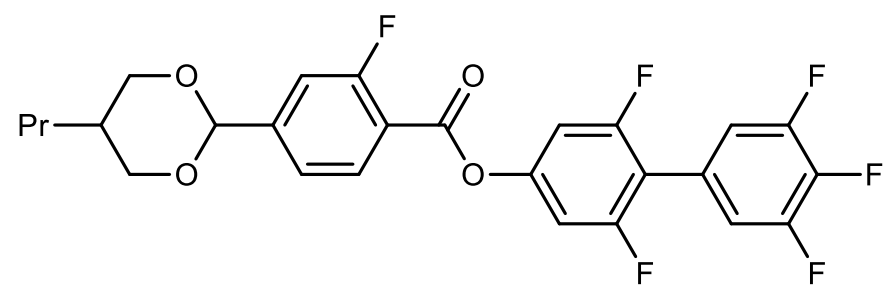

**IDX 519 / CJG 410 / 2,3',4',5',6-pentafluoro-[1,1'-biphenyl]-4-yl 2-fluoro-4-(5-propyl-1,3-dioxan-2-yl)benzoate**

Quantities used: 2-fluoro-4-(5-propyl-1,3-dioxan-2-yl)benzoic acid (321 mg, 1.2 mmol, 1.2 mol eqv.), 4-(3,4,5-trifluorophenyl)-3,5-difluorophenol (260 mg, 1 mmol, 1 mol eqv.), EDC.HCl (501 mg, 2.63 mmol, 1.5 mol eqv.), DMAP (12 mg, 0.10 mmol, 30 mol%), DCM (5 ml). The reaction was performed according to general procedure A. The crude material was purified by flash chromatography over silica gel with a gradient of hexane/DCM, followed by recrystallization from MeOH to give a white solid.

| | |
|---|---|
| Yield: | 428 mg, 84 % |
| $R_F$ (DCM): | 0.83 |
| $^{1}$H NMR (400 MHz): | 7.56 – 7.47 (m, 2H, Ar-**H**)*, 7.22 – 7.09 (m, 4H, Ar-**H**)*, 7.01 – 6.94 (m, 2H, Ar-**H**), 5.75 (s, 1H), 4.25 (dd, *J* = 11.8, 4.7 Hz, 2H, O-C$\mathbf{H_{ax}}$($H_{eq}$)-CH), 3.54 (t, *J* = 11.5 Hz, 2H, O-C$\mathbf{H_{eq}}$($H_{ax}$)-CH), 2.21 – 2.06 (m, 1H, $(CH_2)_2$-C**H**-$CH_2$), 1.33 (h, *J* = 7.3 Hz, 2H, $CH_2$-C$\mathbf{H_2}$-$CH_3$), 1.09 (q, *J* = 7.4 Hz, 2H, $CH_2$-C$\mathbf{H_2}$-$CH_3$), 0.92 (t, *J* = 7.3 Hz, 3H, $CH_2$-C$\mathbf{H_3}$). *Overlapping signals. |
| $^{13}$C{$^{1}$H} NMR (101 MHz): | 163.01, 161.05, 160.18 (dd, *J* = 250.7 Hz, 7.5 Hz), 151.15 (t, *J* = 14.2 Hz), 151.11 (ddd, *J* = 250.2 Hz, 9.8 Hz, 4.2 Hz), 139.50 (dd, *J* = 254.0 Hz, 15.6 Hz), 138.88, 132.31 (d, *J* = 8.6 Hz), 124.29 (t, *J* = 8.9 Hz), 121.85 (d, *J* = 3.3 Hz), 118.69 (d, *J* = 16.6 Hz), 116.41 (d, *J* = 21.6 Hz), 114.83 (dd, *J* = 20.0, 6.0 Hz), 106.68 – 106.23 (m), 98.07 (d, *J* = 2.8 Hz), 72.61, 33.90, 30.21, 19.53, 14.16. |
| $^{19}$F NMR (376 MHz): | -112.28 (d, *J* = 8.8 Hz, 2F, Ar-**F**), -114.40 (dd, *J* = 9.3, *J* = 4.8 Hz, 1F, Ar-**F**), -134.30 (dd, *J* = 20.5, *J* = 8.5 Hz, 2F, Ar-**F**), -159.94 (tt, *J* = 20.9, *J* = 6.6 Hz, 1F, Ar-**F**). |

## 6.6. Scheme 6

**CJG 290 / 4-(3,4,5-trifluorophenyl)-3,5-difluorophenol**

Quantities used: 4-Bromo-3,5-difluorophenol (10.45 g, 50.0 mmol), 3,4,5-trifluorobenzeneboronic acid (9.67 g, 55.0 mmol) THF (125 ml) 2M aqueous $K_2CO_3$ (50 mL), Pd-XPHOS-G3 (~ 1 mol%). General procedure C was followed. The crude material was purified by flash chromatography using a gradient of hexanes:DCM. The chromatographed material was then re-crystallised from hexanes: toluene (10:1) to afford the product as a white solid.

Yield: (White Solid) 11.96 g, 92 %

$R_F$ (DCM): 0.60

$^{1}$H NMR (400 MHz, DMSO): 10.63 (s, 1H, Ar-O**H**), 7.36 – 7.23 (m, 2H, Ar-**H**), 6.59 – 6.51 (m, 2H, Ar-**H**).

$^{13}$C{$^{1}$H} NMR (101 MHz, DMSO): 206.63, 160.19 (dd, *J* = 245.6 Hz, 10.1 Hz) 160.00 (t, *J* = 15.1 Hz), 150.48 (ddd, *J* = 247.2 Hz, 9.7 Hz, 4.2 Hz), 138.29 (dt, *J* = 250.2 Hz, 15.6 Hz), 126.44 – 125.91 (m), 115.27 (dd, *J* = 21.6 Hz, 4.6 Hz), 106.08 (t, *J* = 18.9 Hz), 100.62 (dd, *J* = 25.0, 2.8 Hz), 100.34 – 99.64 (m).

$^{19}$F NMR (376 MHz, DMSO): -115.06 (d, *J* = 10.5 Hz, 2F, Ar-**F**), -135.84 (dd, *J* = 21.3, 9.5 Hz, 2F, Ar-**F**), -162.47 (tt, *J* = 21.6, 6.8 Hz, 1F, Ar-**F**).

## 6.6. Scheme 7

**RM3024 / 2,6-Difluoro-4-(4-hydroxyphenyl)toluene**

Quantities used: 4-Bromo-2,6-difluorotoluene (5.0 g, 24.4 mmol), 4-hydroxybenzene boronic acid pinacol ester (6.2 g, 28 mmol), THF (100 ml), 2M aqueous sodium carbonate (50 ml), Pd-XPHOS-G3 (5 mg). General Procedure C was followed. The crude

| | |
|---|---|
| Yield: | 4.3 g, 80.1% |
| $R_F$ (Hexanes): | 0.65 |
| $^{1}$H NMR (400 MHz): | 9.67 (s, 1H, ArO**H**), 7.53 (ddd, 2H, *J* = 8.9 Hz, *J* = 3.1 Hz, *J* = 2.4 Hz, Ar**H**), 7.33 – 7.22 (m, 2H), 6.84 (ddd, 2H, *J* = 8.6 Hz, *J* = 3.1 Hz, *J* = 2.1 Hz, Ar**H**), 2.14 (t, 3H, *J* = 1.7 Hz, Ar-C$\mathbf{H}_3$) |
| $^{19}$F NMR (376 MHz): | -114.88 (d, *J* = 8.6 Hz, Ar**F**) |

**3',5'-difluoro-4'-methyl-[1,1'-biphenyl]-4-yl 2,6-difluoro-4-((2r,5r)-5-propyl-1,3-dioxan-2-yl)benzoate *(*WCO 4.7 (IDX1013)):** Quantities Used: 2,6-difluoro-4-((2r,5r)-5-propyl-1,3-dioxan-2-yl)benzoic acid (231 mg, 0.807 mmol), 3',5'-difluoro-4'-methyl-[1,1'-biphenyl]-4-ol (157 mg, 0.713 mmol), EDC.HCl (339 mg, 1.768 mmol), DMAP (cat. amount), DCM (ca. 5 ml). The reaction was performed according to general procedure A. The crude material was purified by flash chromatography over silica gel with a gradient of hexane/DCM, followed by recrystallization from ethanol to afford the title compound as colourless needles.

| | |
|---|---|
| Yield: | 66 mg, 19 % |
| $^{1}$H (400 MHz): | 7.58 (m, 2H, Ar-**H**), 7.33 m, 2H, Ar-**H**), 7.18 (d, *J* = 8.7 Hz, 2H, Ar-**H**), 7.12 – 7.03 (m, 2H, Ar-**H**), 5.40 (s, 1H, ArO$_2$C-**H**), 4.25 (dd, *J* = 11.8, *J* = 4.8 Hz, 2H, OC-**H**$_2$), 3.54 (t, *J* = 11.5 Hz, 2H, OC-**H**$_2$), 2.23 (s, 3H, ArC-**H**$_3$), 2.20 – 2.07 (m, 1H, CCC-**H**), 1.35 (h, *J* = 7.5 Hz, 2H, C-**H**$_2$), 1.11 (q, *J* = 8.3 Hz, 2H, C-**H**$_2$), 0.94 (t, *J* = 7.3 Hz, 3H, C-**H**$_3$) |
| $^{13}$C{$^{1}$H} (101 MHz): | 162.00 (dd, *J* = 245.9 Hz, *J* = 9.6 Hz), 160.98 (dd *J* = 258.0 Hz, *J* = 5.4 Hz), 159.91, 150.45, 145.35, 139.80, 137.38, 128.12, 122.25, 112.42, 110.31 (dd, *J* = 23.9 Hz, 2.8 Hz), 109.64 (d, *J* = 8.6 Hz), 109.37 (d, *J* = 8.9 Hz), 99.01, 72.72, 34.04, 30.38, 19.67, 14.33, 7.13 (t, *J* = 3.6 Hz) |
| $^{19}$F NMR(376 MHz): | -108.81 (d, *J* = 9.7 Hz, Ar-**F**), -114.52 (d, *J* = 7.6 Hz, Ar-**F**) |
| HRMS (ESI+): | 489.1683 ([M+H], calcd. for ($C_{27}H_{25}F_4O_4$): 489.1683, err = 0.1 ppm) |
| HPLC: | Assay (HPLC, C18): 99.96 % (254 nm, 5.308 min) |

## 6.8. Scheme 8

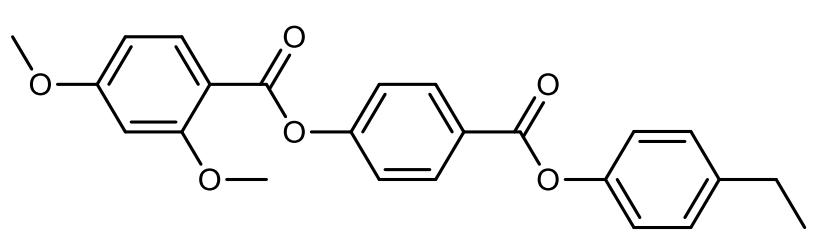

**4-((4-ethylphenoxy)carbonyl)phenyl-2,4-dimethoxybenzoate (COB30 IDX 1014)**

Quantities used: 4-ethylphenol (0.18 g, 1.5 mmol), 4-((2,4-dimethoxybenzoyl)oxy)benzoic acid (0.50 g, 1.65 mmol), **[4]** EDC.HCl (0.40 g, 2.1 mmol), DMAP (cat. amount), DCM (30 mL). The reaction was performed according to general procedure A. The crude material was purified by flash chromatography over silica gel, using a gradient of hexane/ethyl acetate, followed by recrystallisation from acetonitrile to give white crystals.

Yield: 0.39 g (64 %).

$R_F$: 0.55 (DCM)

$^{1}$H NMR (400 MHz): 8.25 (m, 2H, r-**H**), 8.09 (d, 1H, *J* = 8.7 Hz, Ar-**H**), 7.35 (m 2H, Ar-**H**), 7.25 m, 2H, Ar-**H**), 7.12 (m, 2H, Ar-**H**), 6.58 (dd, 1H, *J* = 8.7 Hz, *J* = 2.3 Hz, Ar-**H**), 6.54 (d, 1H, *J* = 2.3 Hz, Ar-**H**), 3.94 (s, 3H, R-O-**CH$_3$**), 3.90 (s, 3H, R-O-**CH$_3$**), 2.68 (q, 2H, *J* = 7.6 Hz, R-**CH$_2$**-CH$_3$), 1.26 (t, 3H, *J* = 7.6 Hz, R-CH$_2$-**CH$_3$**).

$^{13}$C{$^{1}$H} NMR (101 MHz): 165.41, 164.90, 163.07, 162.59, 155.52, 148.95, 141.98, 134.76, 131.80, 128.96, 126.95, 122.34, 121.54, 110.76, 105.07, 99.17, 56.19, 55.76, 28.46, 15.73.

HRMS: @caitlin

HPLC: @caitlin

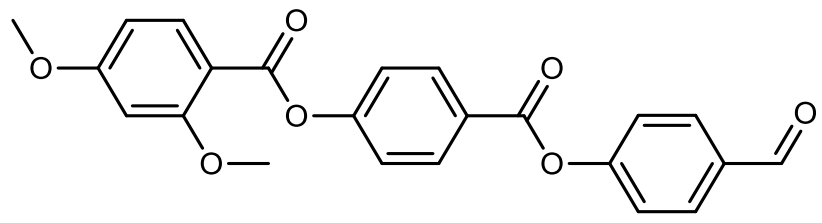

**4-((4-formylphenyloxy)carbonyl)phenyl 2,4-dimethoxybenzoate (COB27 IDX 1407)**

Quantities used: 4-hydroxybenzalehdye (0.18 g, 1.5 mmol), 4-((2,4-dimethoxybenzoyl)oxy)benzoic acid (0.50 g, 1.65 mmol), **[REF: RM2017 Chem Eur J]** EDC.HCl (0.40 g, 2.1 mmol), DMAP (cat. amount), DCM (30 mL). The reaction was performed according to general procedure A. The crude material was purified by flash chromatography over silica gel, using a gradient of hexane/ethyl acetate, followed by recrystallisation from acetonitrile to give a white solid.

Yield: 0.09 g (15 %).

$R_F$: 0.20 (DCM)

$^{1}$H NMR (400 MHz): 10.03 (s, 1H, R-C**H**O), 8.26 (m, 2H, Ar-**H**), 8.10 (d, 1H, *J* = 8.8 Hz, AR-**H**), 7.98 (m, 2H, Ar-**H**), 7.42 (m, 2H, *J* = 8.6 Hz, Ar-**H**), 7.38 (m, 2H, Ar-**H**), 6.58 (dd, 1H, *J* = 8.7, *J* = 2.3 Hz, Ar-**H**), 3.94 (s, 3H, R-O-$\mathbf{CH_3}$), 3.90 (s, 3H, R-O-$\mathbf{CH_3}$).

$^{13}$C{$^{1}$H} NMR (101 MHz): 191.10, 165.50, 164.03, 162.98, 162.65, 155.95, 155.80, 134.77, 134.23, 131.96, 131.43, 126.10, 122.70, 122.57, 110.59, 105.12, 99.17, 56.20, 55.77.

HRMS: @caitlin

HPLC: @caitlin

## 7. Analytical Data

### 7.1. Analytical Data for Scheme 1

#### 7.1.1. 3,5-Difluorobenzyl propyl ether

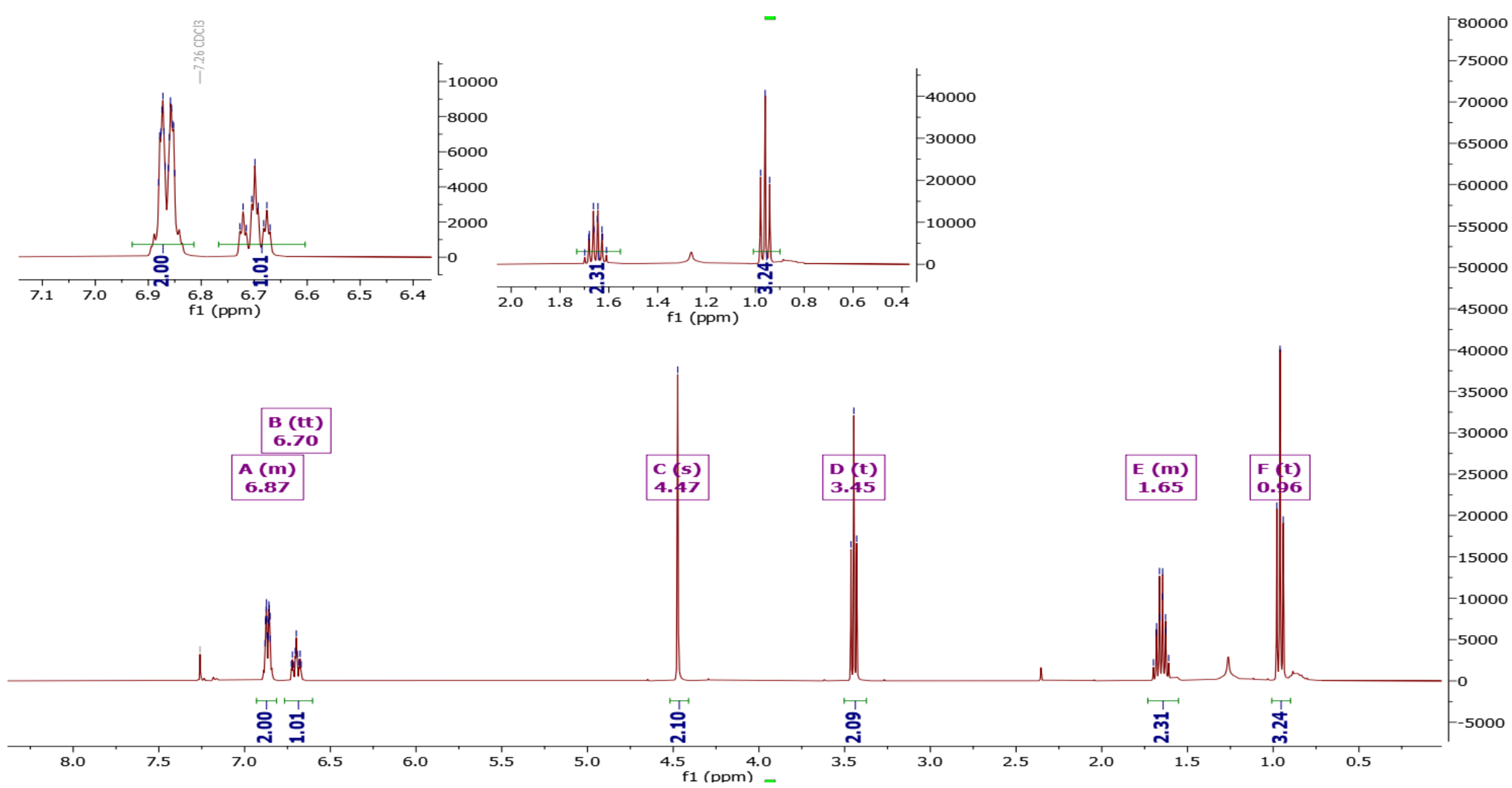

**Figure SI-18:** $^{1}$H NMR of 3,5-difluorobenzyl propyl ether

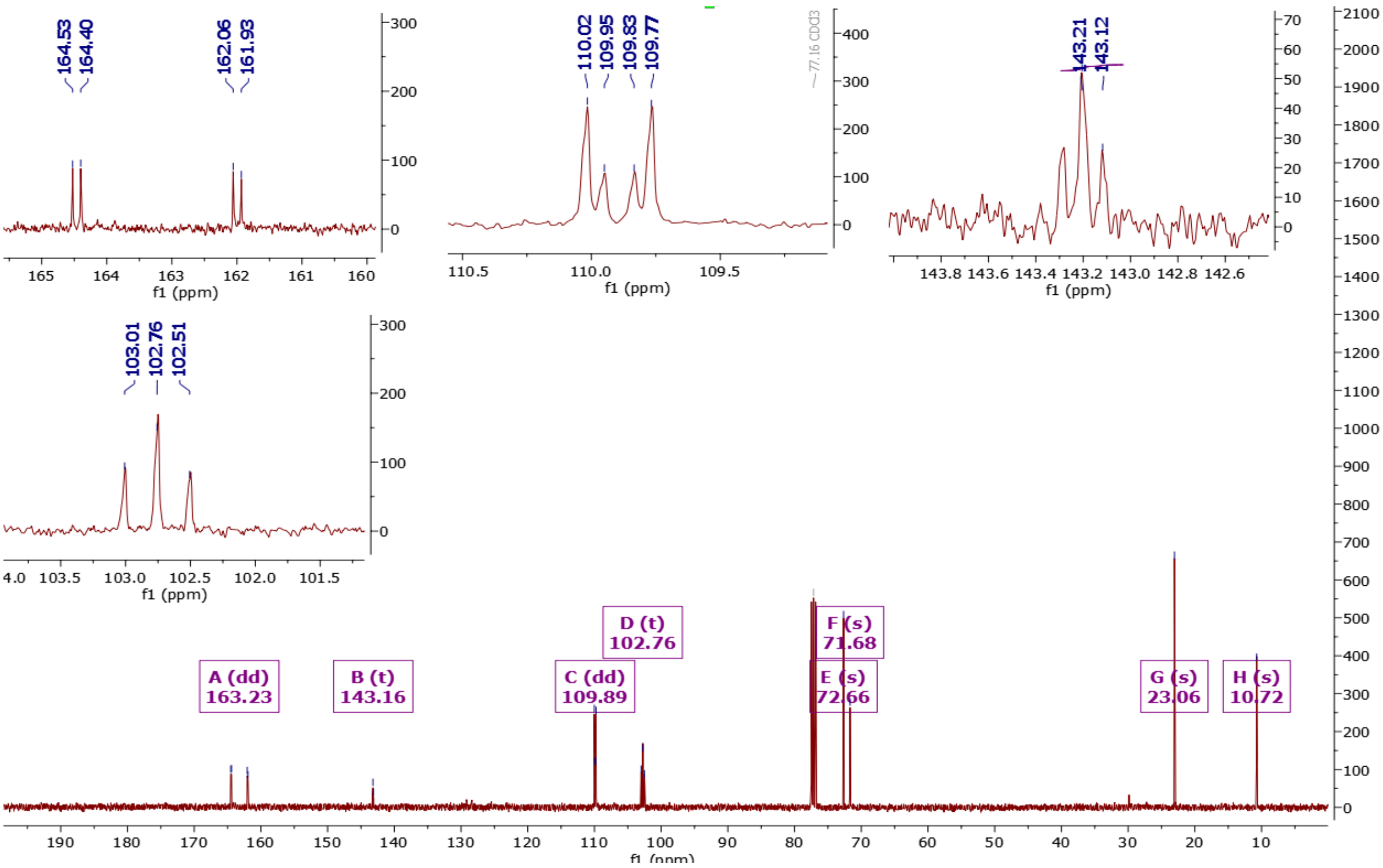

**Figure SI-23:** $^{13}$C{$^{1}$H} NMR of 3,5-difluorobenzyl propyl ether

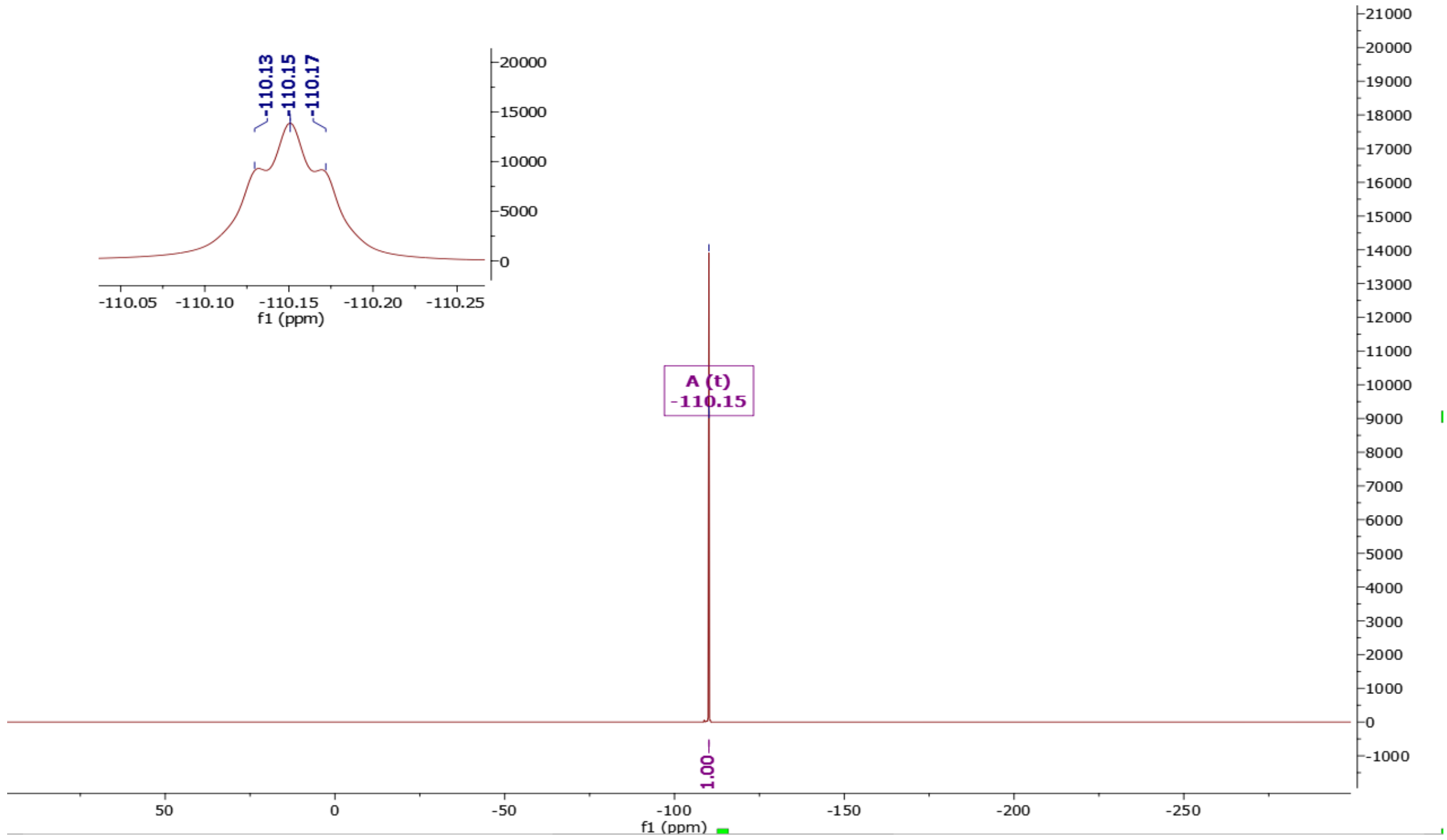

**Figure SI-24:** $^{19}F$ NMR of 3,5-difluorobenzyl propyl ether

### 7.1.2. 2,6-difluoro-2-(propan-1-yloxy)benzoic acid

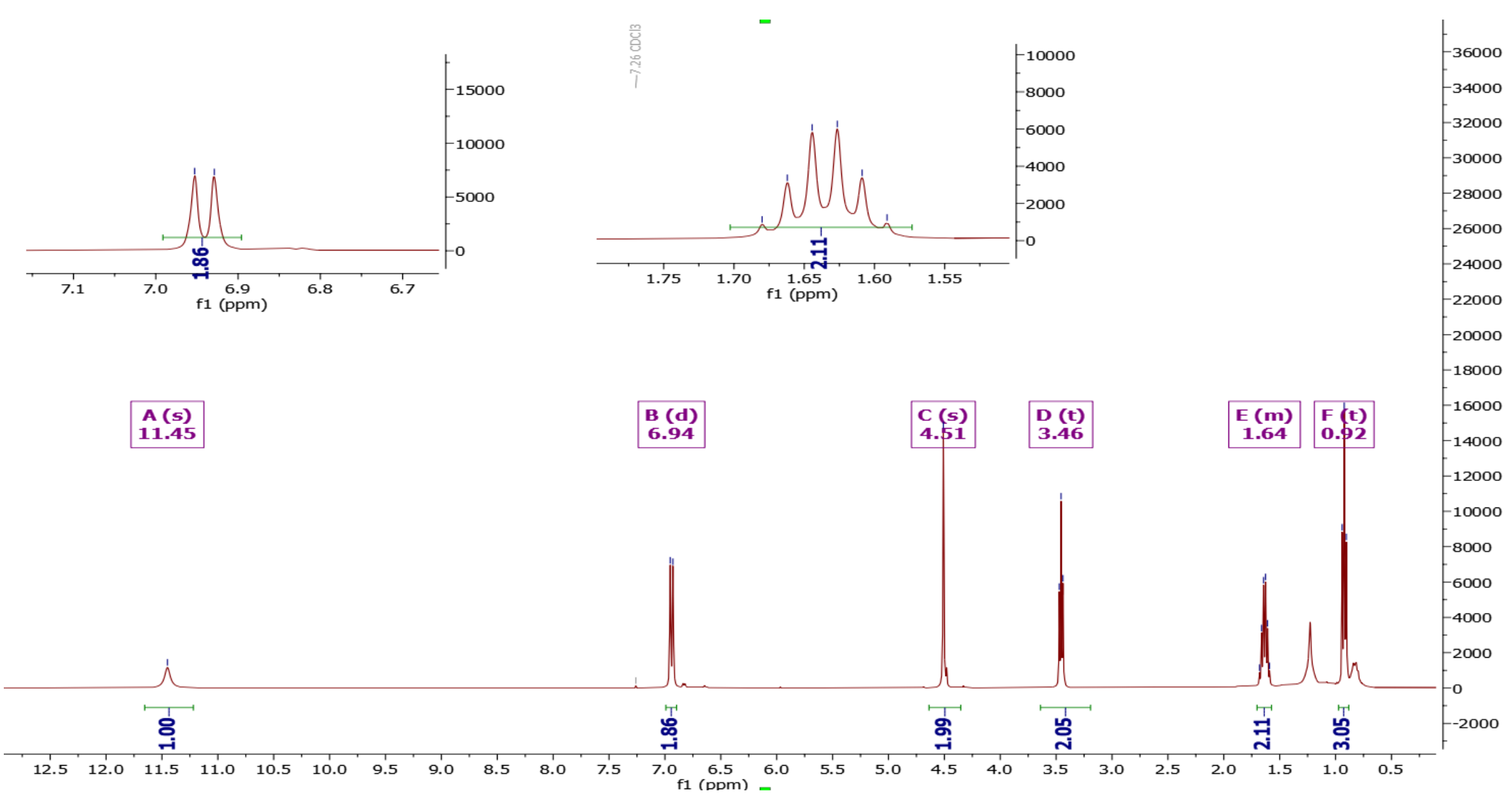

**Figure SI-21:** 1H NMR of 2,6-difluoro-2-(propan-1-yloxy)benzoic acid

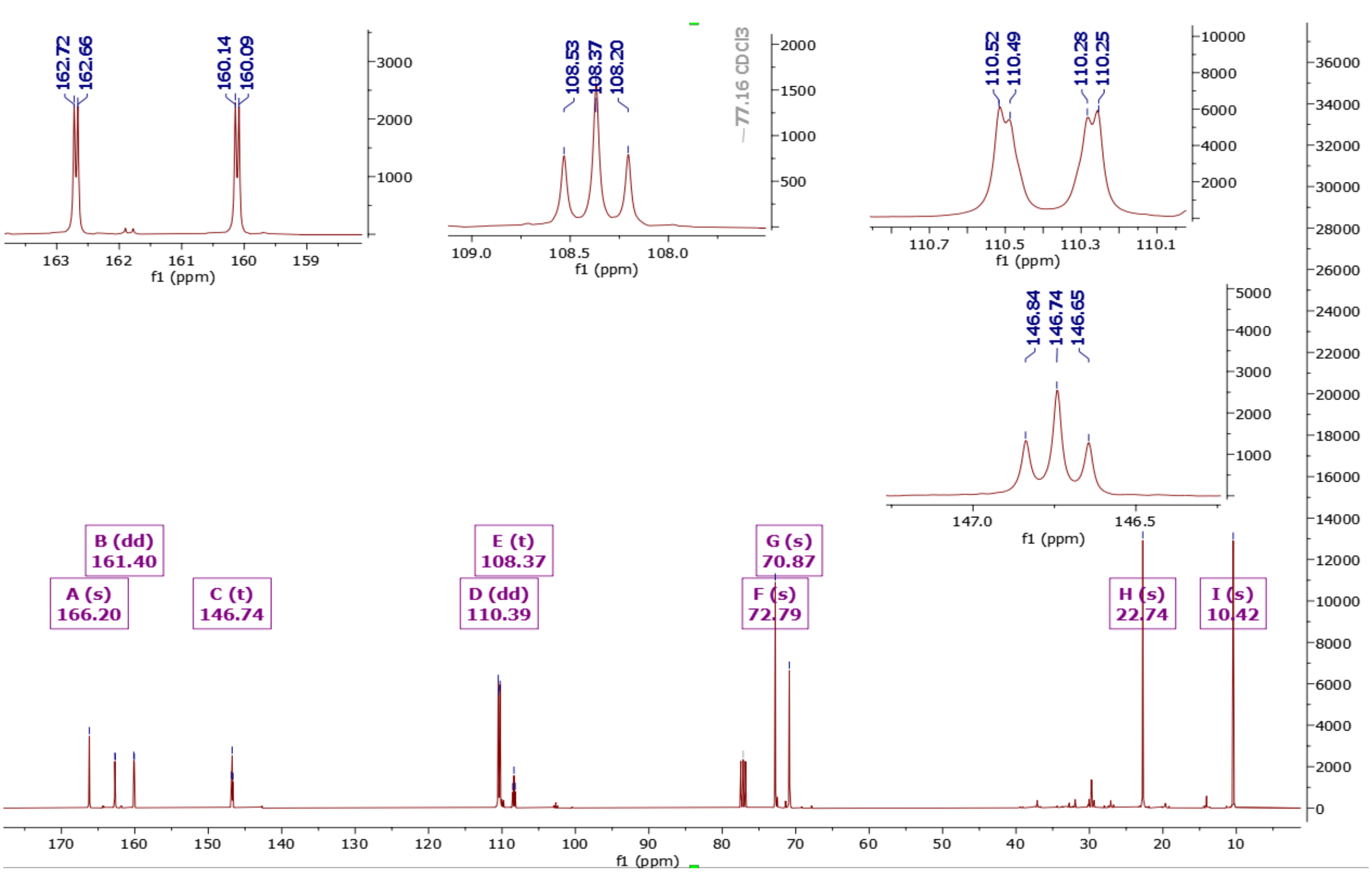

**Figure SI-22:** $^{13}C\{^{1}H\}$ NMR of 2,6-difluoro-2-(propan-1-yloxy)benzoic acid

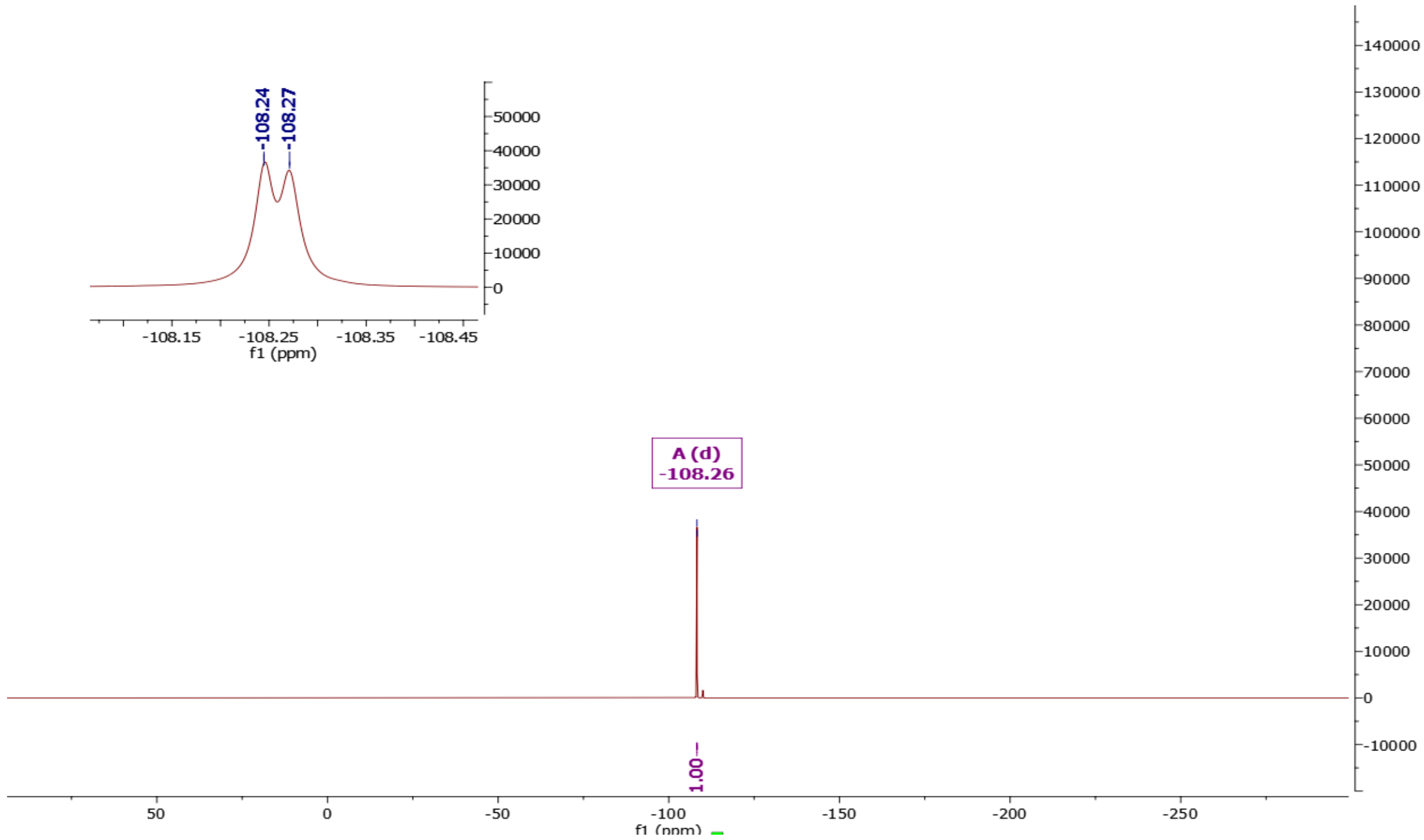

**Figure SI-23:** $^{19}$F NMR of 2,6-difluoro-2-(propan-1-yloxy)benzoic acid

### 7.1.3. (2,3',4',5',6-pentafluoro-[1,1'-biphenyl]-4-yl)methyl 2,6-difluoro-4-(propoxymethyl)benzoate

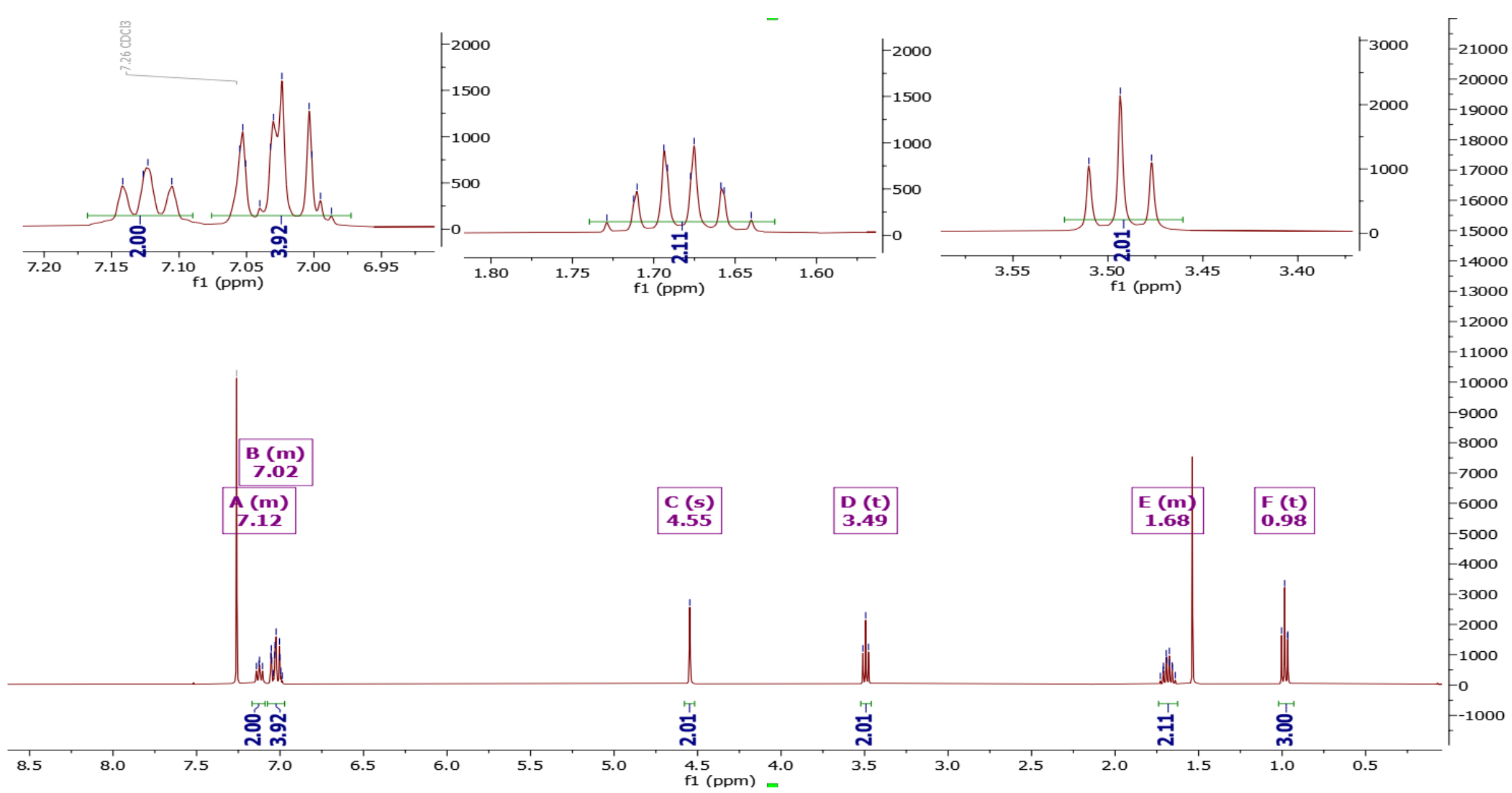

**Figure SI-24:** $^{1}$H NMR for (2,3',4',5',6-pentafluoro-[1,1'-biphenyl]-4-yl)methyl 2,6-difluoro-4-(propoxymethyl)benzoate

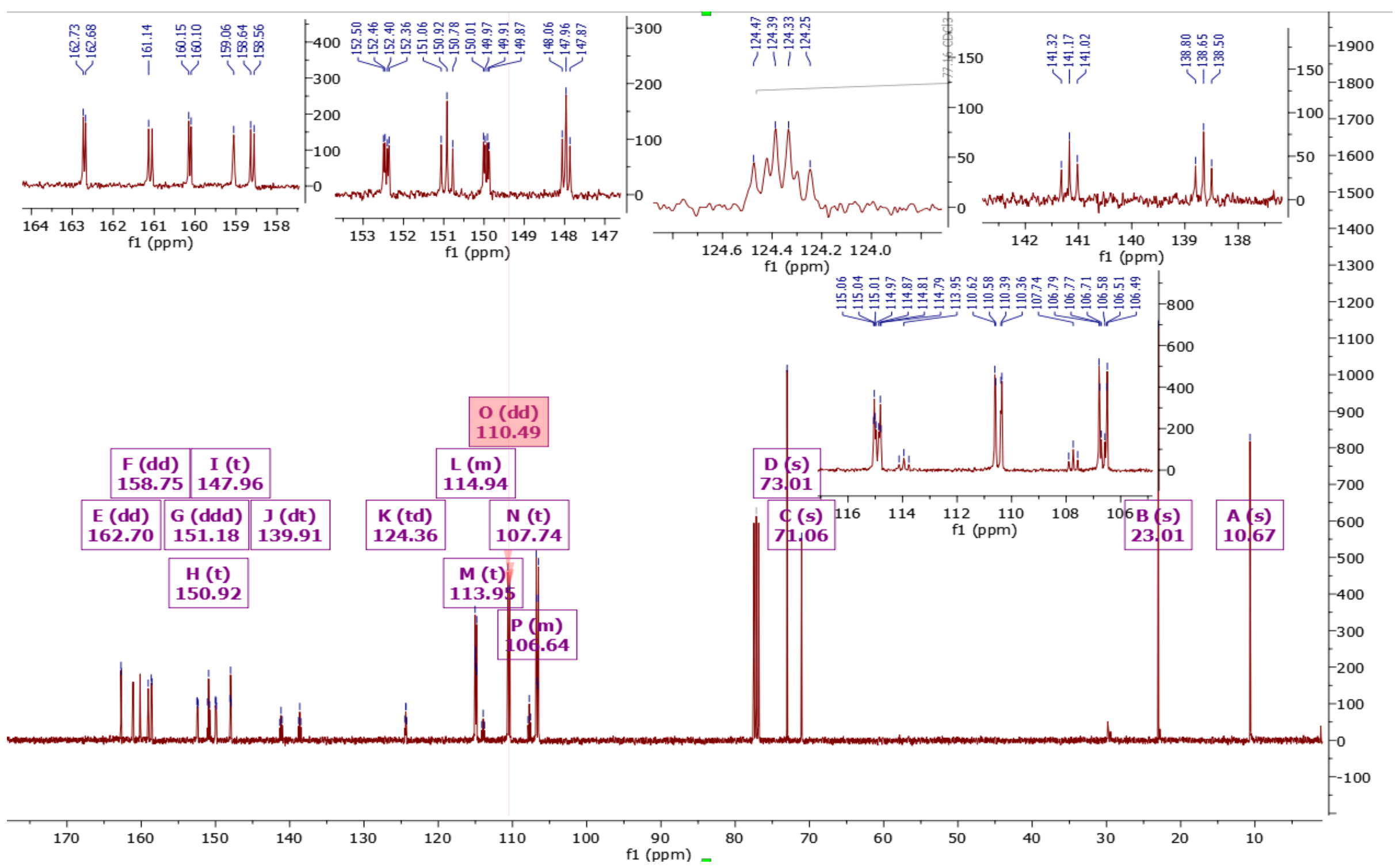

**Figure SI-25:** $^{13}$C{$^{1}$H} NMR for (2,3',4',5',6-pentafluoro-[1,1'-biphenyl]-4-yl)methyl 2,6-difluoro-4-(propoxymethyl)benzoate

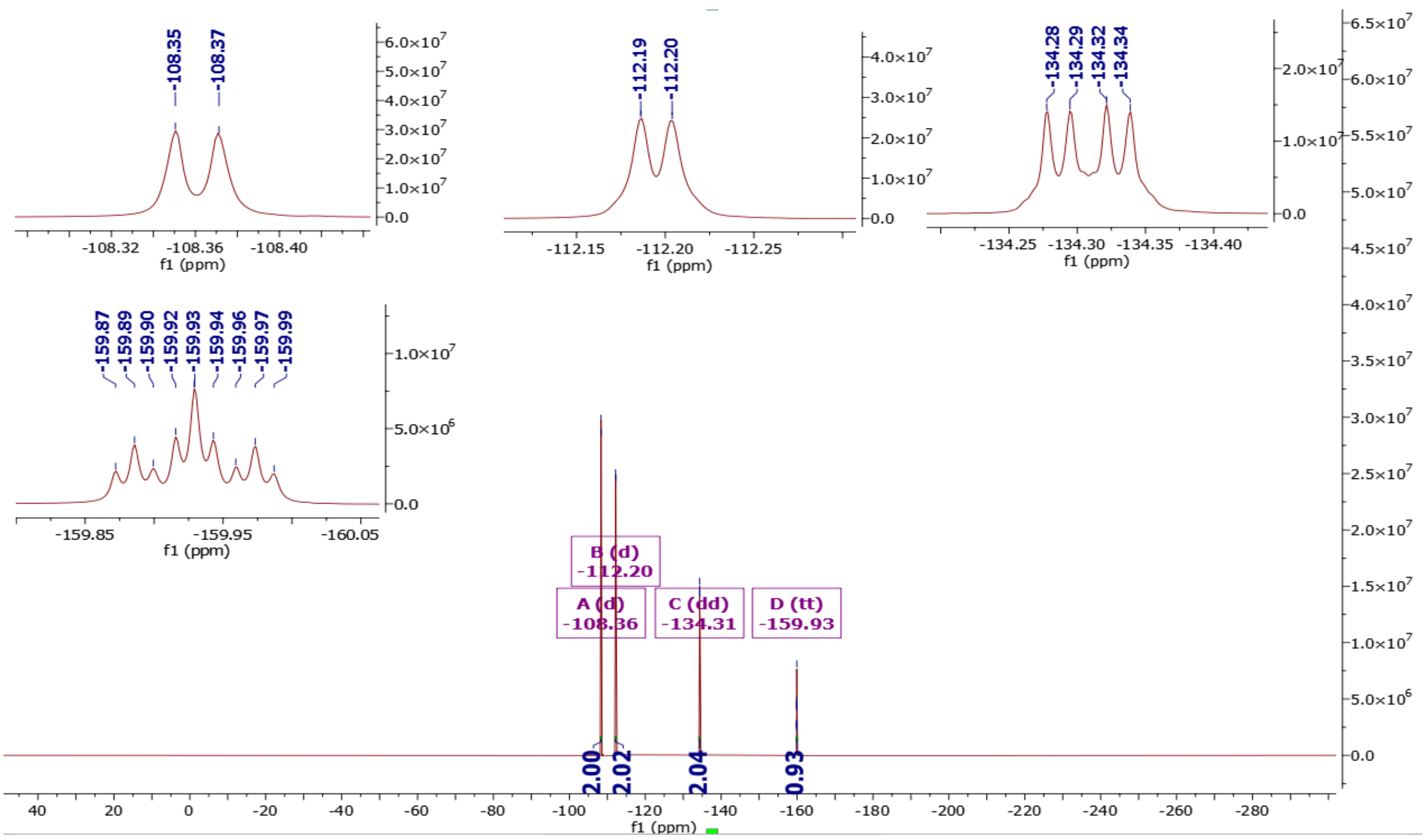

**Figure SI-26:** $^{19}$F NMR for (2,3',4',5',6-pentafluoro-[1,1'-biphenyl]-4-yl)methyl 2,6-difluoro-4-(propoxymethyl)benzoate

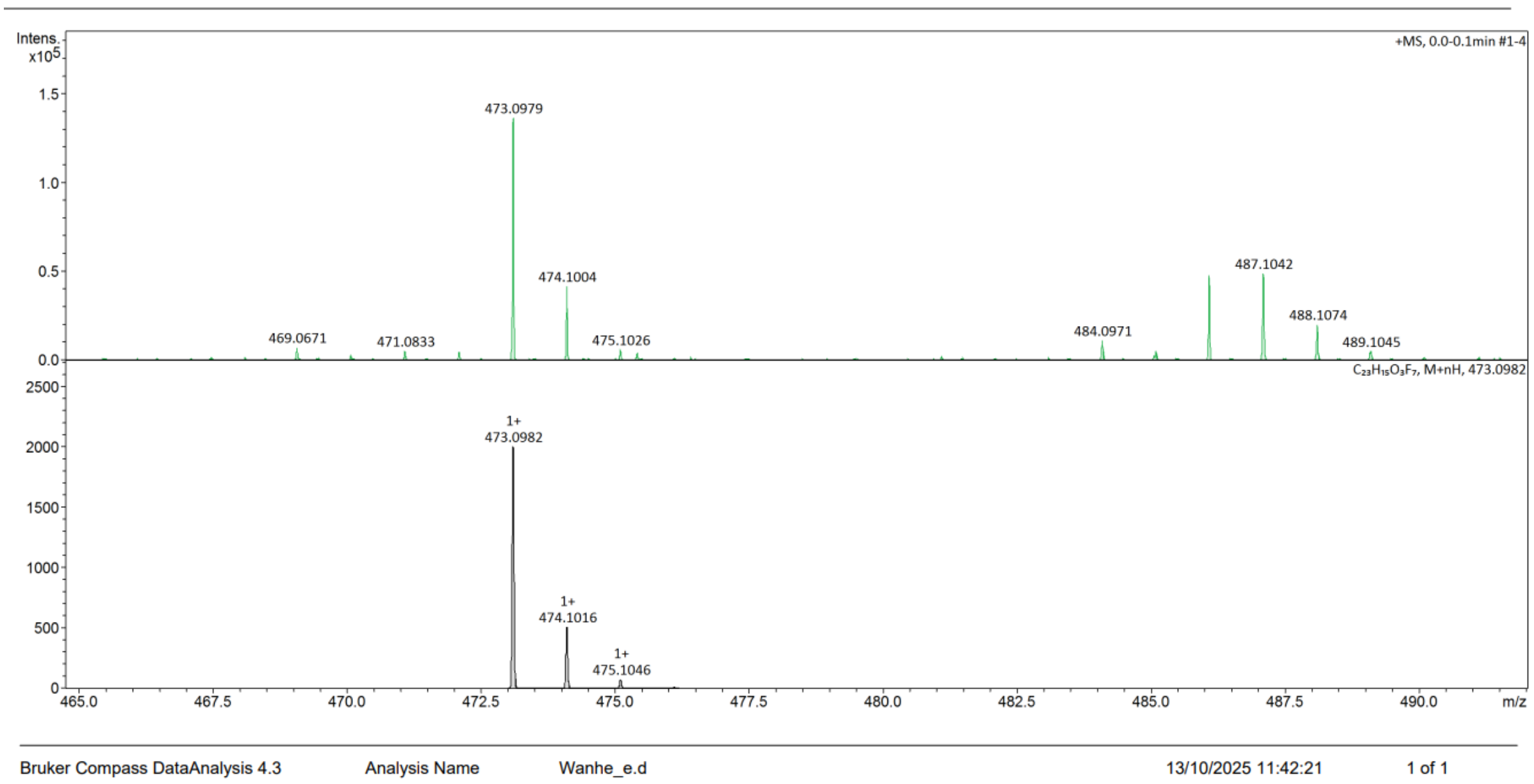

**Figure SI-27:** HRMS for (2,3',4',5',6-pentafluoro-[1,1'-biphenyl]-4-yl)methyl 2,6-difluoro-4-(propoxymethyl)benzoate

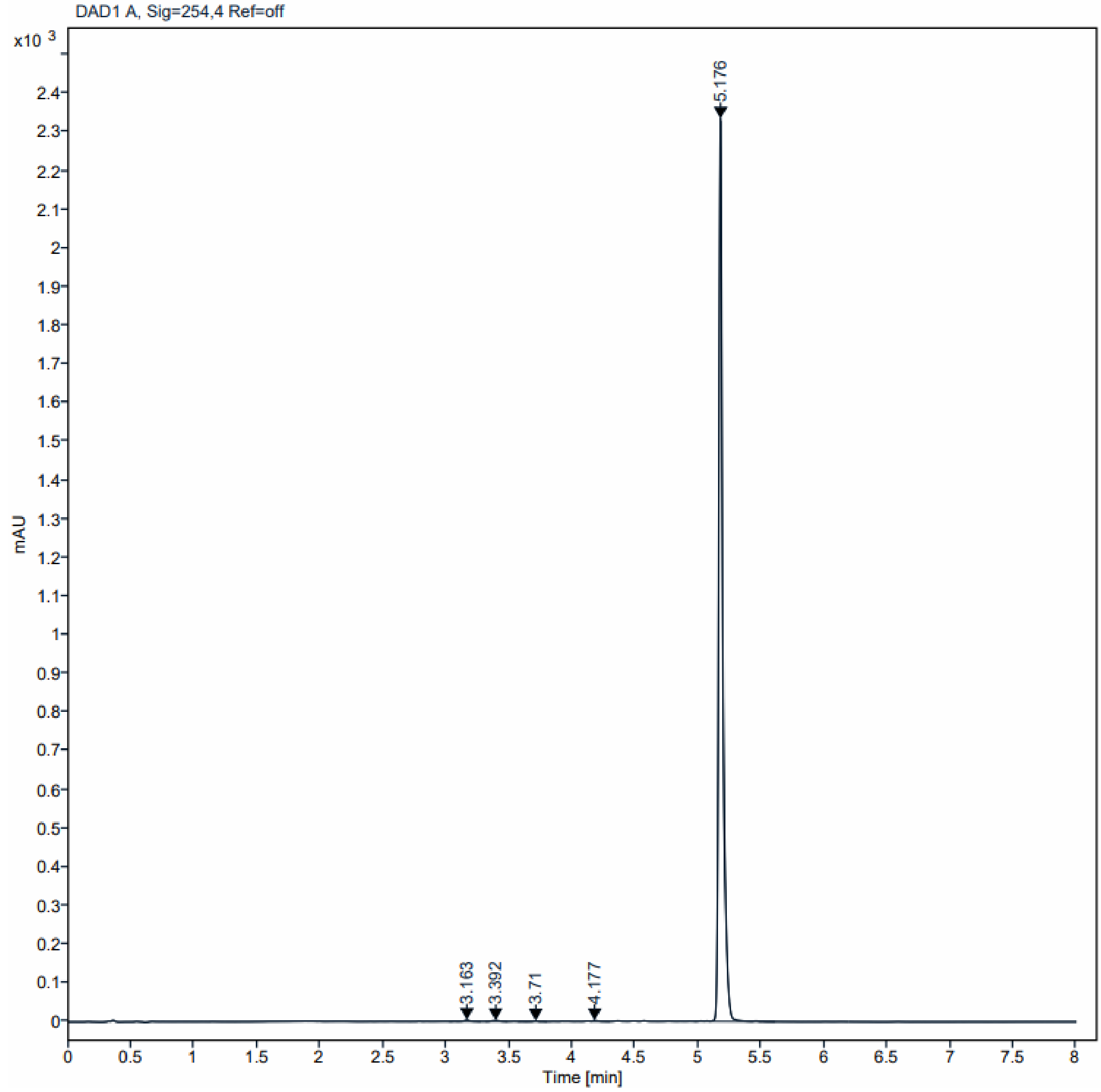

**Figure SI-28:** RP-HPLC for (2,3',4',5',6-pentafluoro-[1,1'-biphenyl]-4-yl)methyl 2,6-difluoro-4-(propoxymethyl)benzoate

## 7.2.1. Analytical data for IDX2538

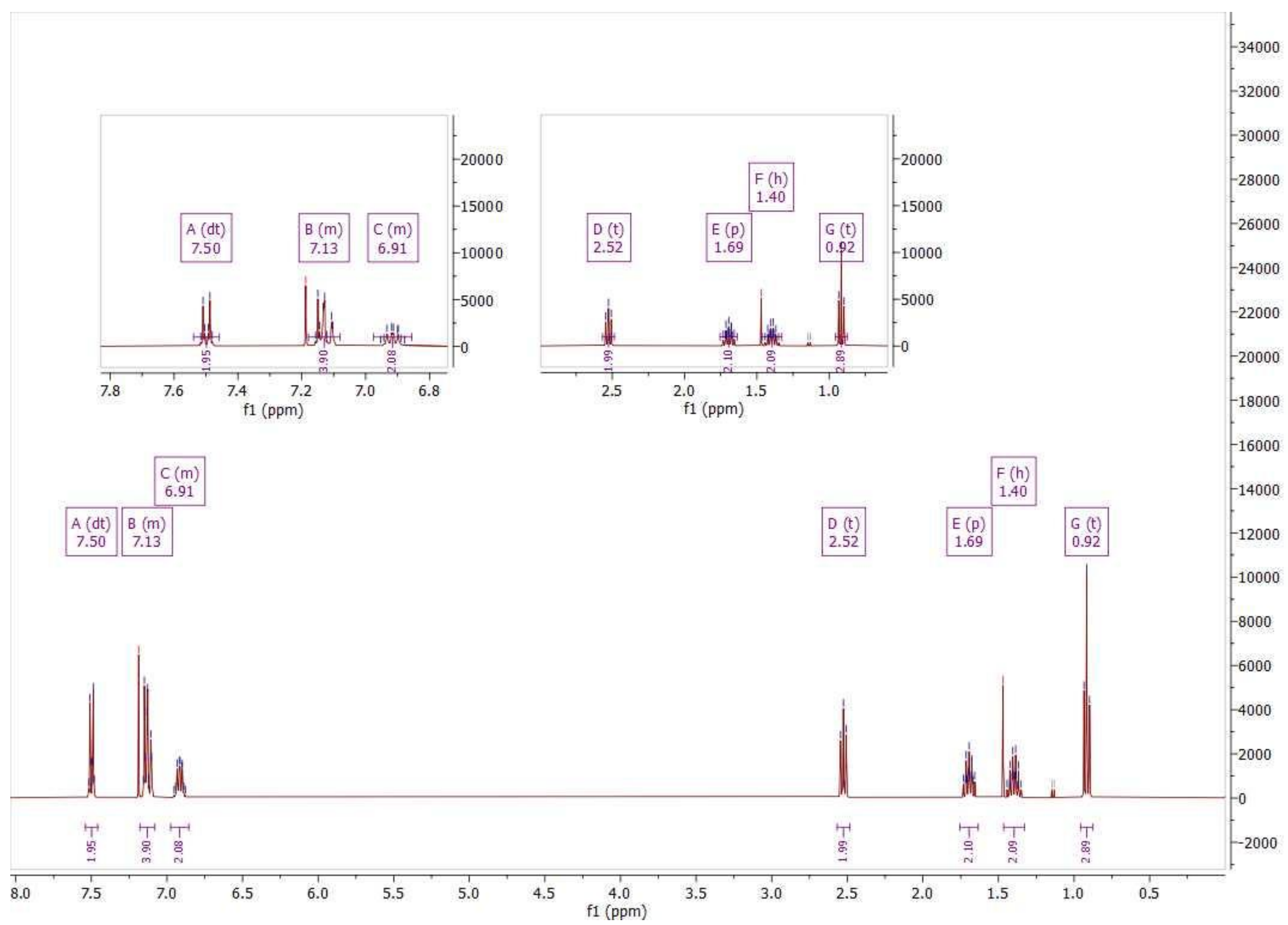

**Figure SI-29:** $^{1}$H NMR for IDX2538

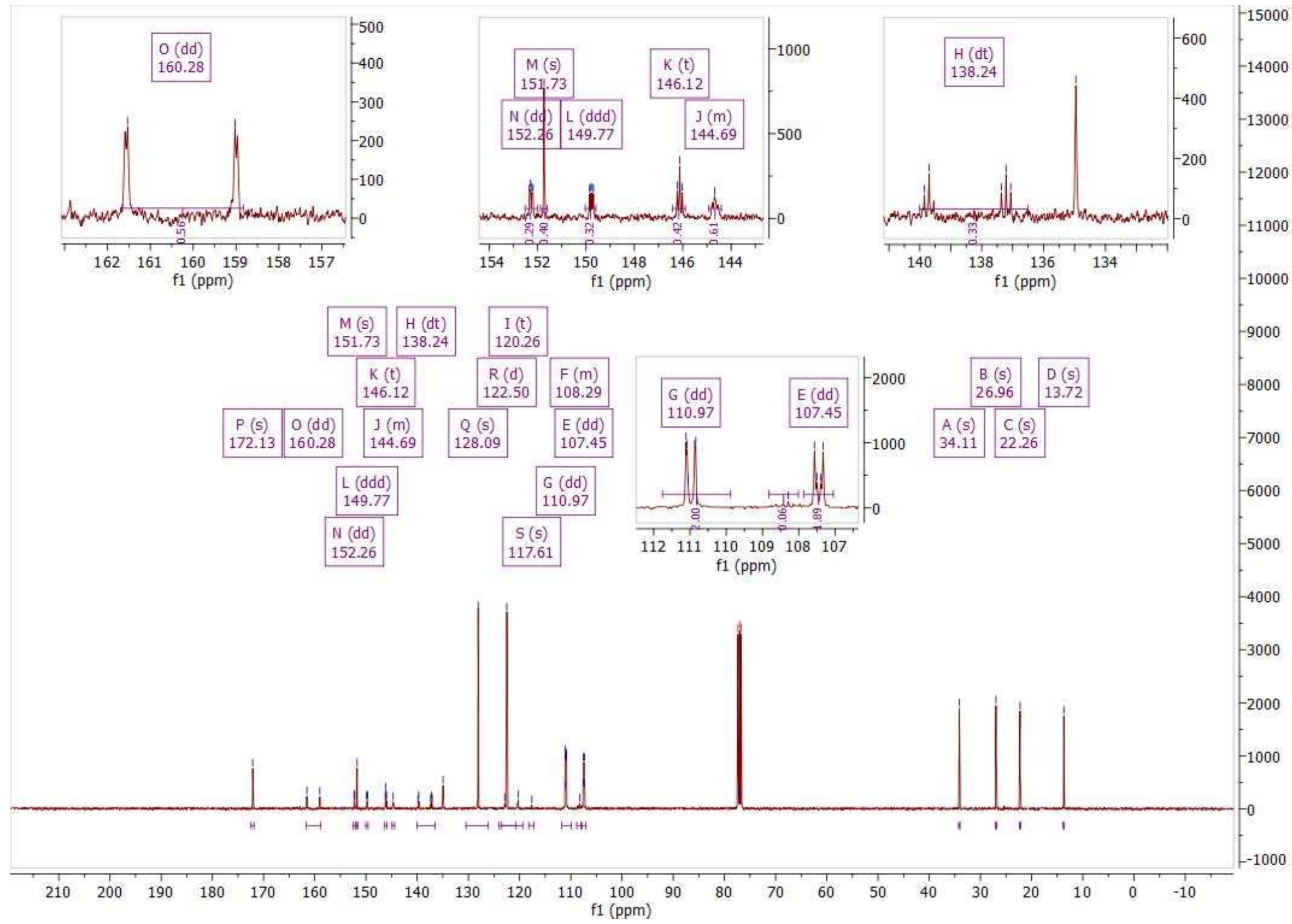

**Figure SI-30:** $^{13}C\{^{1}H\}$ NMR for IDX2538

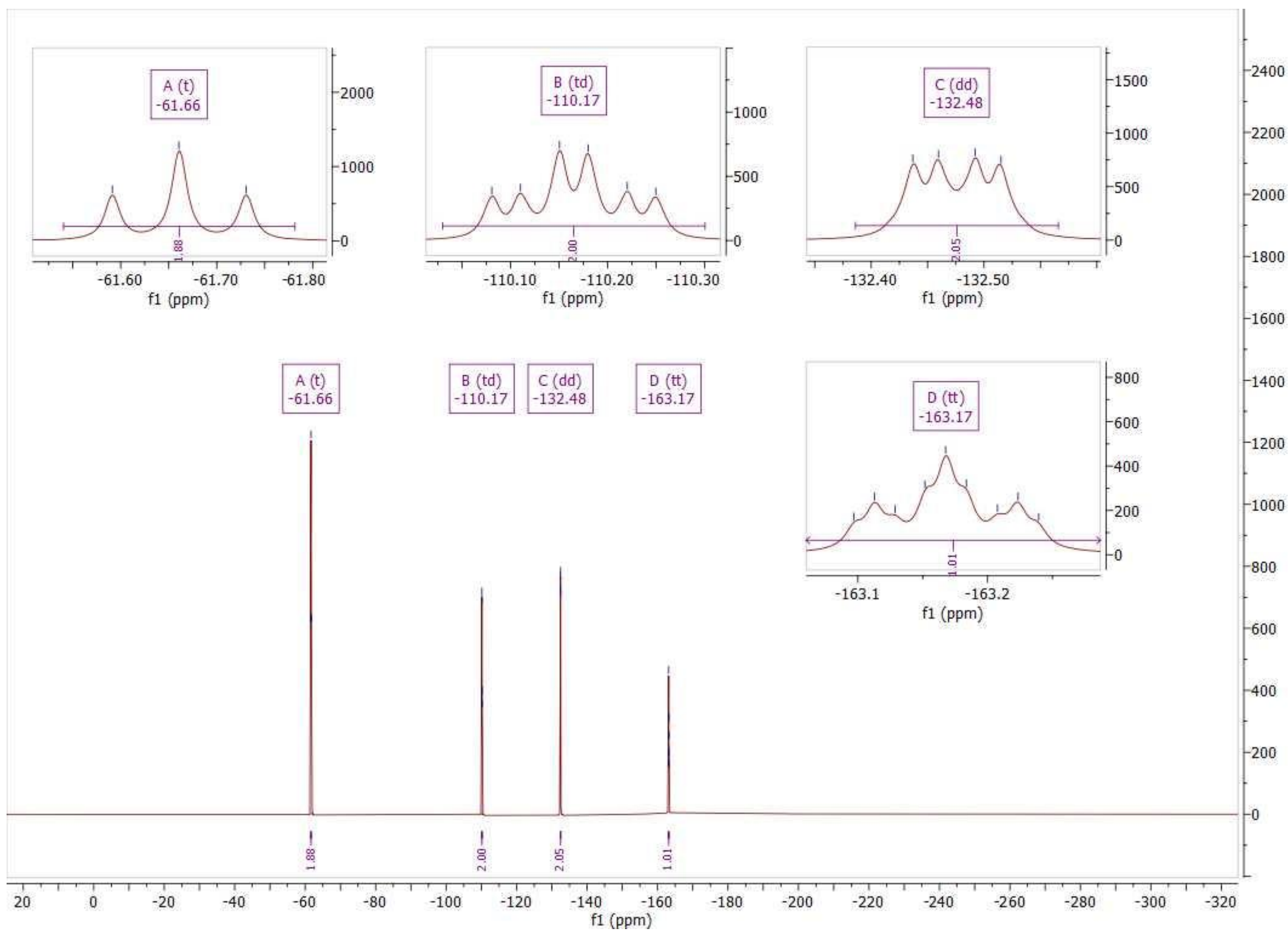

**Figure SI-31:** $^{19}F$ NMR for IDX2538

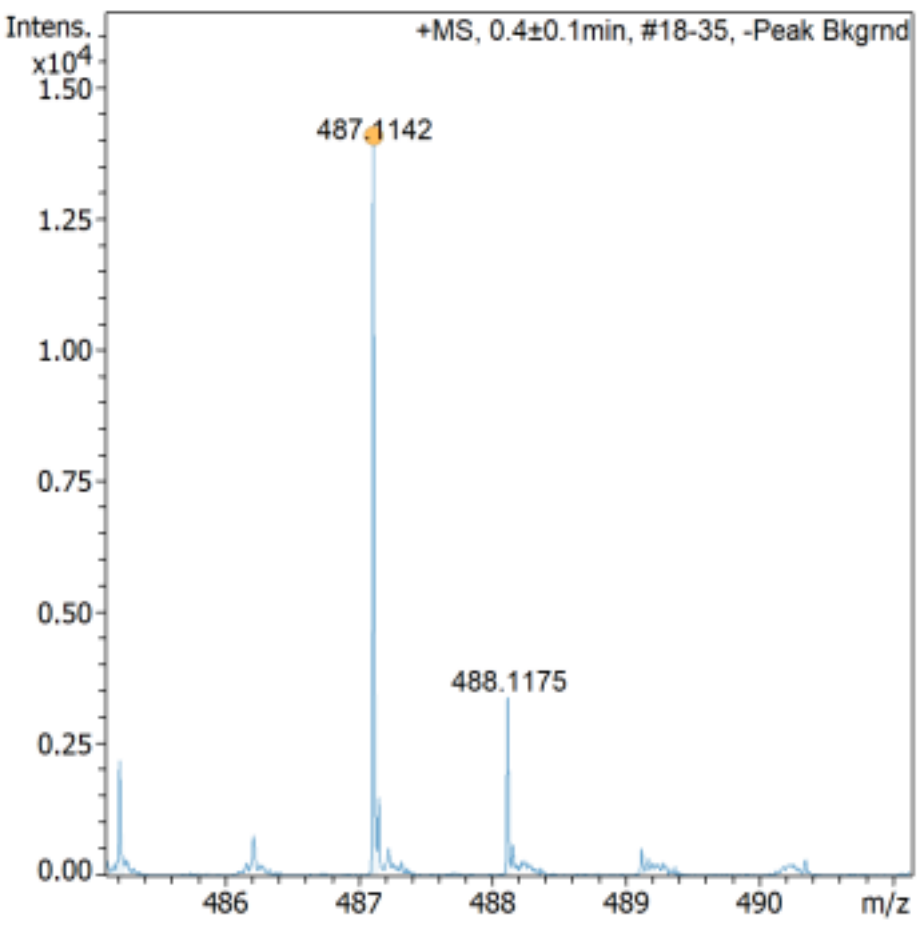

**Figure SI-32:** HRMS for IDX2538

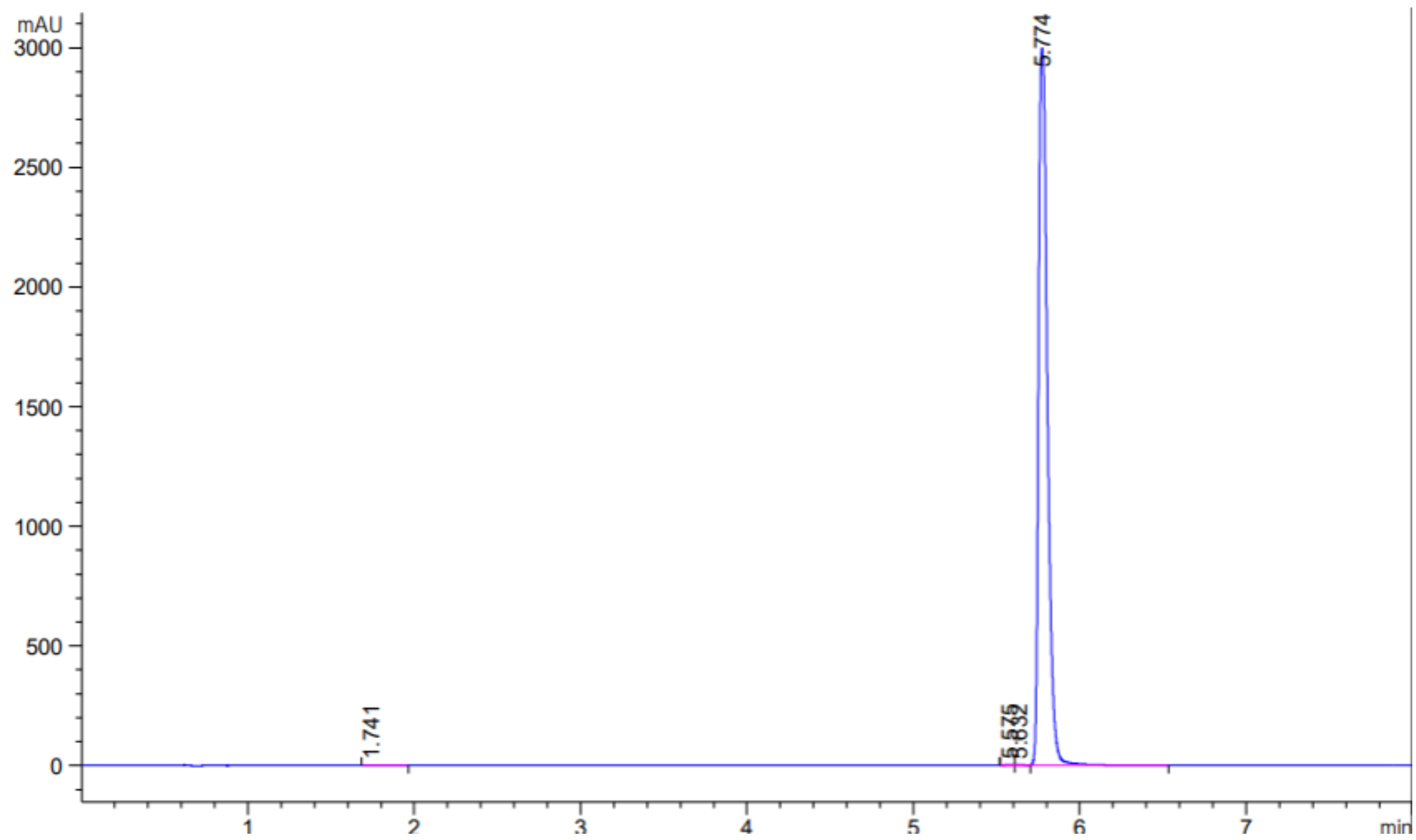

**Figure SI-33:** RP-HPLC for IDX2538

### 7.2.2. Analytical data for IDX1646

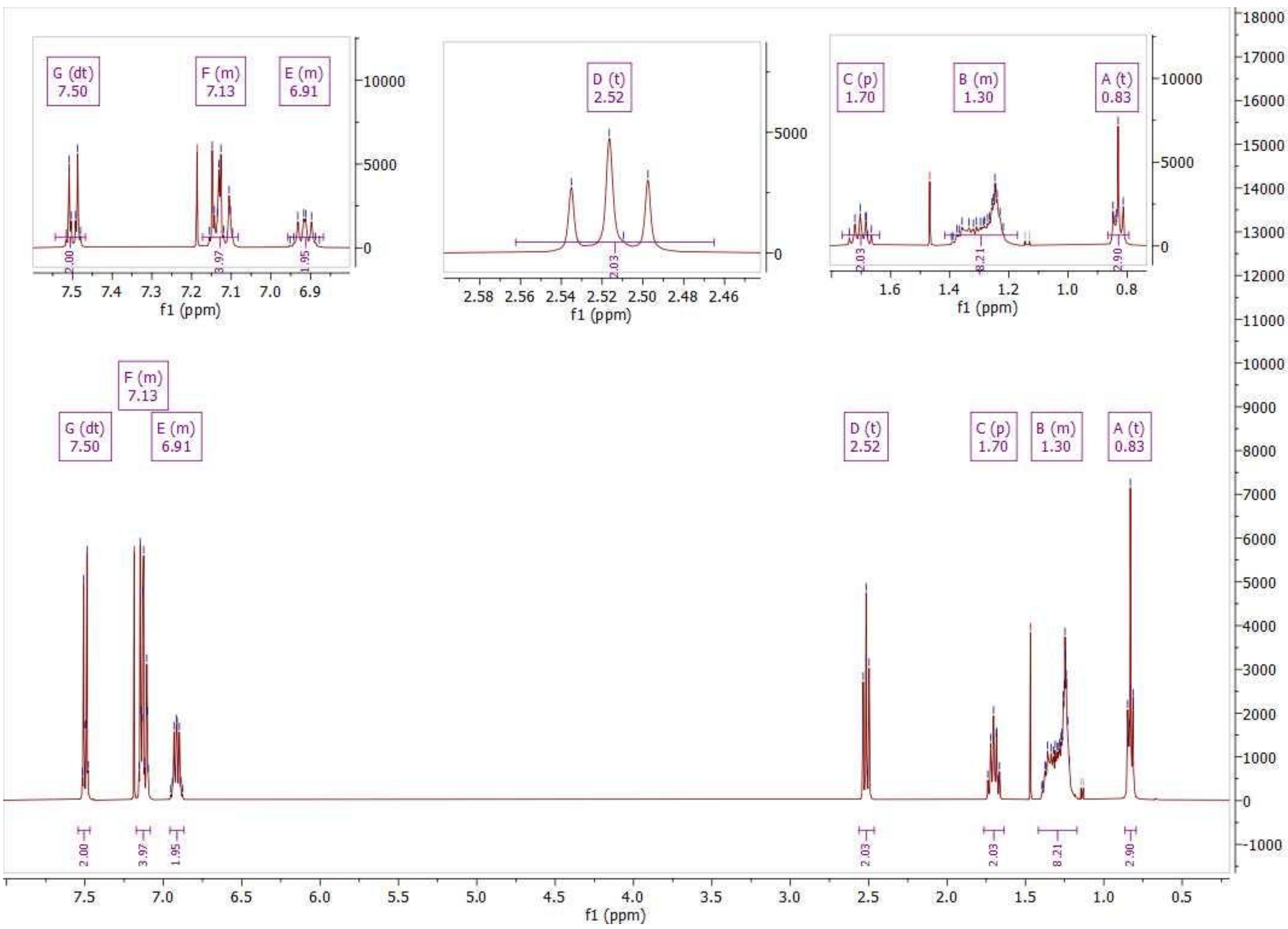

**Figure SI-34:** $^{1}$H NMR for IDX1646

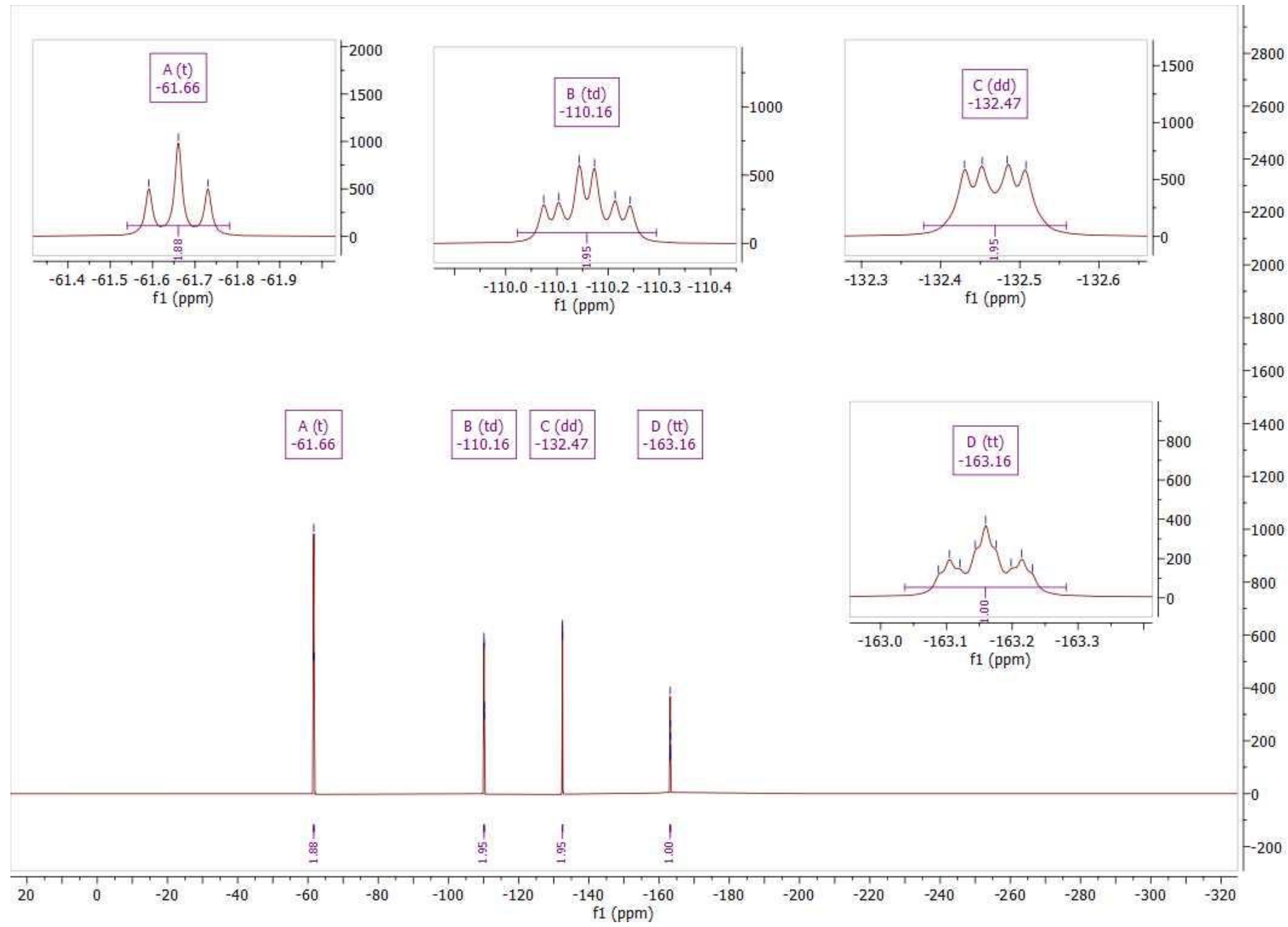

**Figure SI-35:** $^{19}F$ NMR for IDX1646

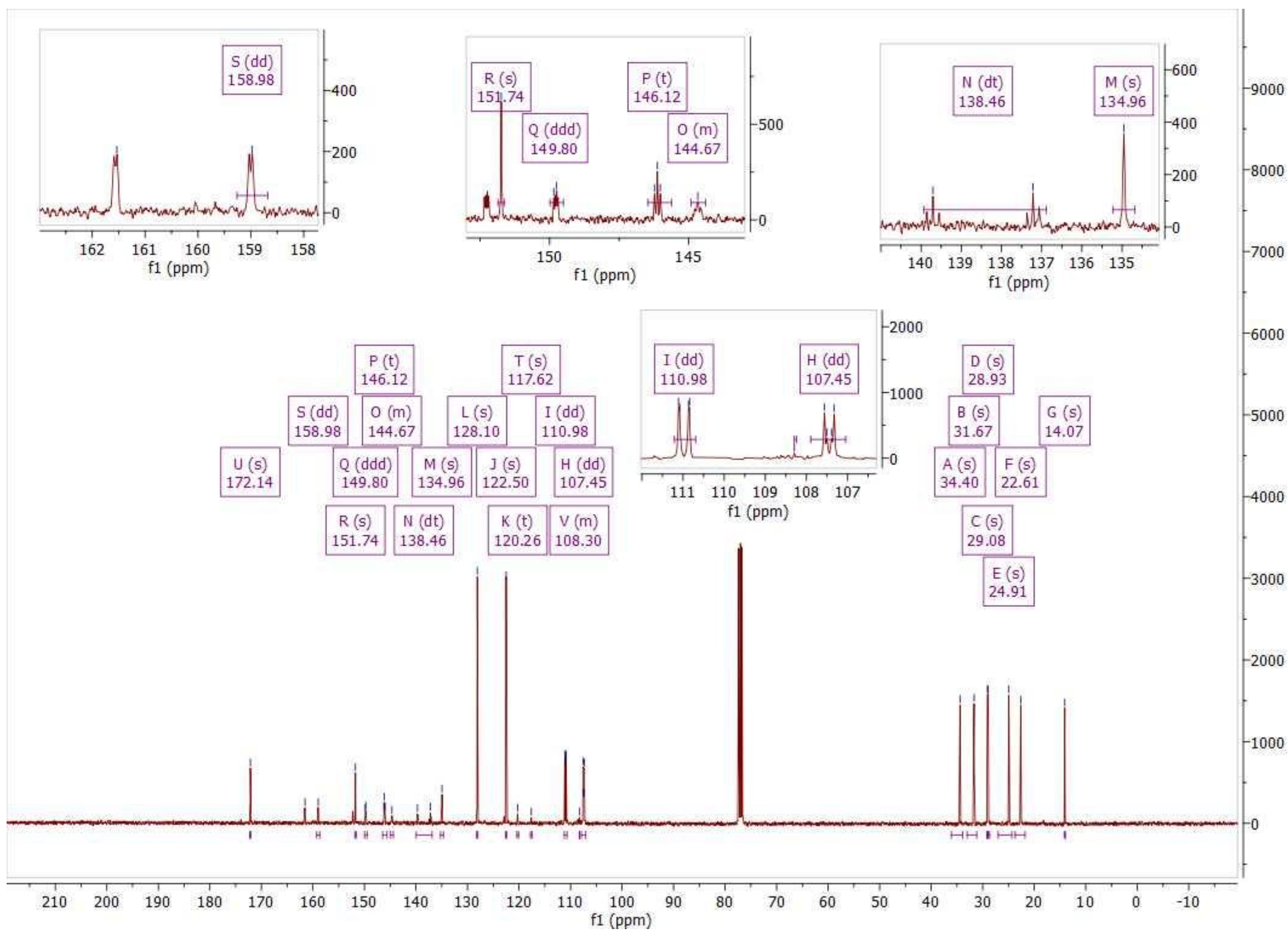

**Figure SI-36:** $^{13}C\{^{1}H\}$ NMR for IDX1646

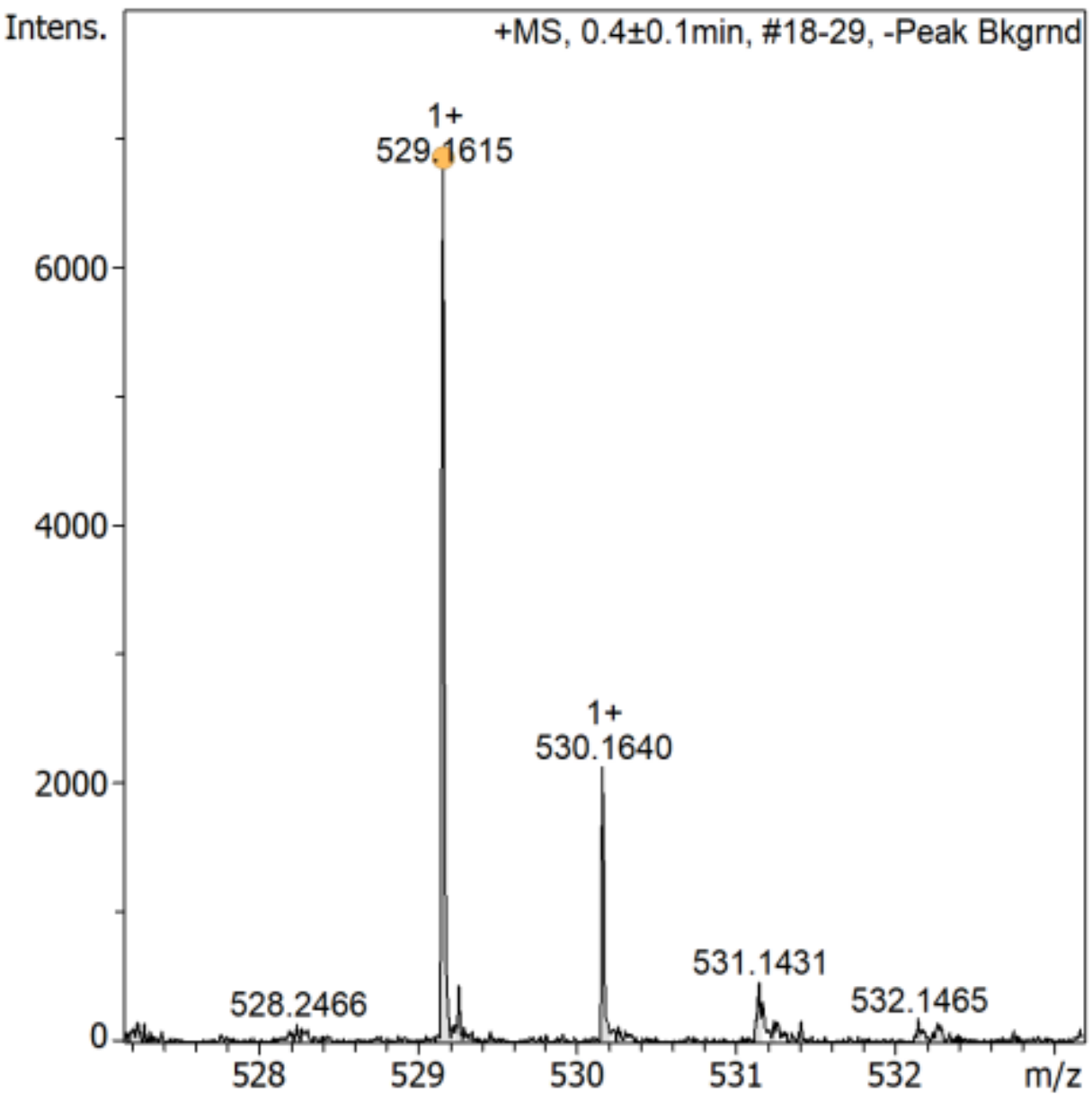

**Figure SI-37:** HRMS for IDX1646

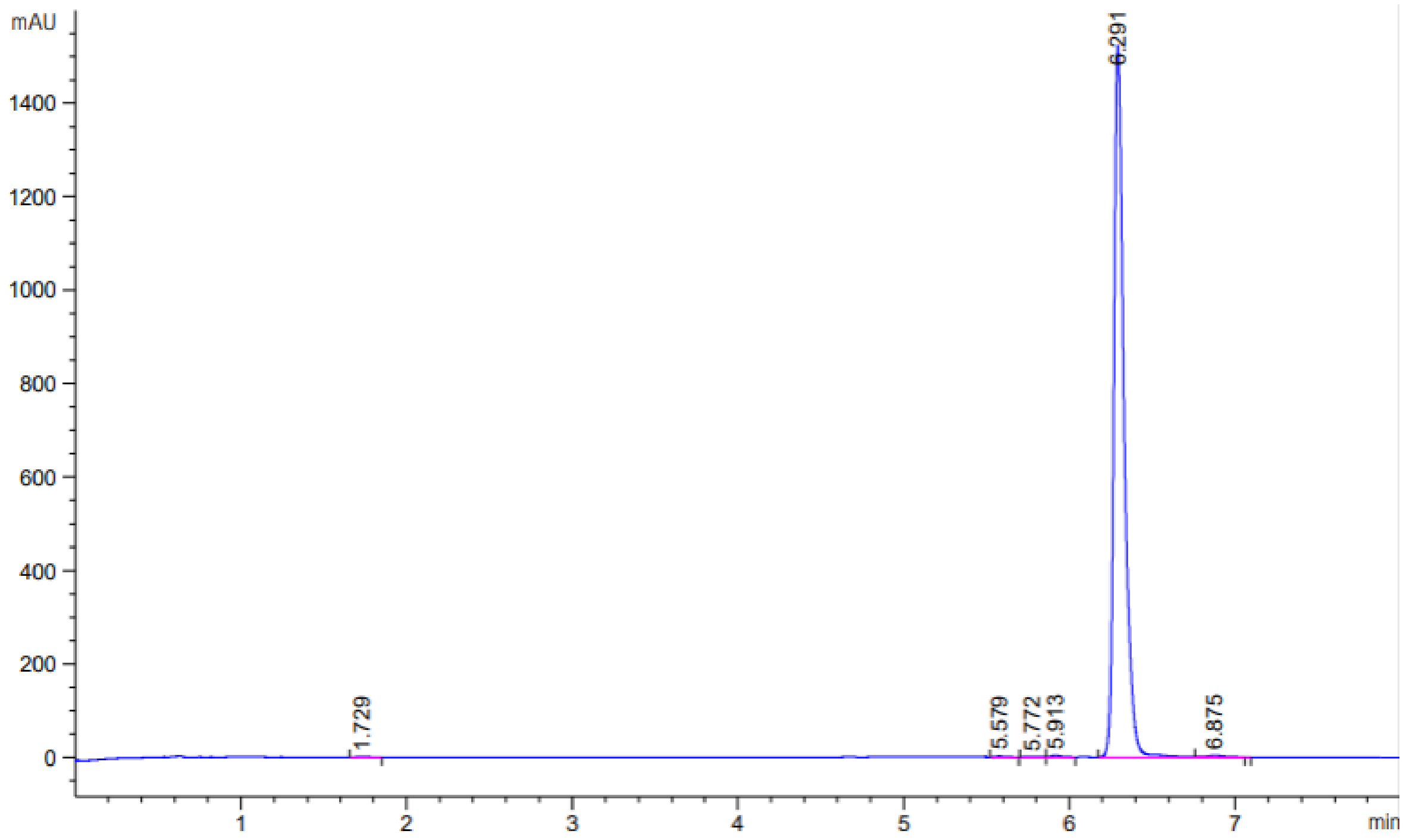

**Figure SI-38:** RP-HPLC for IDX1646

## 7.3. Analytical data for Scheme 3

### 7.3.1. IDX323 / CJG 412 / 2-bromo-5-ethoxy-1,3-diflurobenzene

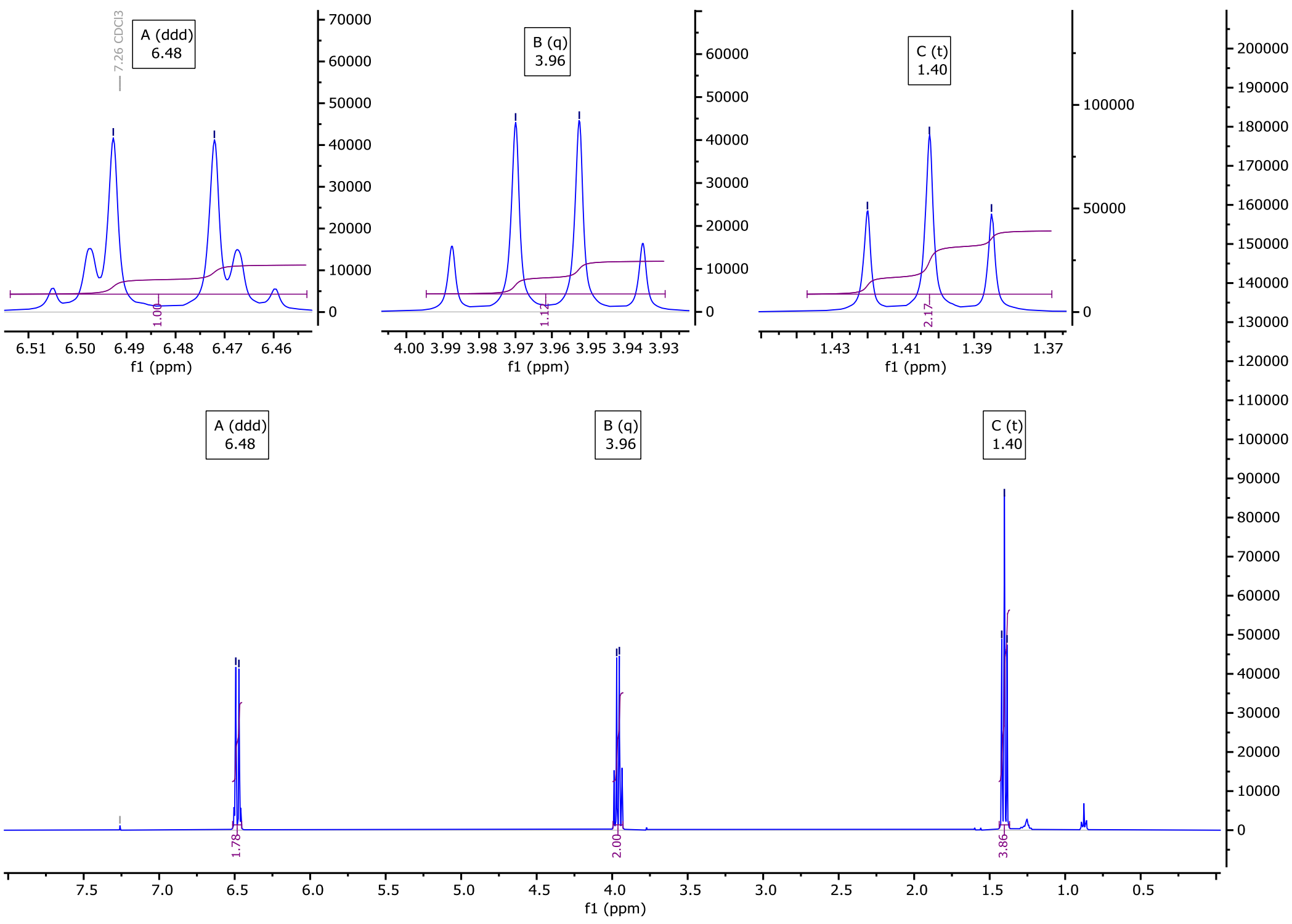

**Figure SI-39:** $^{1}$H NMR of 2-bromo-5-ethoxy-1,3-diflurobenzene

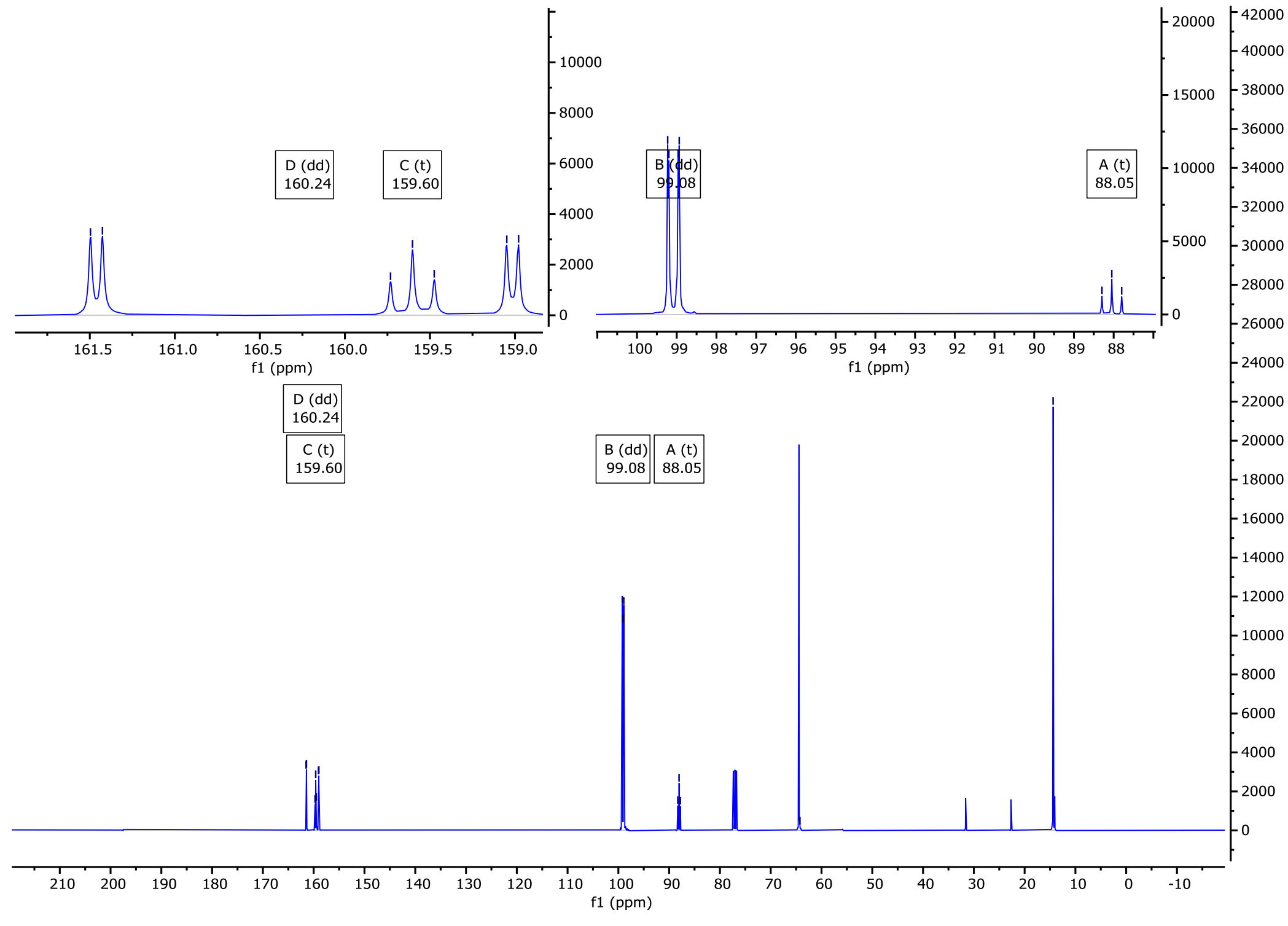

**Figure SI-40:** $^{13}$C{$^{1}$H} NMR of 2-bromo-5-ethoxy-1,3-diflurobenzene

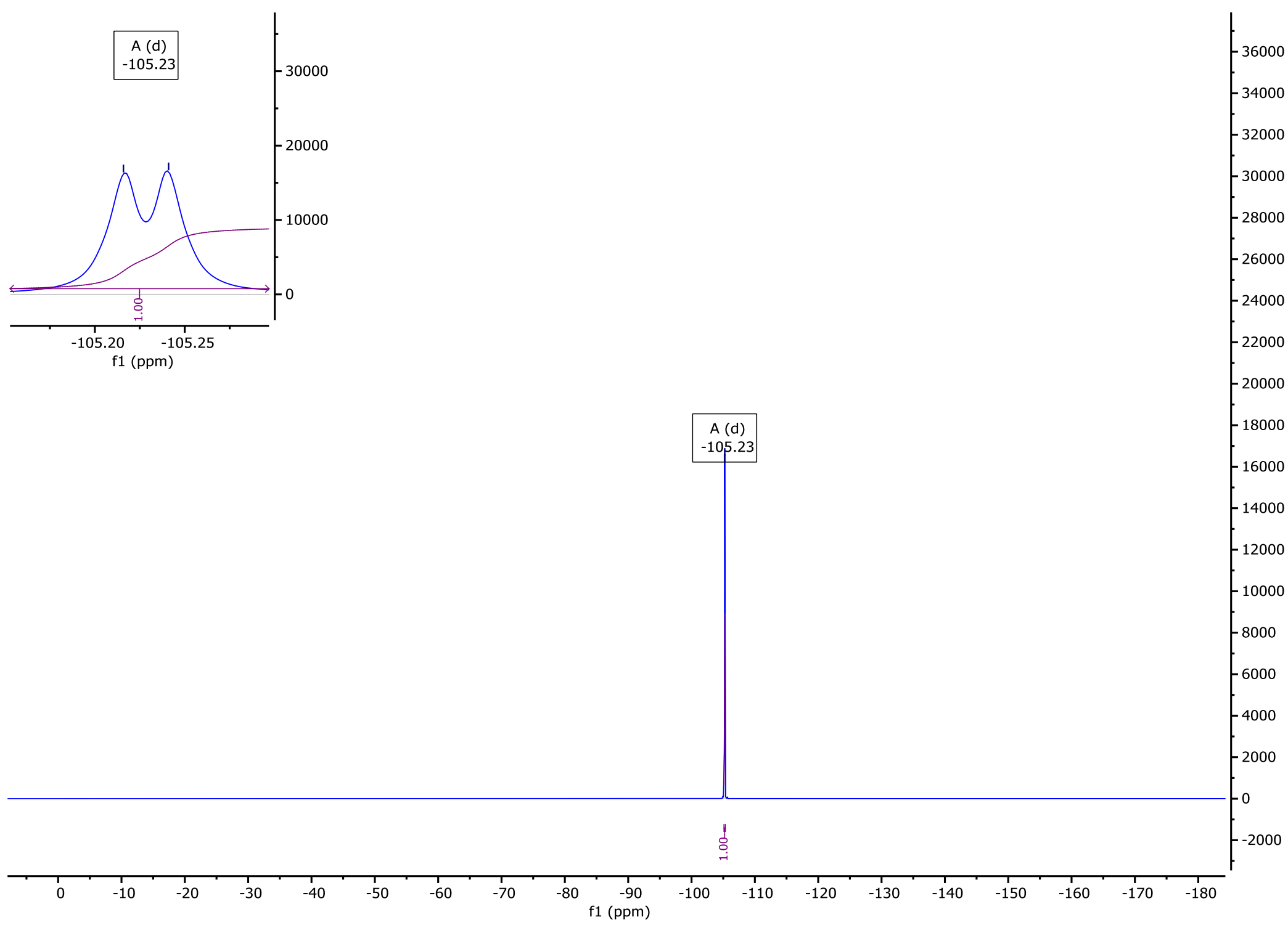

**Figure SI-41:** $^{19}$F NMR of 2-bromo-5-ethoxy-1,3-diflurobenzene

### 7.3.2. Analytical Data for IDX??? / CJG 483 / 4-ethoxy-2,6-diflurobenzoic acid

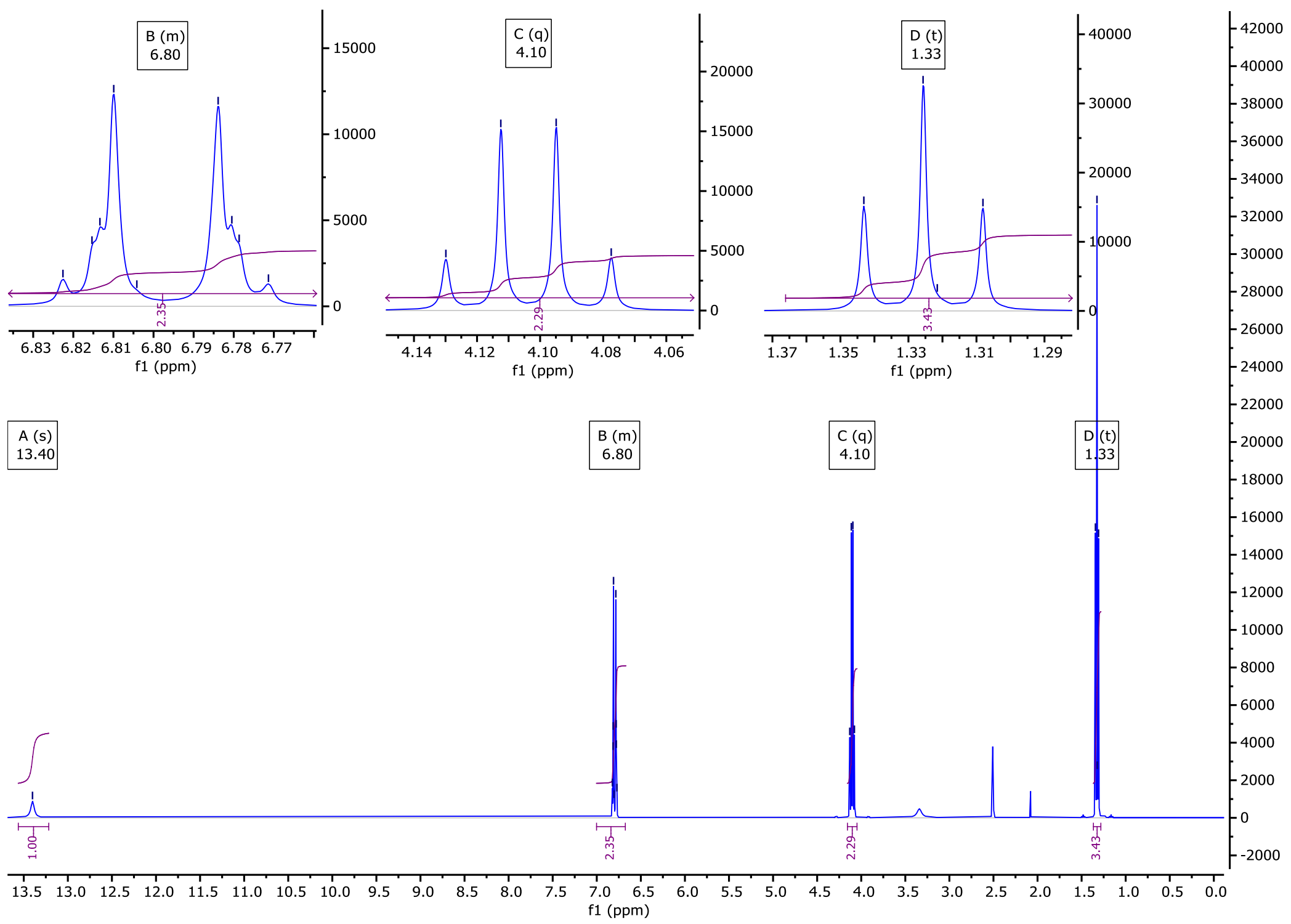

**Figure SI-42:** $^{1}$H NMR of 4-ethoxy-2,6-diflurobenzoic acid

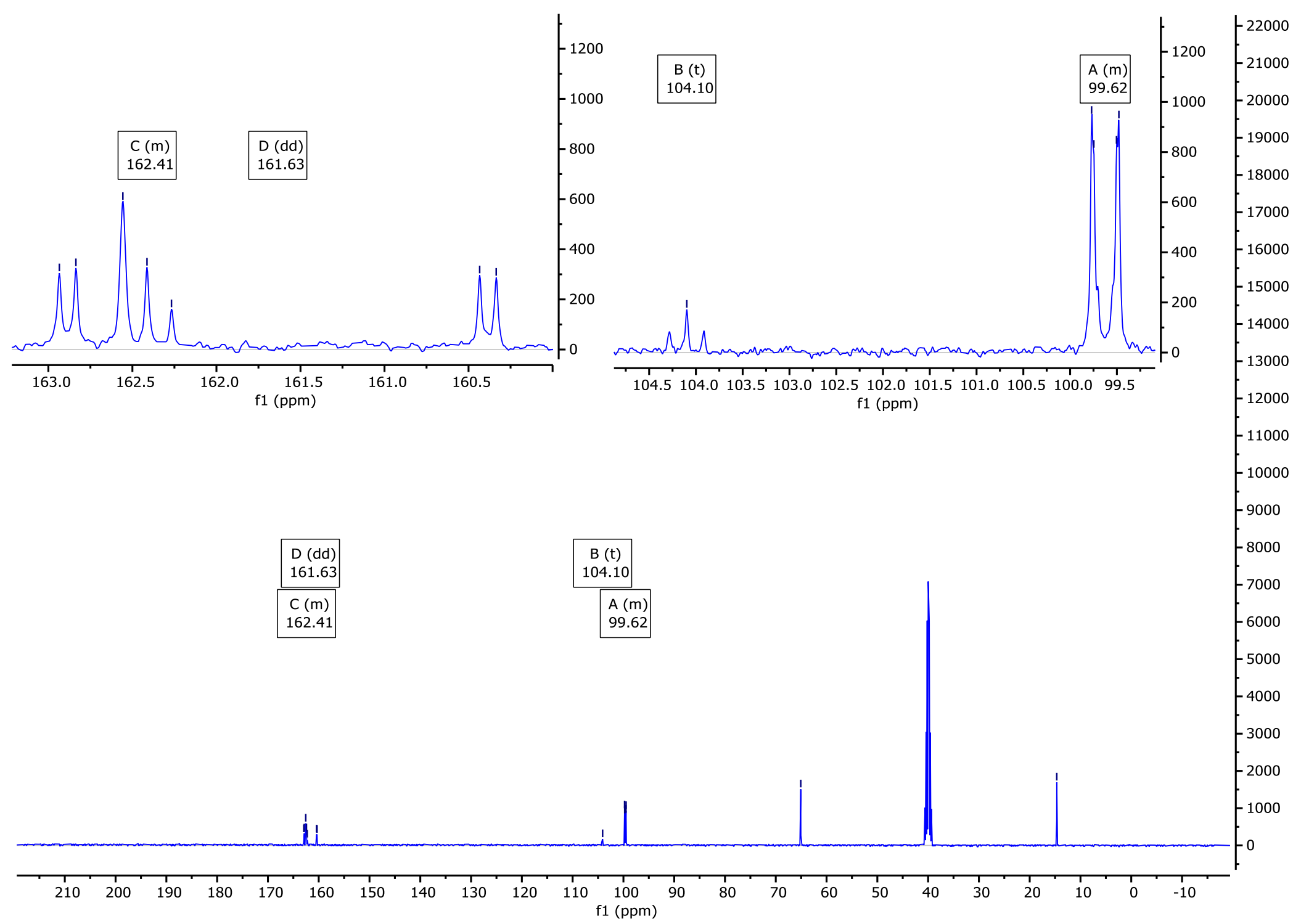

**Figure SI-43:** $^{13}C\{^{1}H\}$ NMR of 4-ethoxy-2,6-diflurobenzoic acid

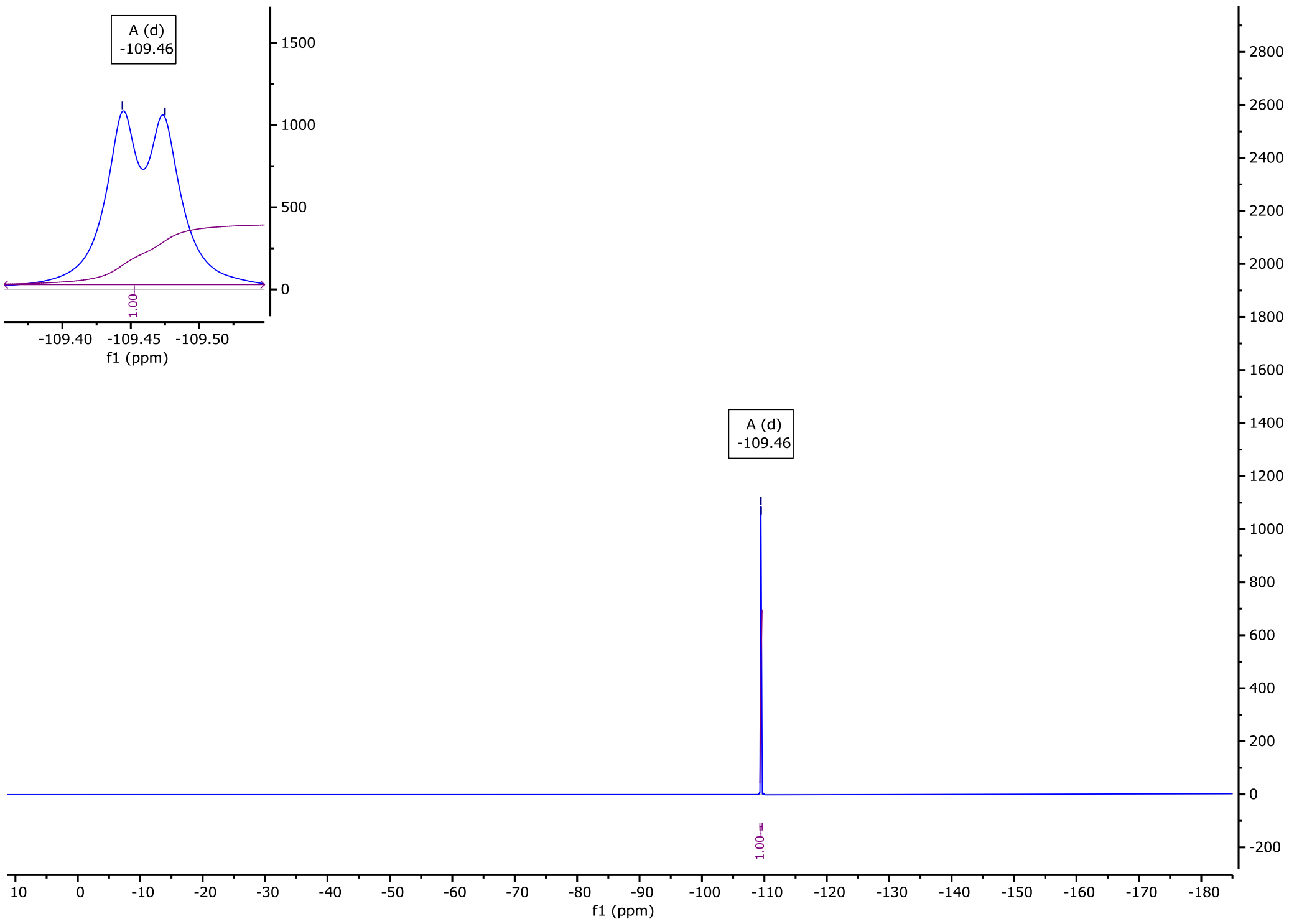

**Figure SI-44:** $^{19}F$ NMR of 4-ethoxy-2,6-diflurobenzoic acid

### 7.3.4. Analytical data for IDX 232 / CJG 484 / 2,3',4',5'-tetrafluoro-[1,1'-biphenyl]-4-yl 4-ethoxy-2,6-difluorobenzoate:

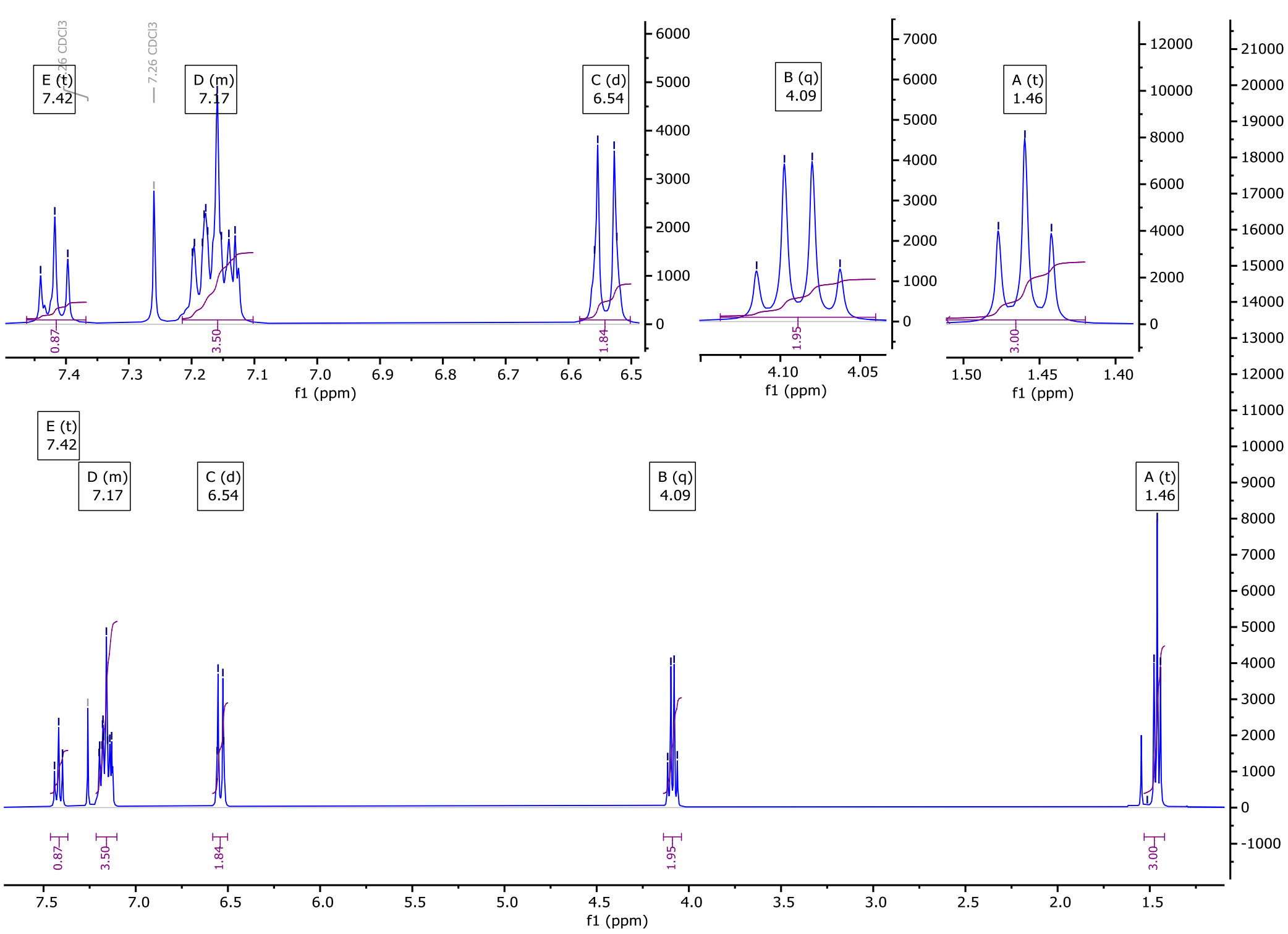

**Figure SI-45:** $^{1}$H NMR for 2,3',4',5'-tetrafluoro-[1,1'-biphenyl]-4-yl 4-ethoxy-2,6-difluorobenzoate

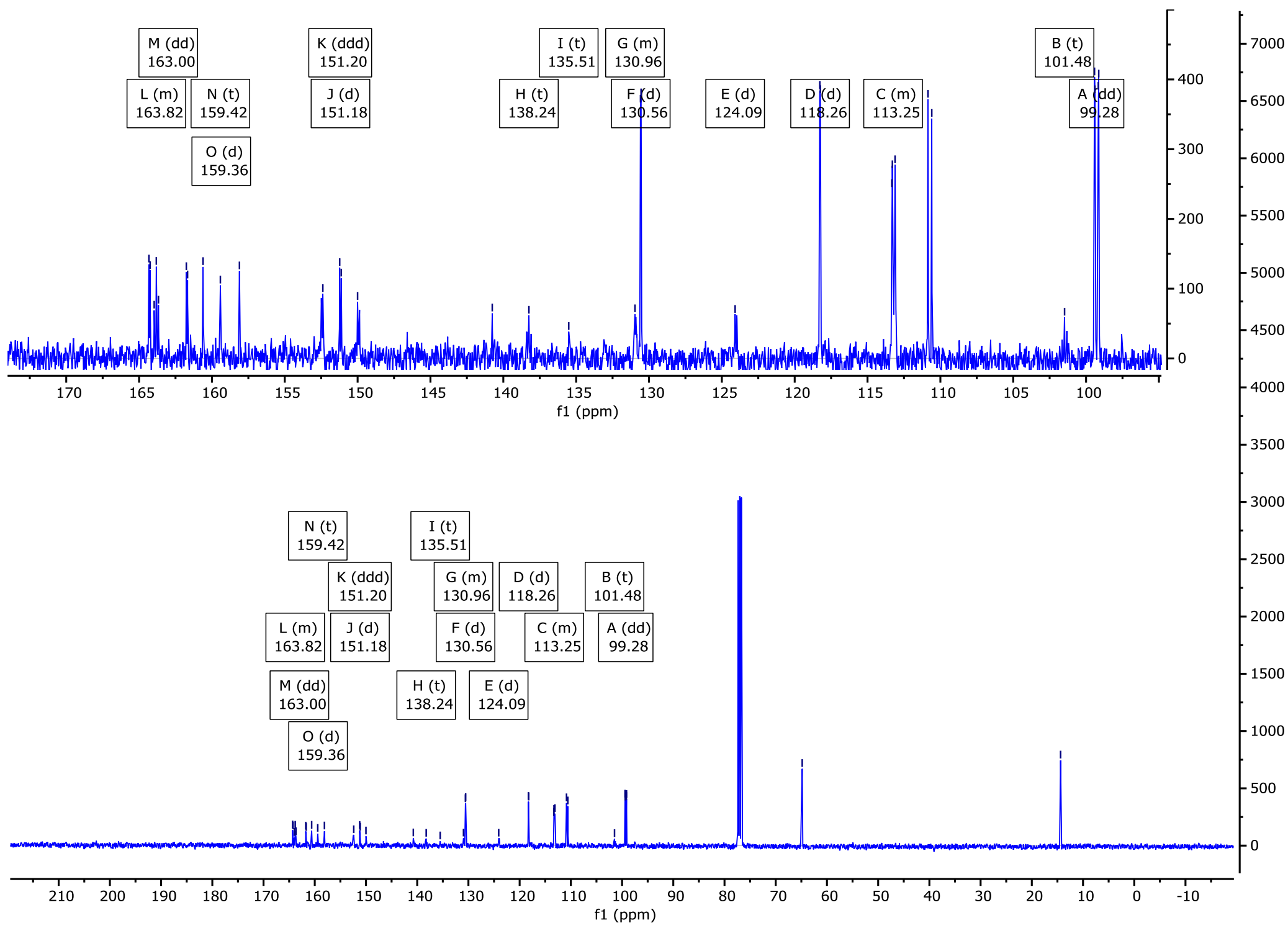

**Figure SI-46:** $^{13}$C{$^{1}$H} NMR for 2,3',4',5'-tetrafluoro-[1,1'-biphenyl]-4-yl 4-ethoxy-2,6-difluorobenzoate

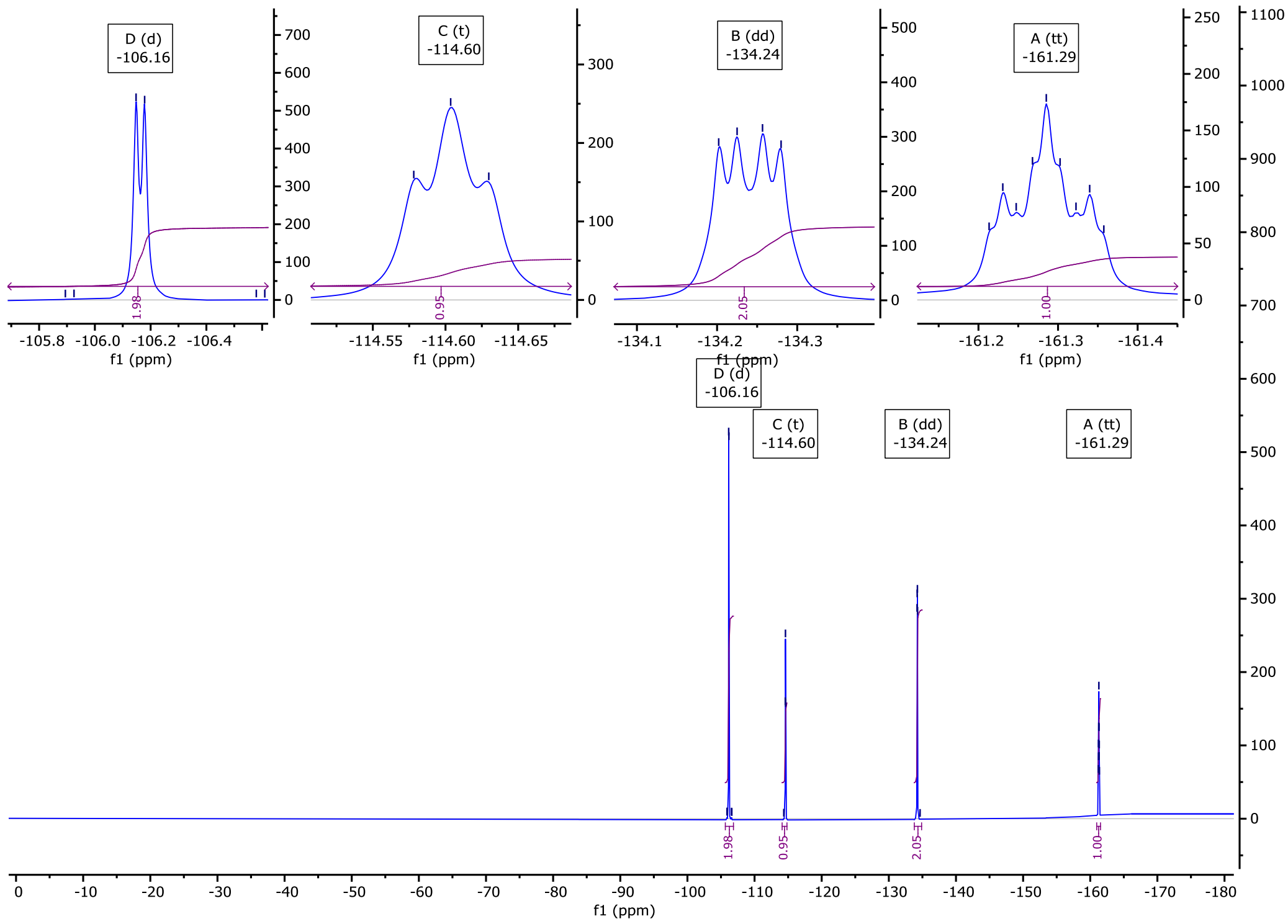

**Figure SI-47:** $^{19}$F NMR 2,3',4',5'-tetrafluoro-[1,1'-biphenyl]-4-yl 4-ethoxy-2,6-difluorobenzoate

## 7.4. Scheme 4

### 7.4.1. 3,5-difluorobenzyl butyl ether

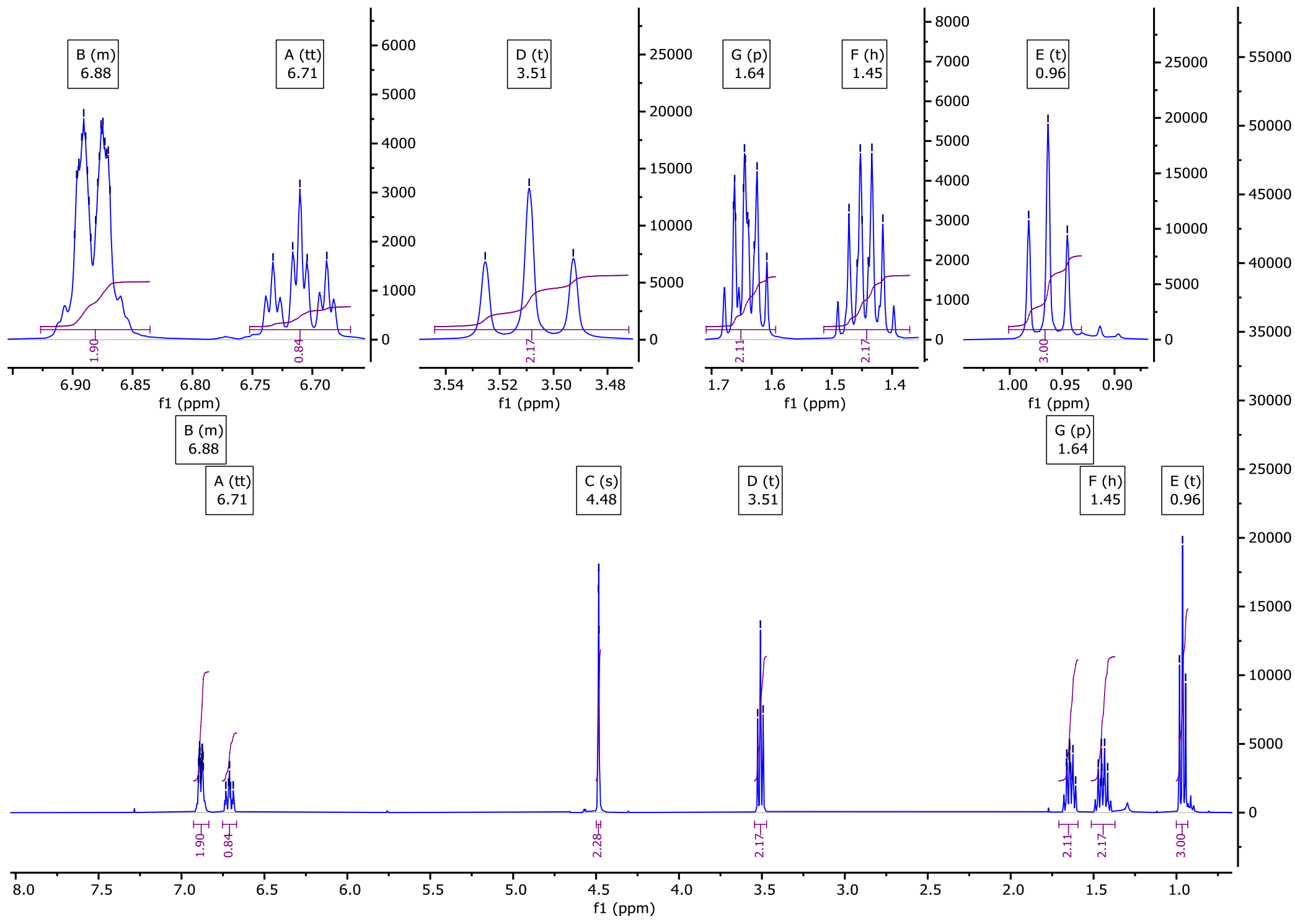

**Figure SI-48:** $^{1}$H NMR for 3,5-difluorobenzyl butyl ether

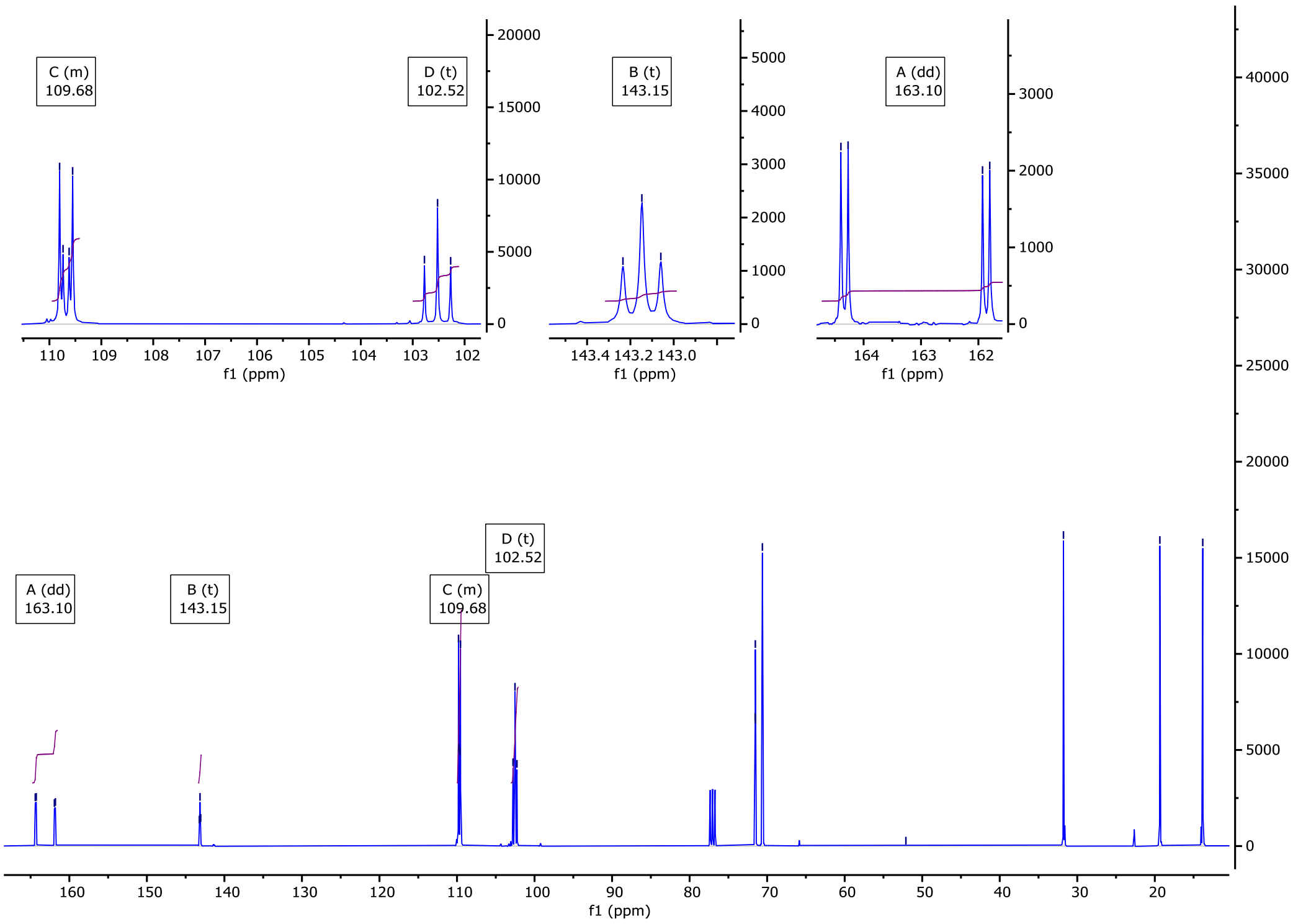

**Figure SI-49:** $^{13}$C{$^{1}$H} NMR for 3,5-difluorobenzyl butyl ether

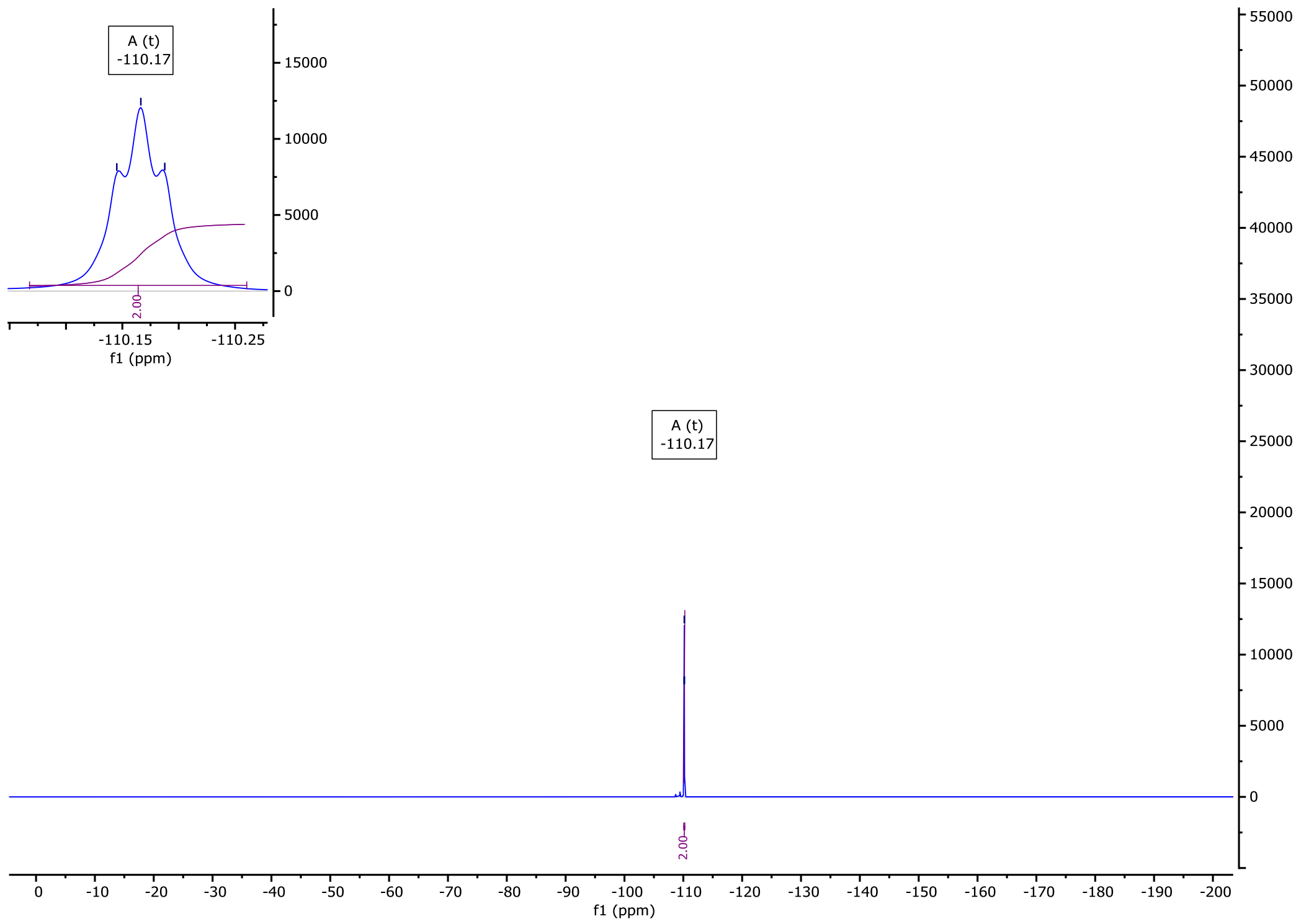

**Figure SI-50:** $^{19}F$ NMR for 3,5-difluorobenzyl butyl ether

### 7.4.2. 2,6-difluoro-2-(butyl-1-yloxy)benzoic acid

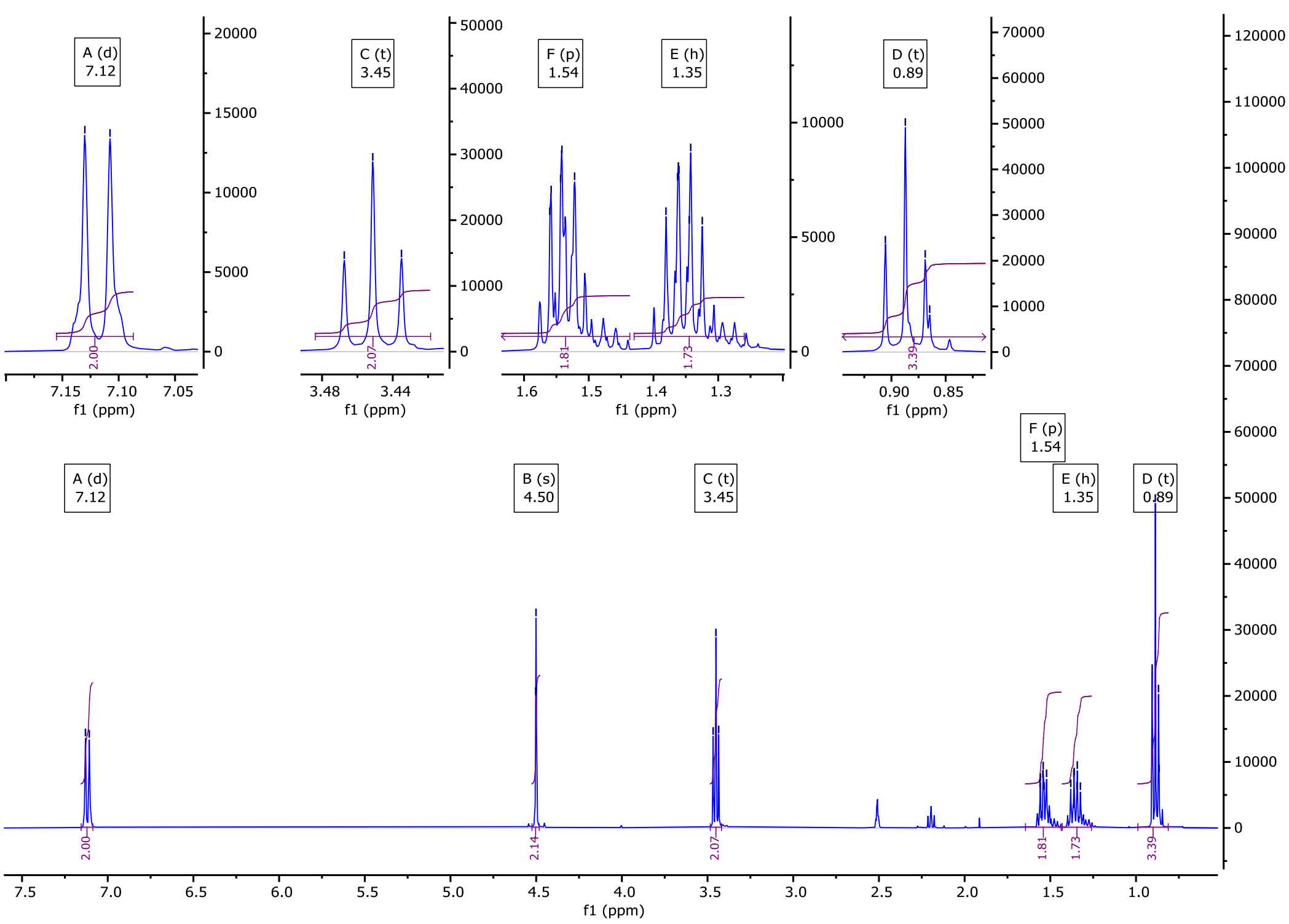

**Figure SI-51:** $^{1}$H NMR for 2,6-difluoro-2-(butyl-1-yloxy)benzoic acid

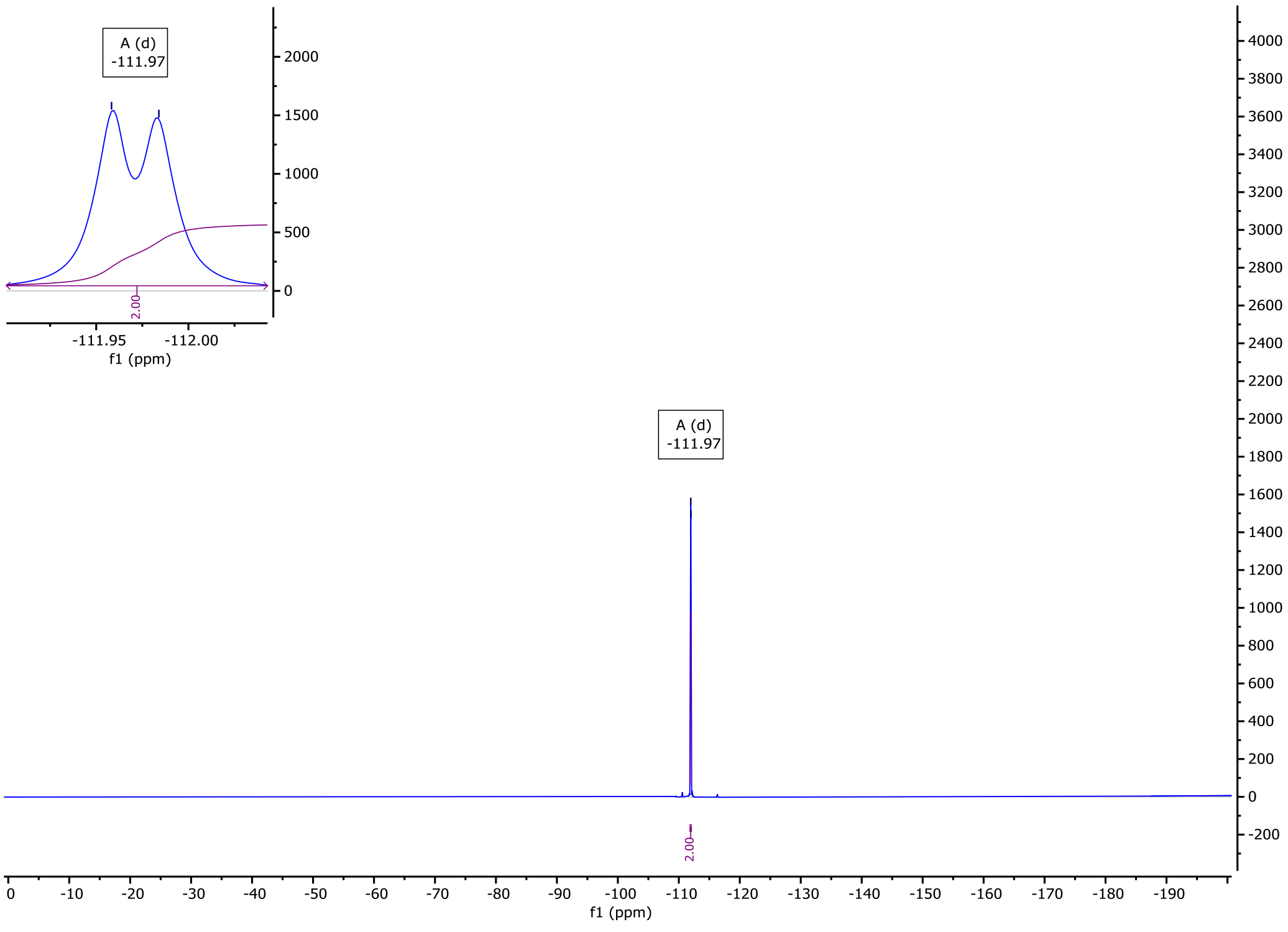

**Figure SI-52:** $^{19}$F NMR for 2,6-difluoro-2-(butyl-1-yloxy)benzoic acid

### 7.4.3. IDX 59 / CJG 480 / 2,3',4',5',6-pentafluoro-[1,1'-biphenyl]-4-yl 4-(butoxymethyl)-2,6-difluorobenzoate

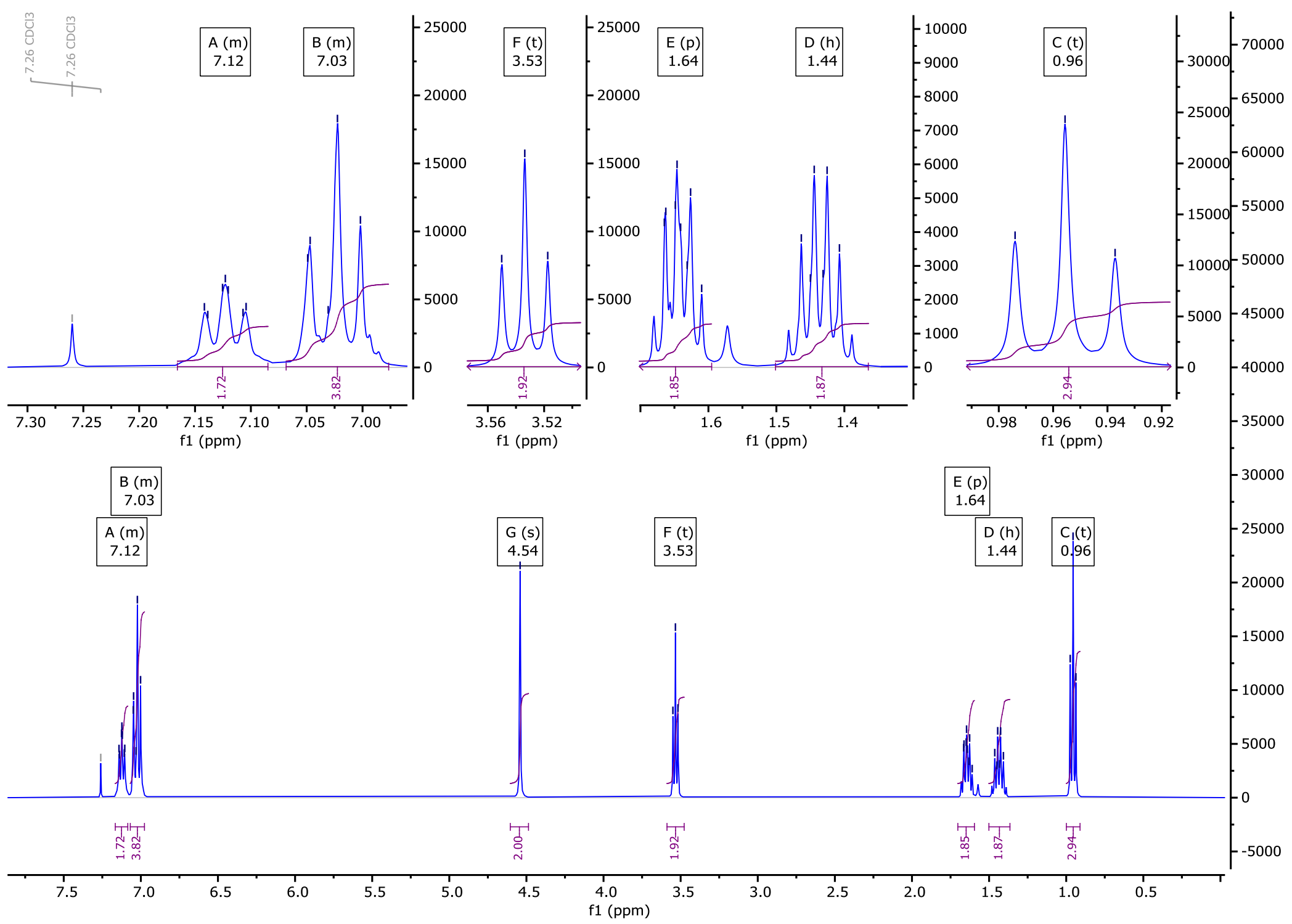

**Figure SI-53:** [1]H NMR of 2,3',4',5',6-pentafluoro-[1,1'-biphenyl]-4-yl 4-(butoxymethyl)-2,6-difluorobenzoate.

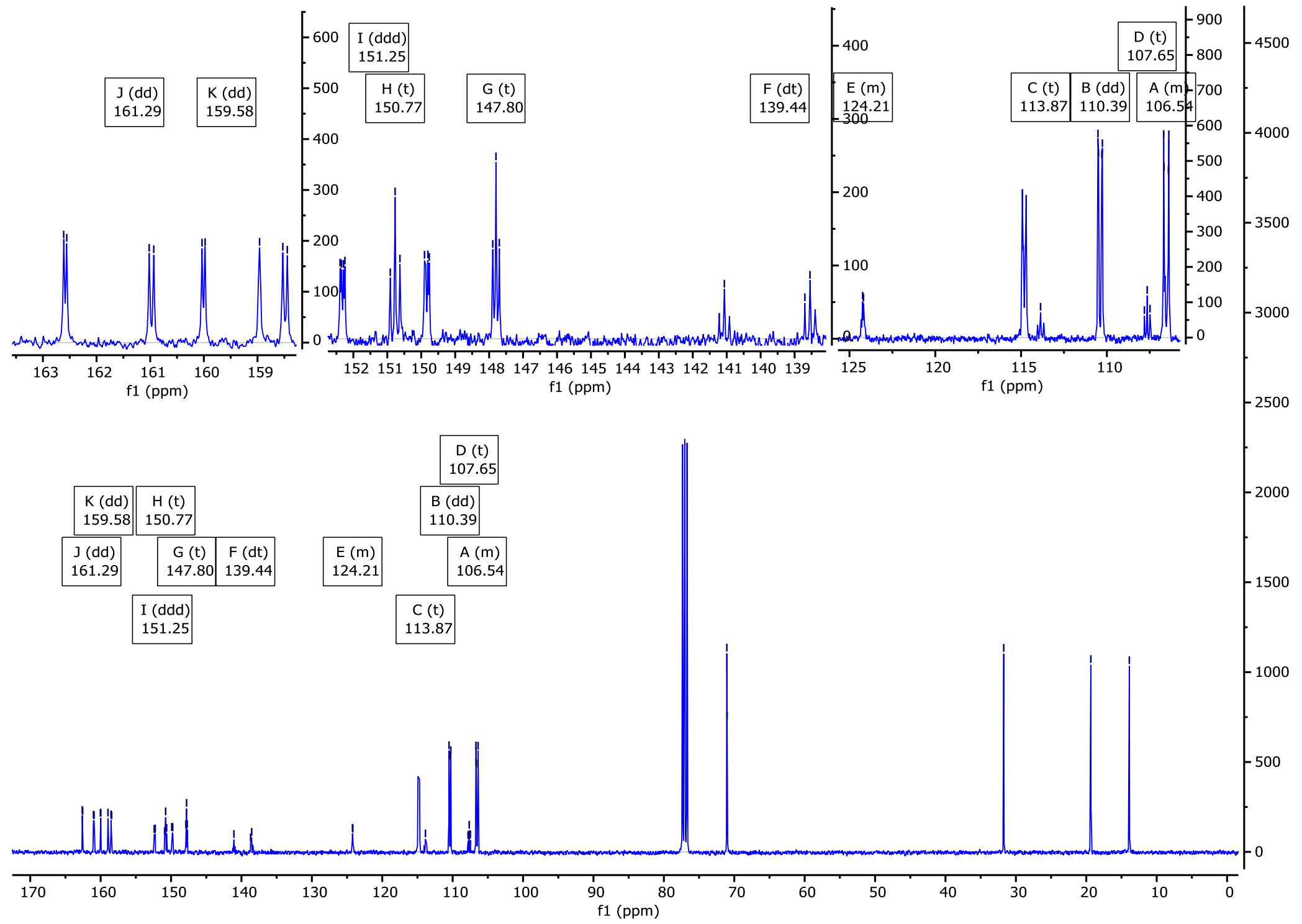

**Figure SI-54:** $^{13}C\{^{1}H\}$ NMR of 2,3',4',5',6-pentafluoro-[1,1'-biphenyl]-4-yl 4-(butoxymethyl)-2,6-difluorobenzoate.

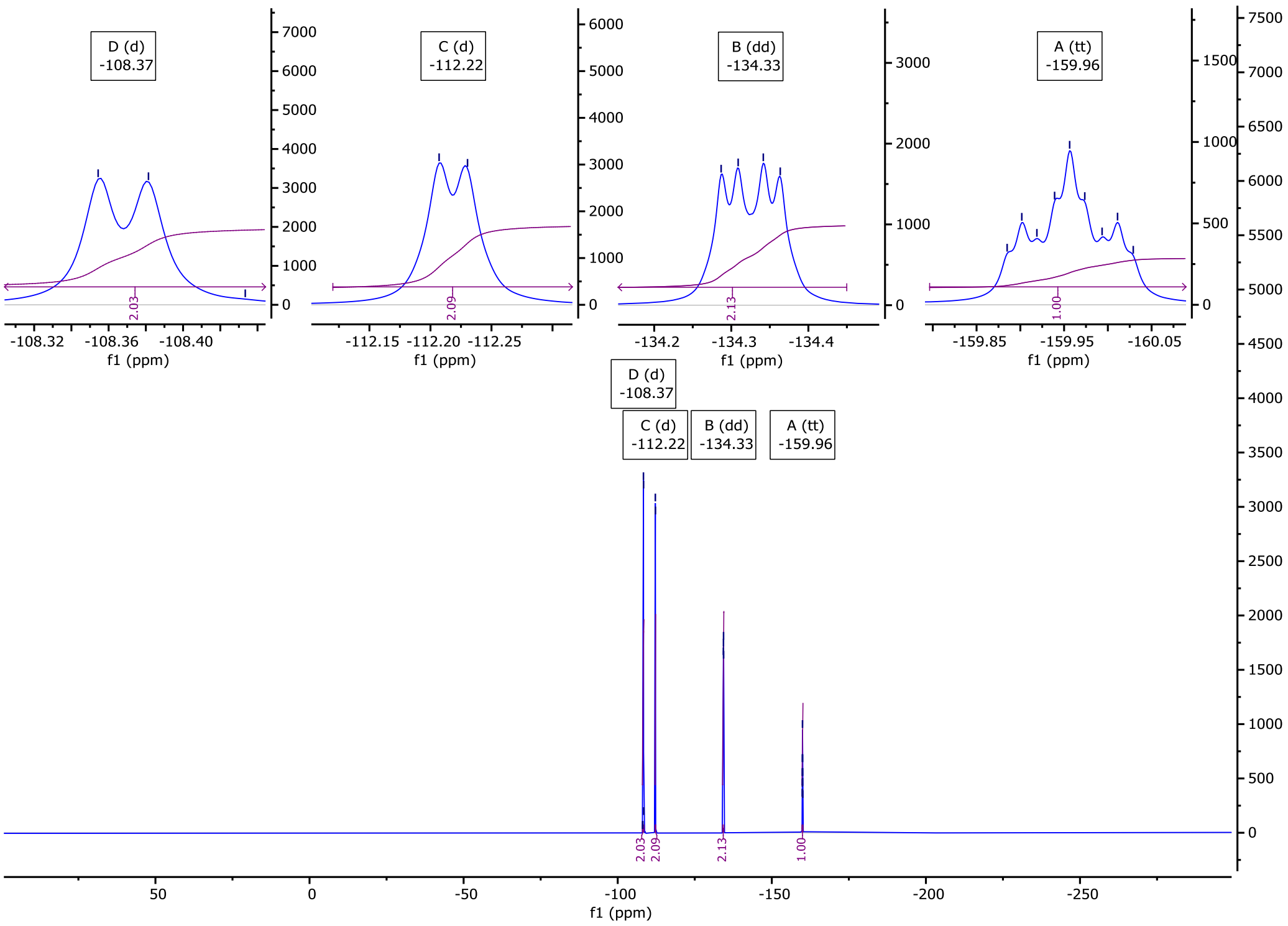

**Figure SI-55:** $^{19}F$ NMR of 2,3',4',5',6-pentafluoro-[1,1'-biphenyl]-4-yl 4-(butoxymethyl)-2,6-difluorobenzoate.

## 7.5. Scheme 5

### 7.5.1. IDX 519 / CJG 410 / 2,3',4',5',6-pentafluoro-[1,1'-biphenyl]-4-yl 2-fluoro-4-(5-propyl-1,3-dioxan-2-yl)benzoate

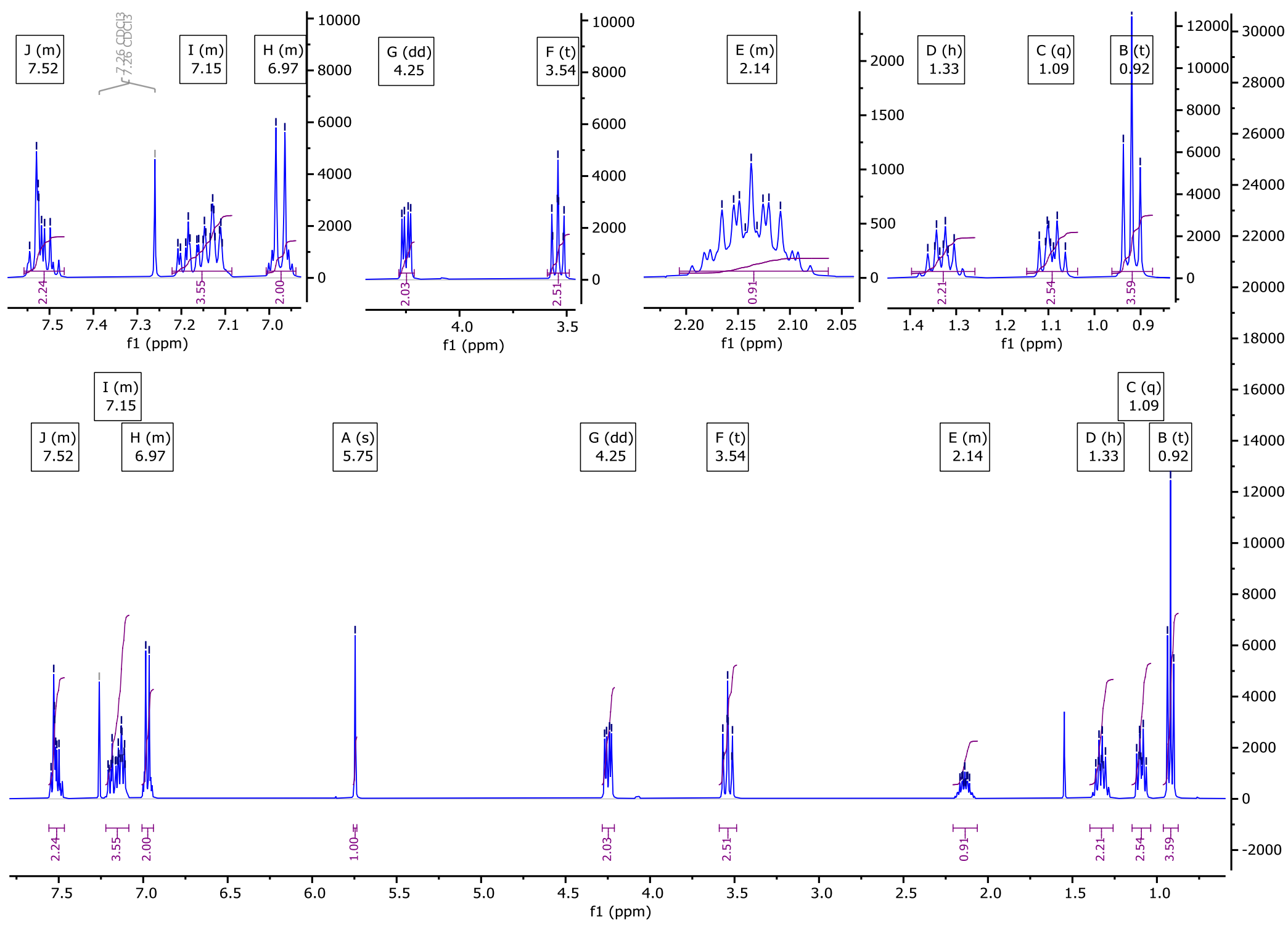

**Figure SI-56:** $^{1}$H NMR of 2,3',4',5',6-pentafluoro-[1,1'-biphenyl]-4-yl 2-fluoro-4-(5-propyl-1,3-dioxan-2-yl)benzoate

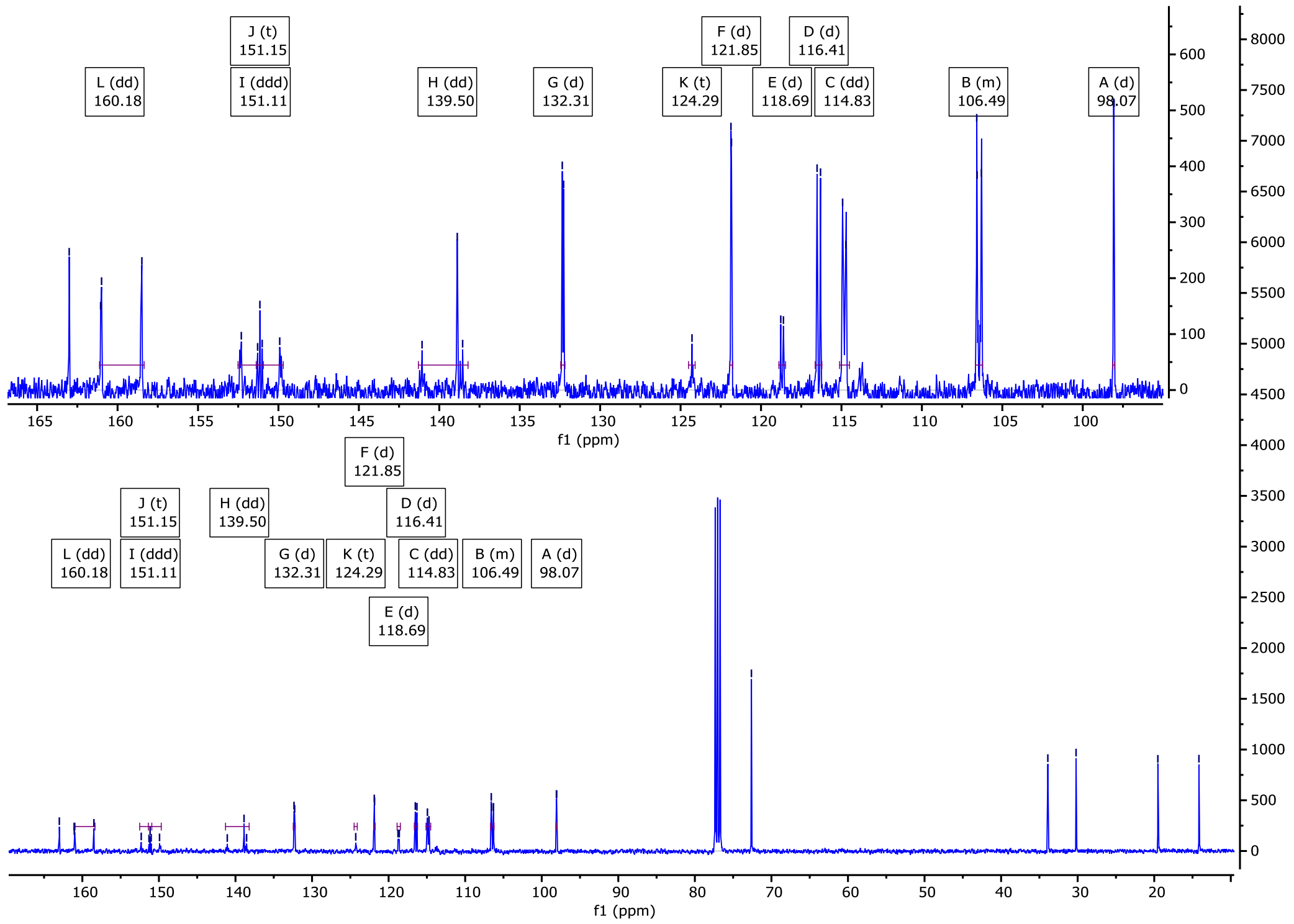

**Figure SI-57:** $^{13}C\{^{1}H\}$ NMR of 2,3',4',5',6-pentafluoro-[1,1'-biphenyl]-4-yl 2-fluoro-4-(5-propyl-1,3-dioxan-2-yl)benzoate

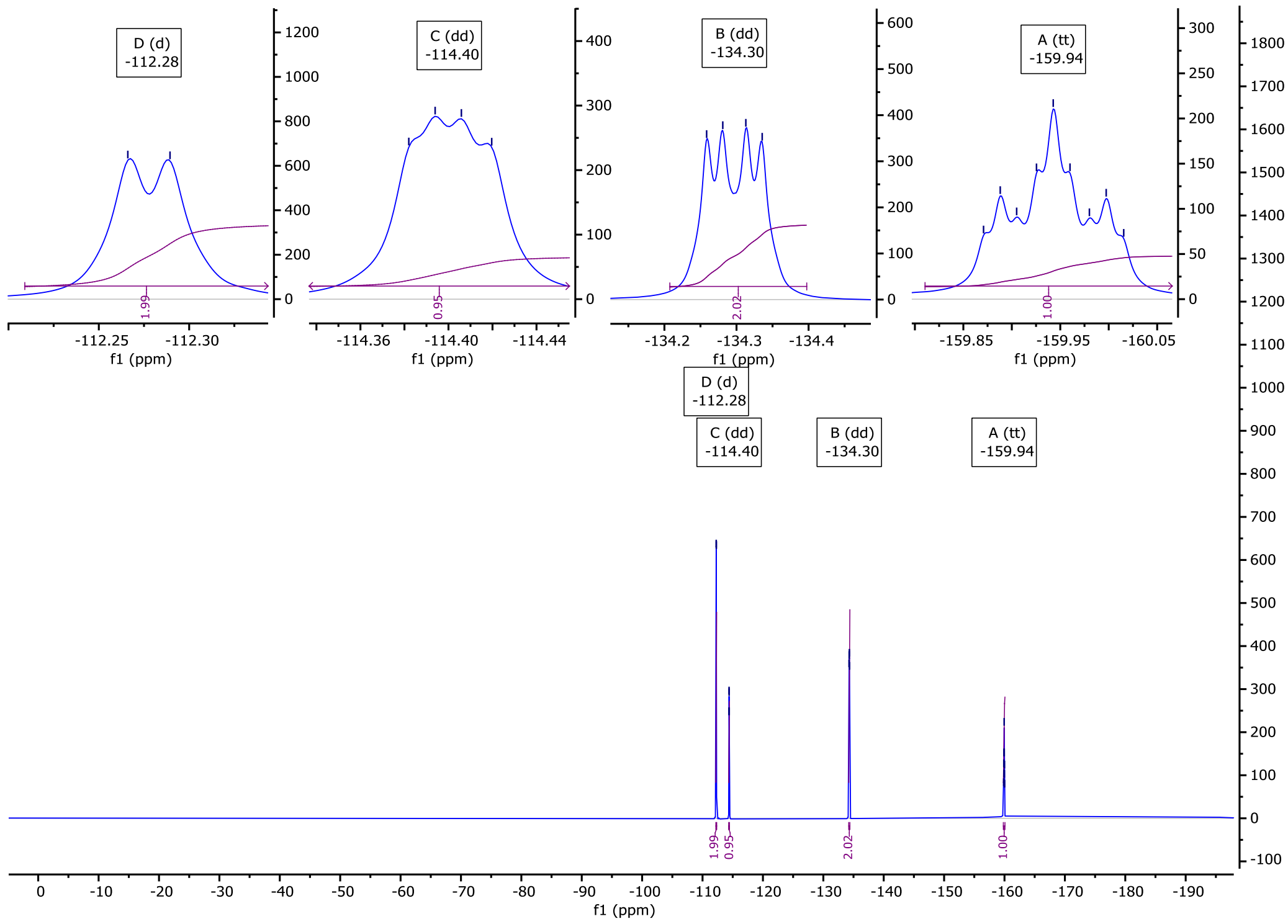

**Figure SI-58:** $^{19}F$ NMR of 2,3',4',5',6-pentafluoro-[1,1'-biphenyl]-4-yl 2-fluoro-4-(5-propyl-1,3-dioxan-2-yl)benzoate

## 7.6. Scheme 6

### 7.6.1. 4-(2,6-Difluoro-4-hydroxyphenyl)-3,4,5-trifluorobenzene

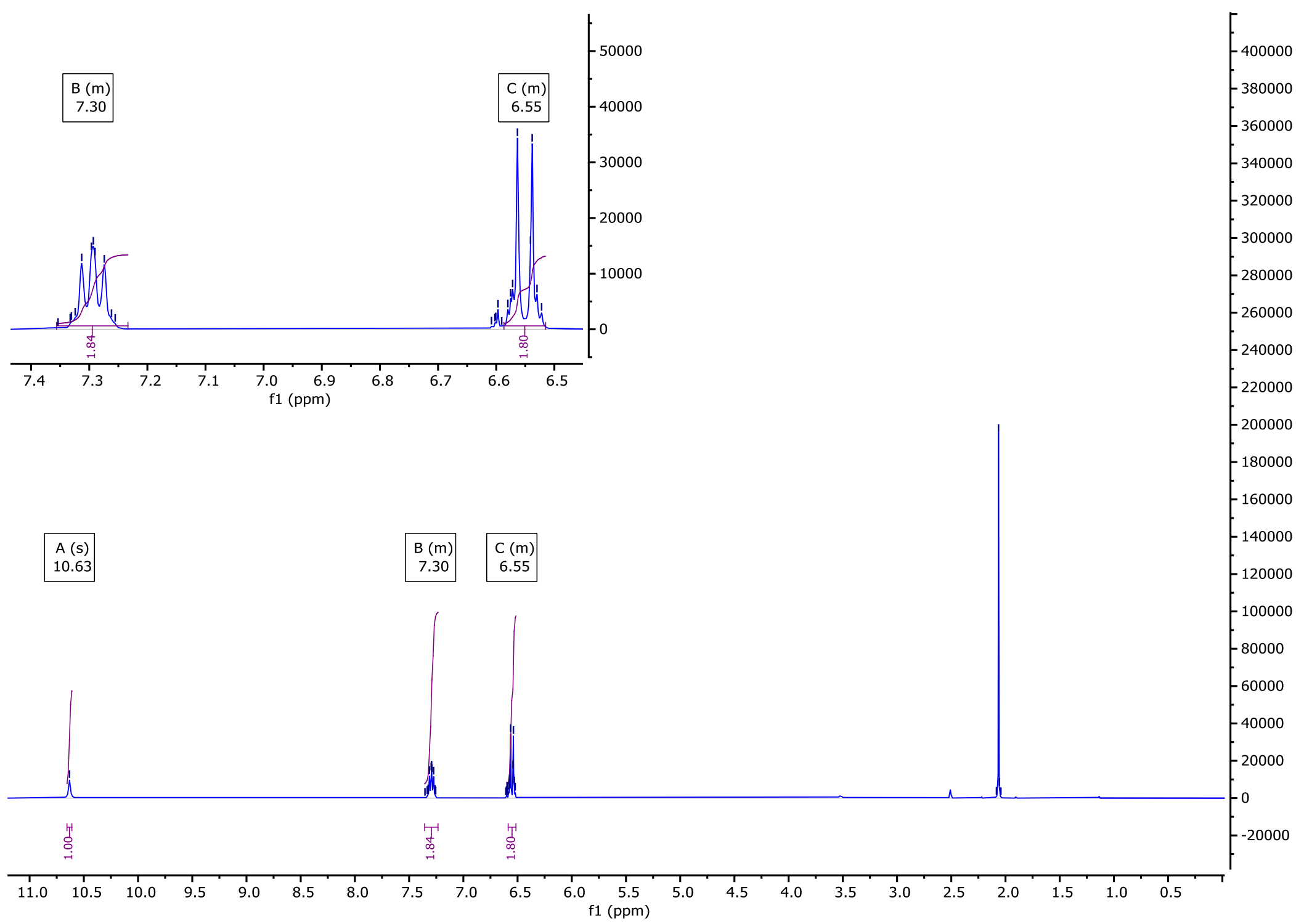

**Figure SI-59:** $^{1}$H NMR of 4-(2,6-Difluoro-4-hydroxyphenyl)-3,4,5-trifluorobenzene

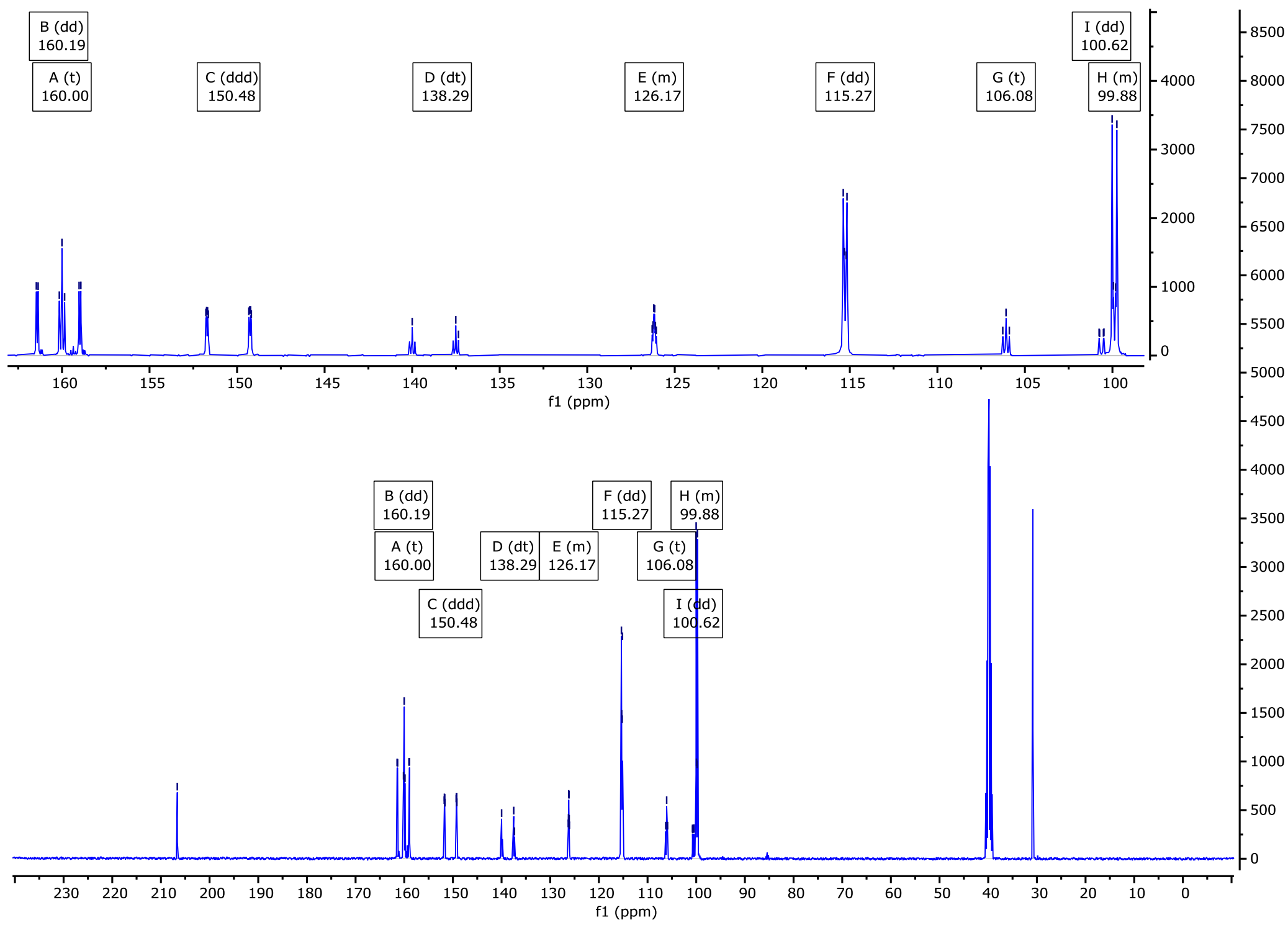

**Figure SI-60:** $^{13}$C{$^{1}$H} NMR of 4-(2,6-Difluoro-4-hydroxyphenyl)-3,4,5-trifluorobenzene

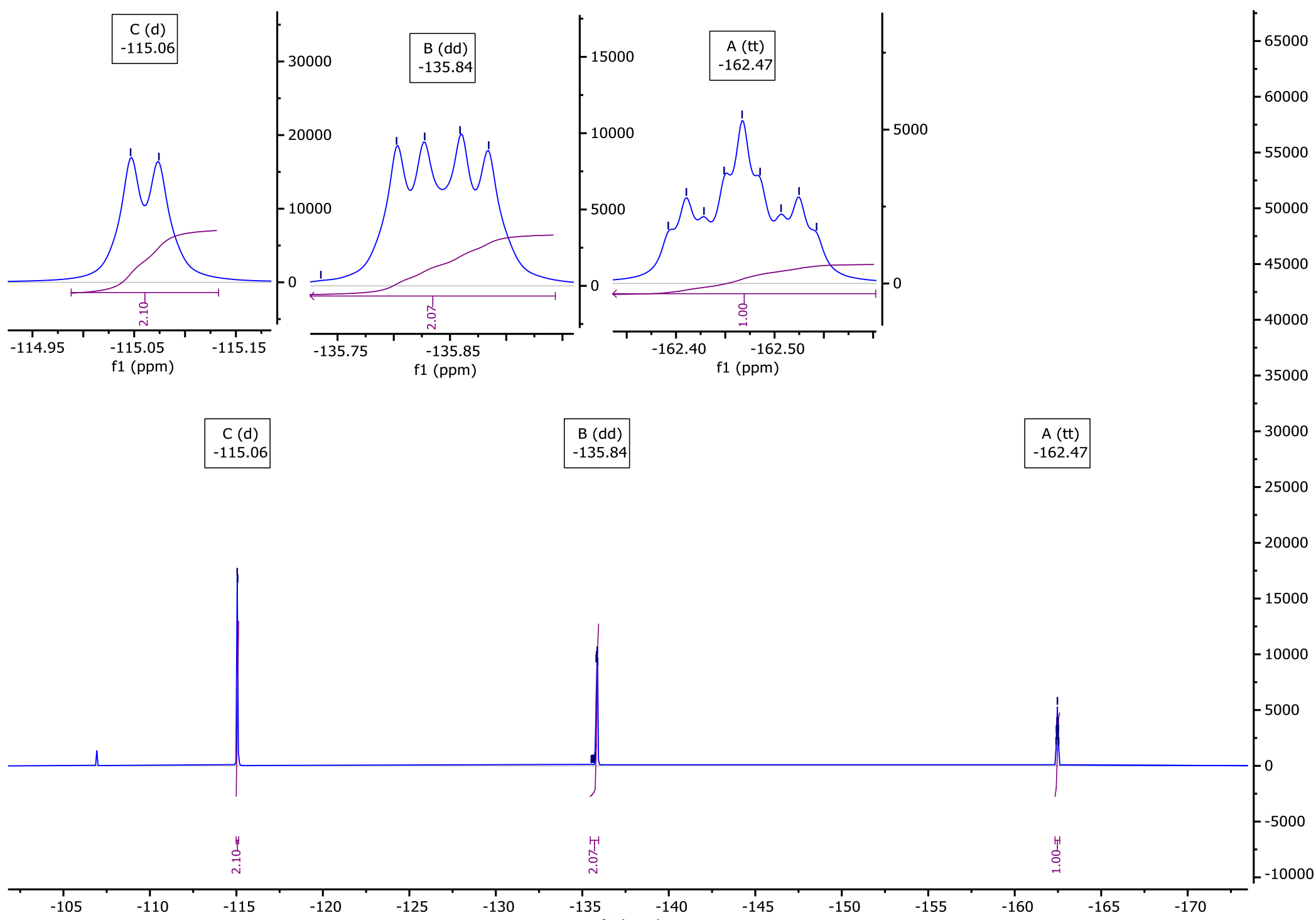

**Figure SI-61:** $^{19}$F NMR of 4-(2,6-Difluoro-4-hydroxyphenyl)-3,4,5-trifluorobenzene

## 7.7. Scheme 7

### 7.7.1. 2,6-Difluoro-4-(4-hydroxyphenyl)toluene

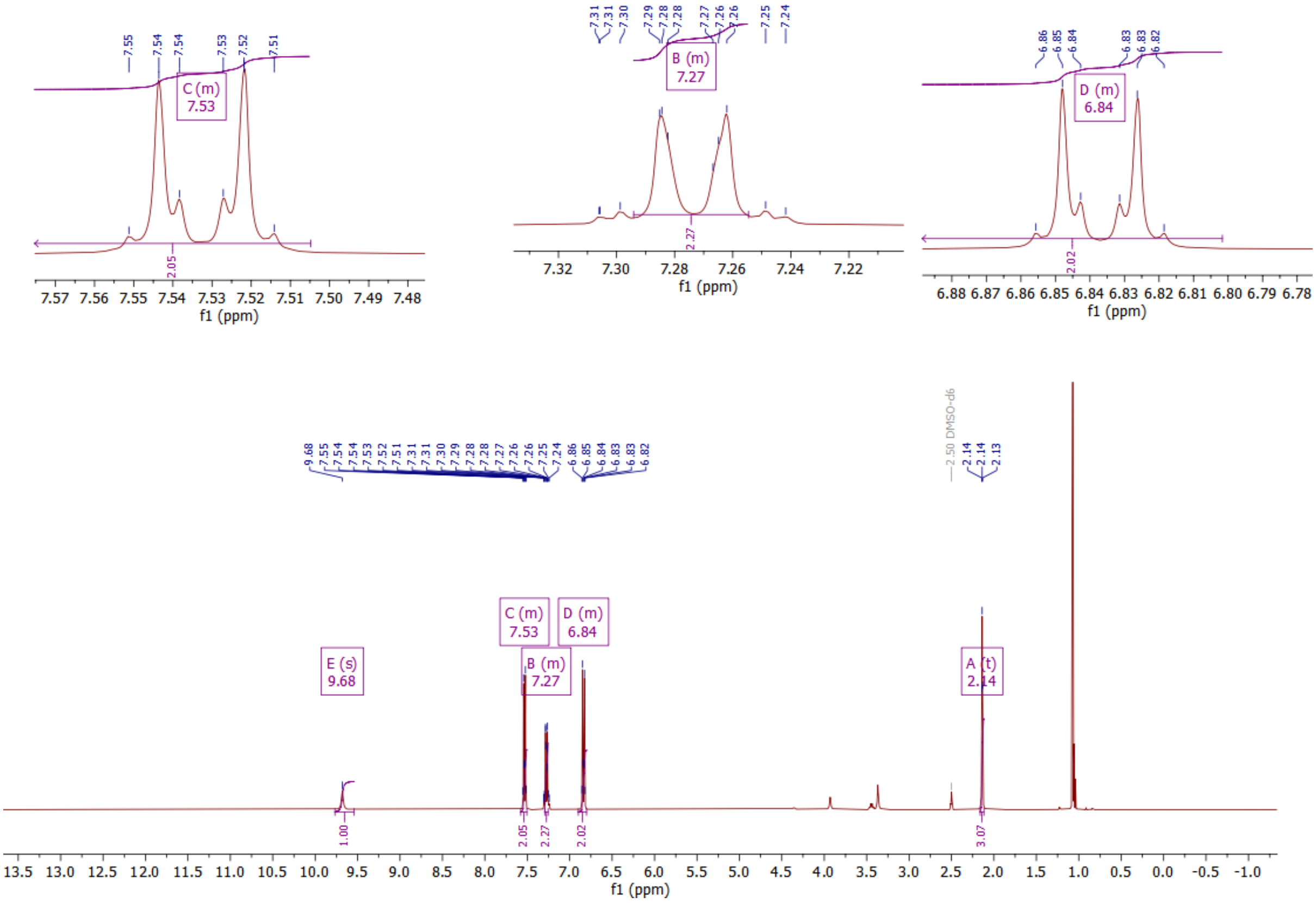

**Figure SI-62:** $^{1}$H NMR of 2,6-Difluoro-4-(4-hydroxyphenyl)toluene)

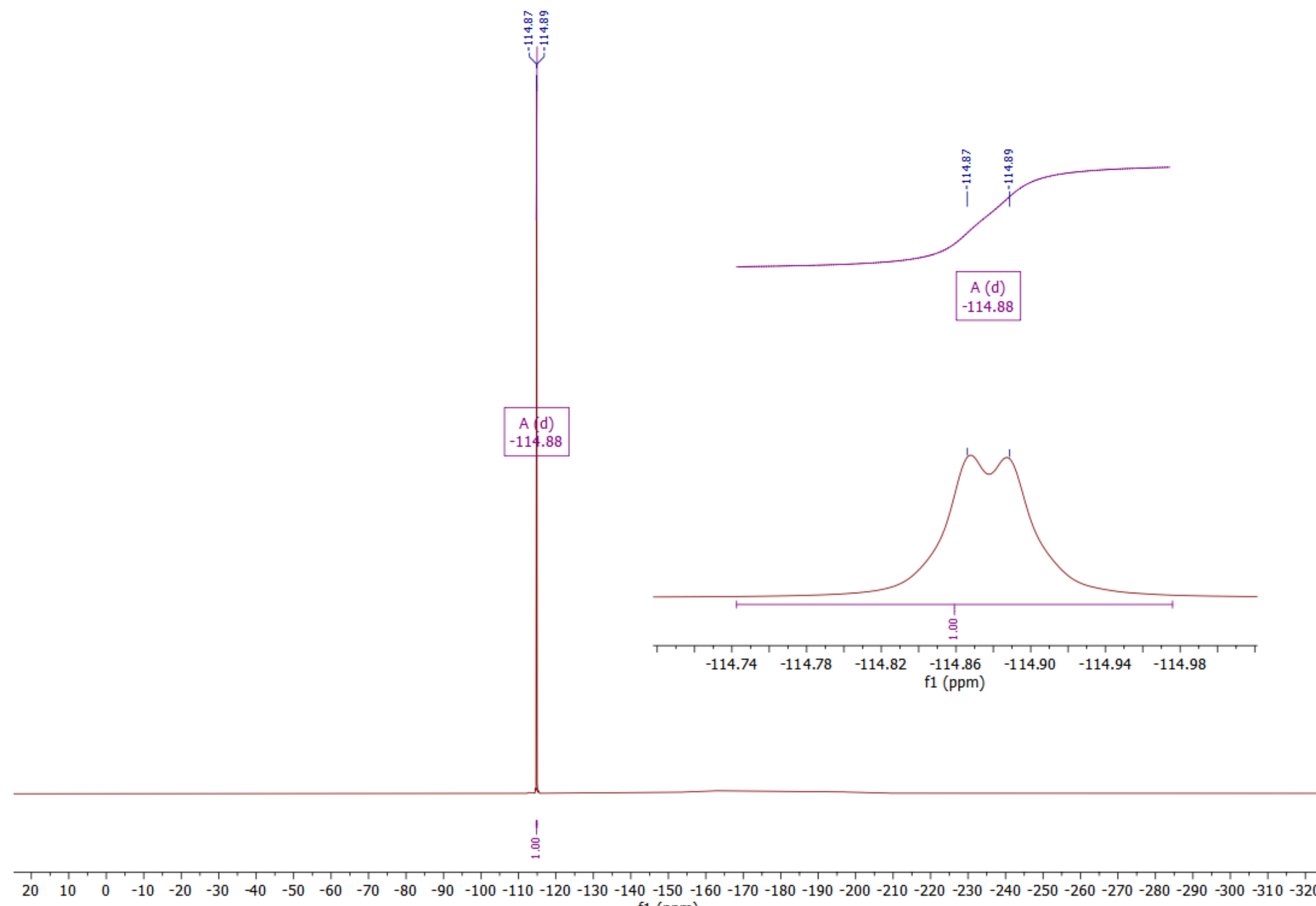

**Figure SI-63:** $^{19}$F NMR of 2,6-difluoro-4-(4-hydroxyphenyl)toluene)

### 7.7.2. 3',5'-difluoro-4'-methyl-[1,1'-biphenyl]-4-yl 2,6-difluoro-4-((2r,5r)-5-propyl-1,3-dioxan-2-yl)benzoate

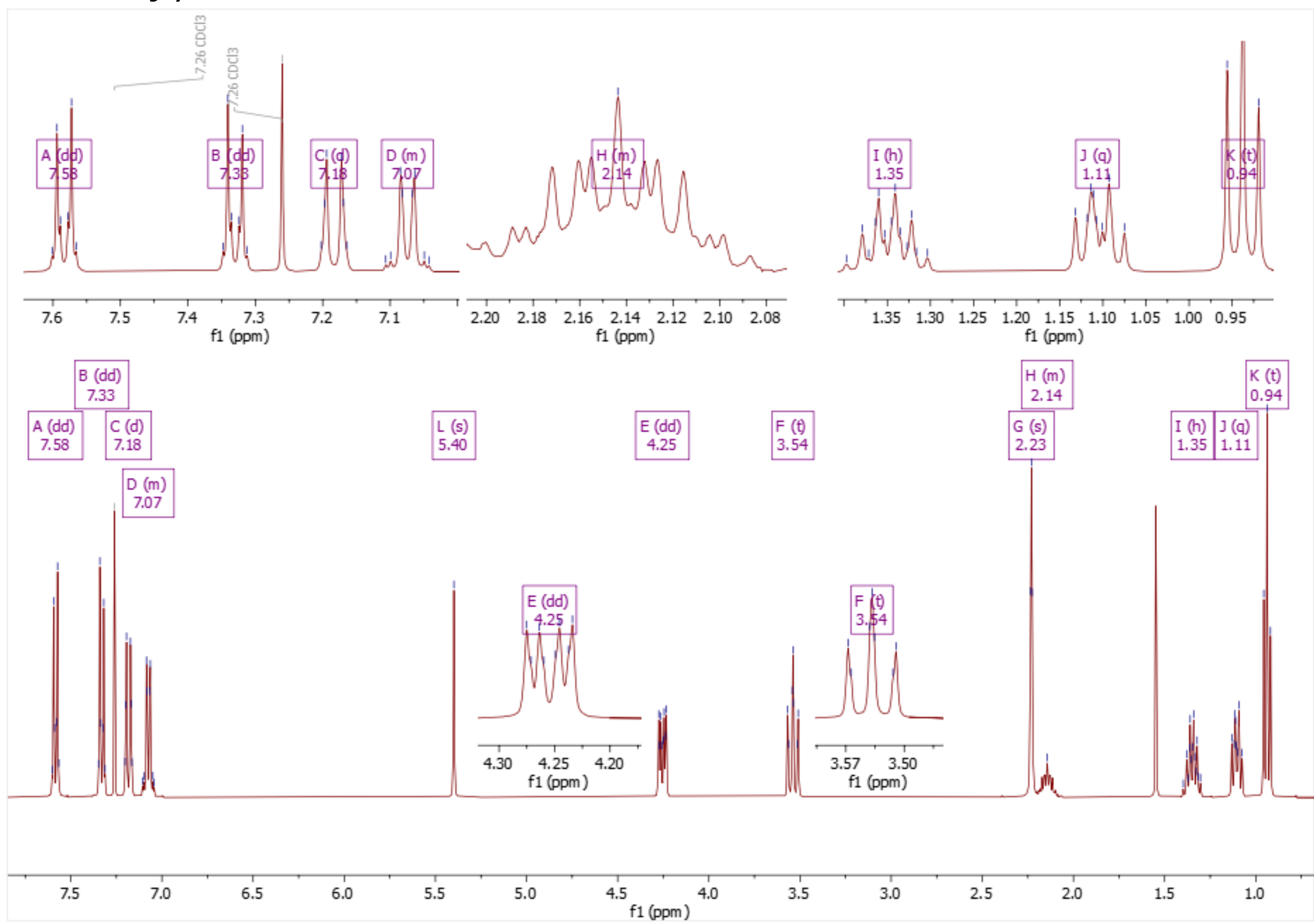

**Figure SI-64:** $^{1}$H NMR for 3',5'-difluoro-4'-methyl-[1,1'-biphenyl]-4-yl 2,6-difluoro-4-((2r,5r)-5-propyl-1,3-dioxan-2-yl)benzoate

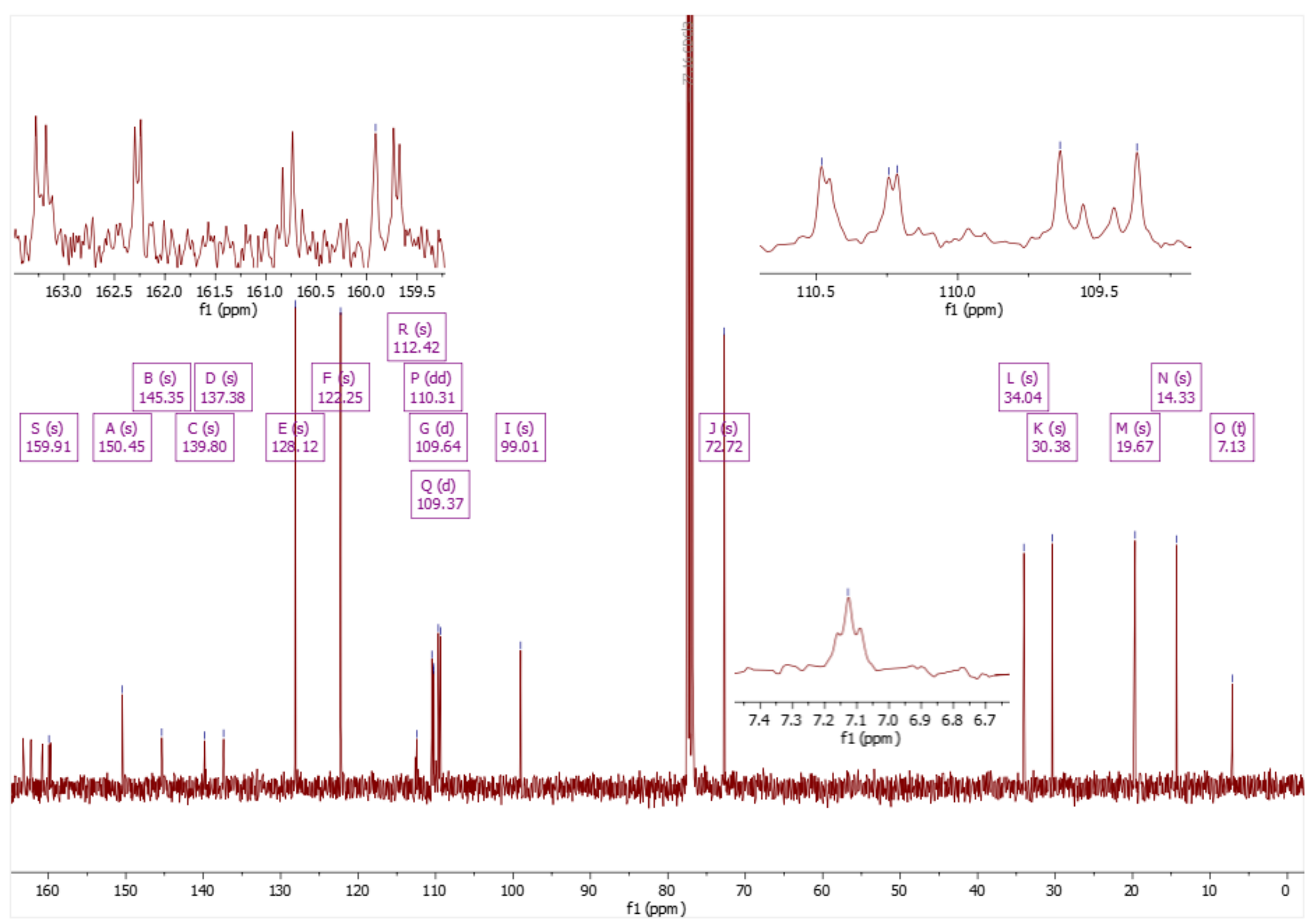

**Figure SI-65:** $^{13}C\{^{1}H\}$ NMR for 3',5'-difluoro-4'-methyl-[1,1'-biphenyl]-4-yl 2,6-difluoro-4-((2r,5r)-5-propyl-1,3-dioxan-2-yl)benzoate

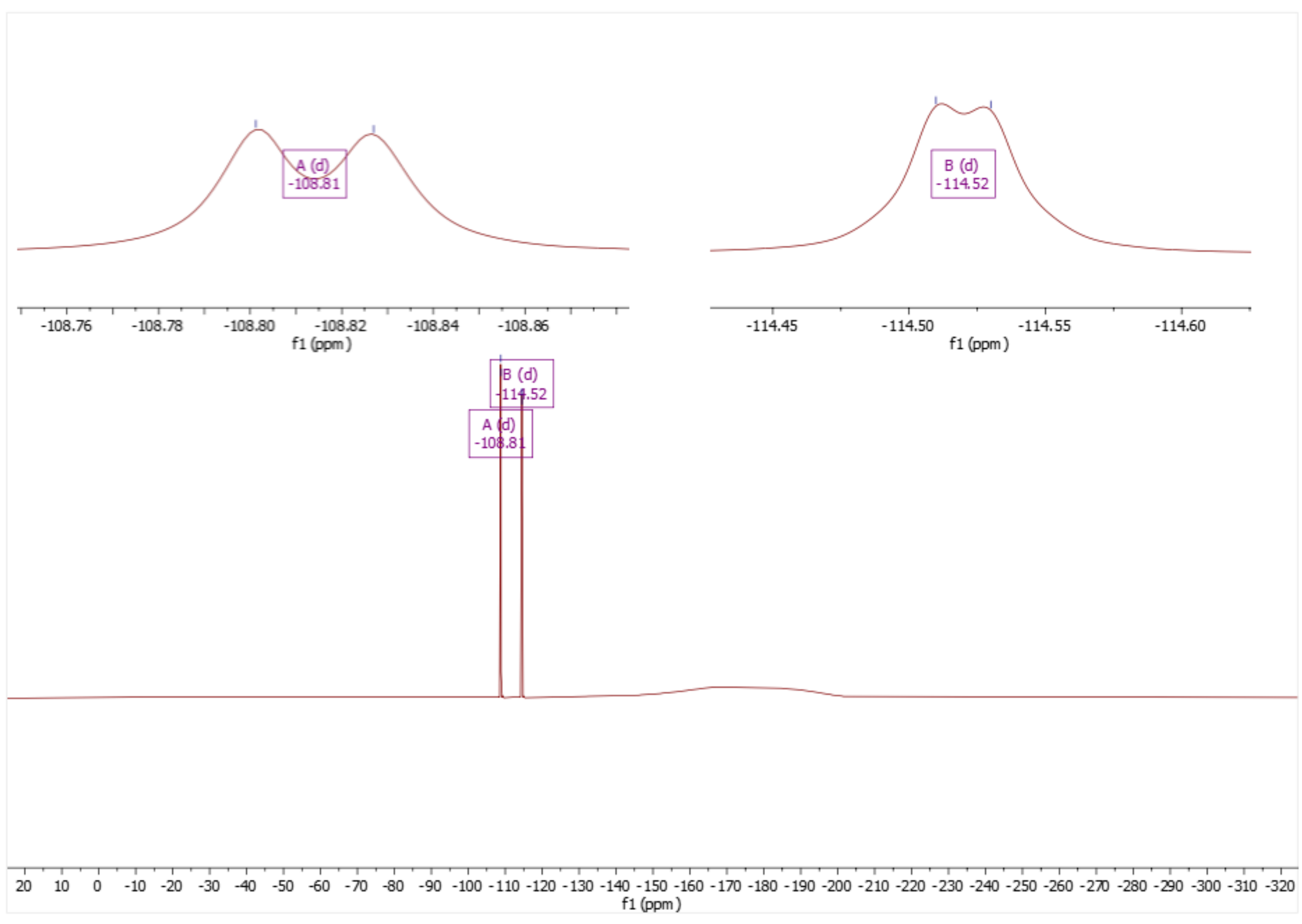

**Figure SI-66:** $^{19}F$ NMR for 3',5'-difluoro-4'-methyl-[1,1'-biphenyl]-4-yl 2,6-difluoro-4-((2r,5r)-5-propyl-1,3-dioxan-2-yl)benzoate

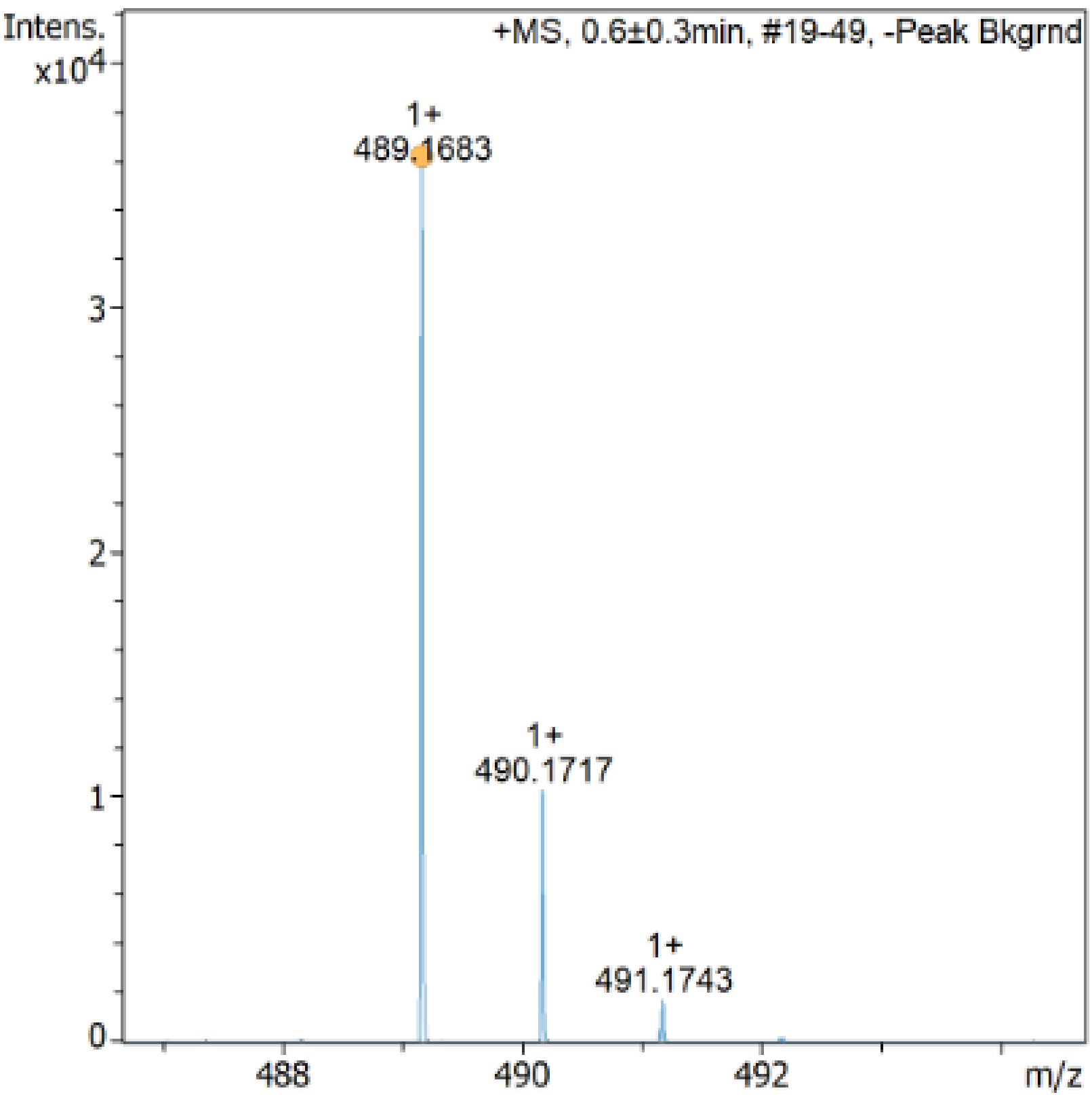

**Figure SI-67:** HRMS for 3',5'-difluoro-4'-methyl-[1,1'-biphenyl]-4-yl 2,6-difluoro-4-((2r,5r)-5-propyl-1,3-dioxan-2-yl)benzoate

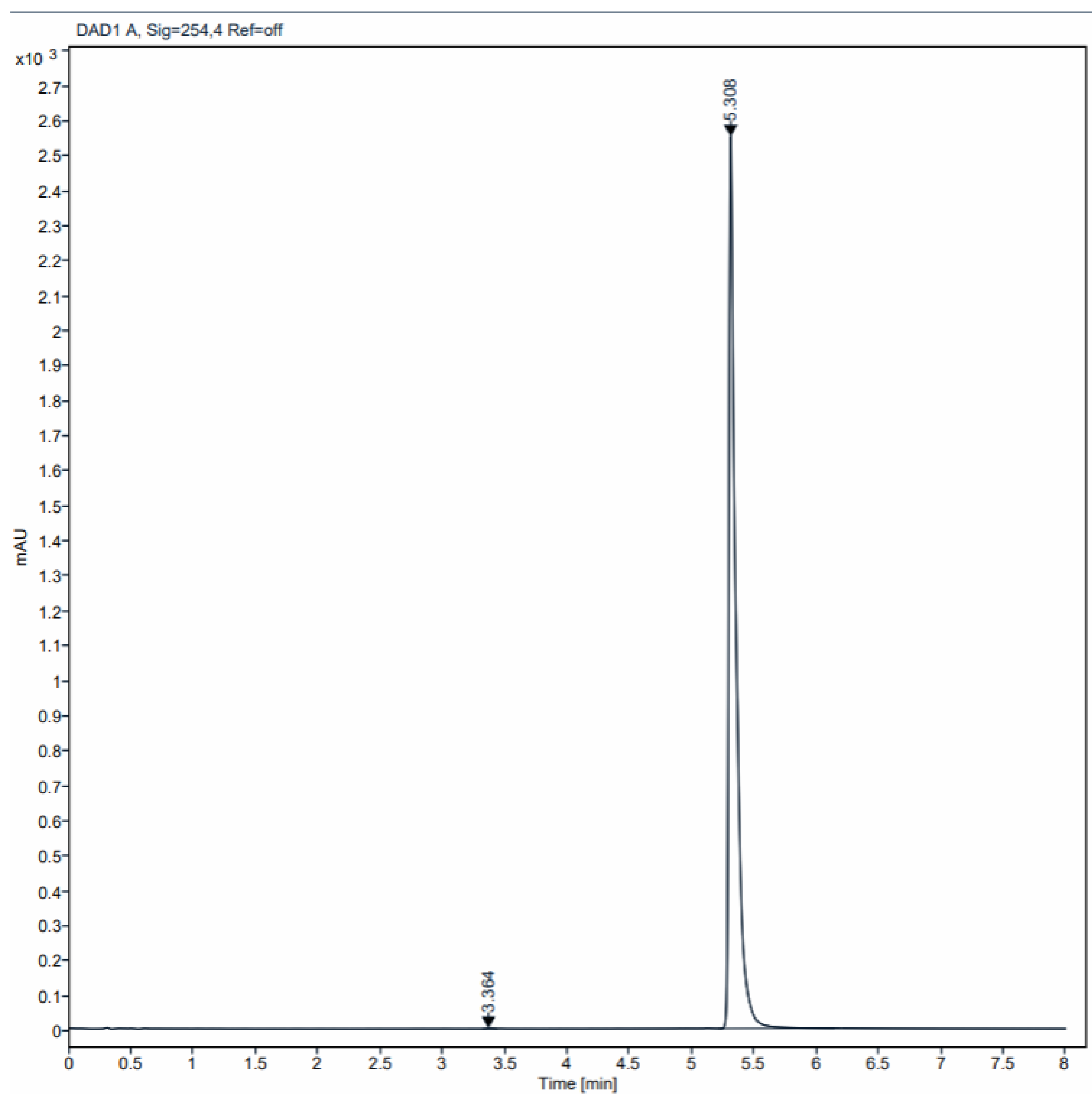

**Figure SI-68:** HPLC for 3',5'-difluoro-4'-methyl-[1,1'-biphenyl]-4-yl 2,6-difluoro-4-((2r,5r)-5-propyl-1,3-dioxan-2-yl)benzoate

### 7.8.1. 4-((4-Ethylphenoxy)carbonyl)phenyl-2,4-dimethoxybenzoate (COB30 IDX 1014)

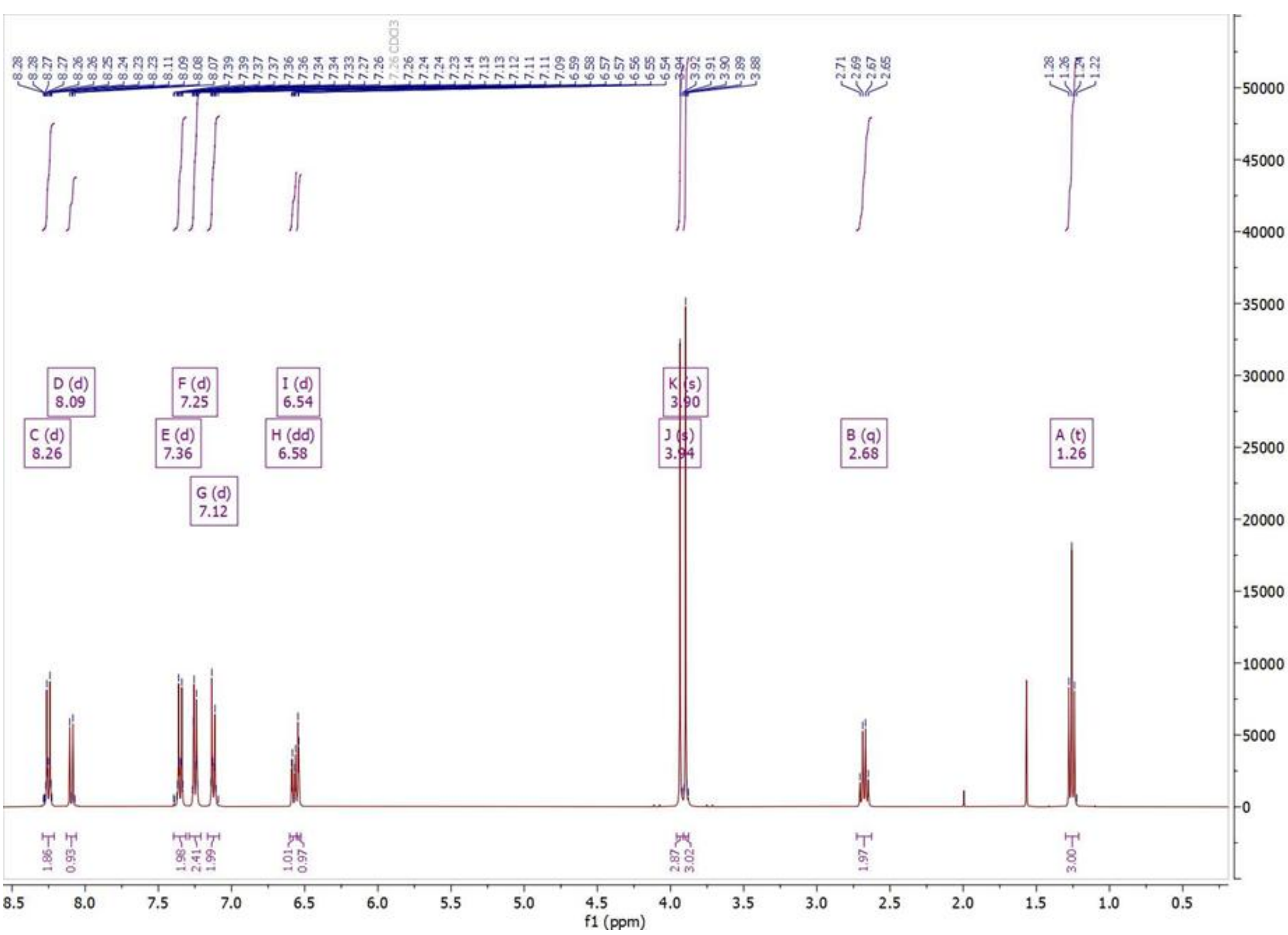

**Figure SI-69:** $^{1}H$ NMR for 4-((4-Ethylphenoxy)carbonyl)phenyl-2,4-dimethoxybenzoate (COB30 IDX 1014)

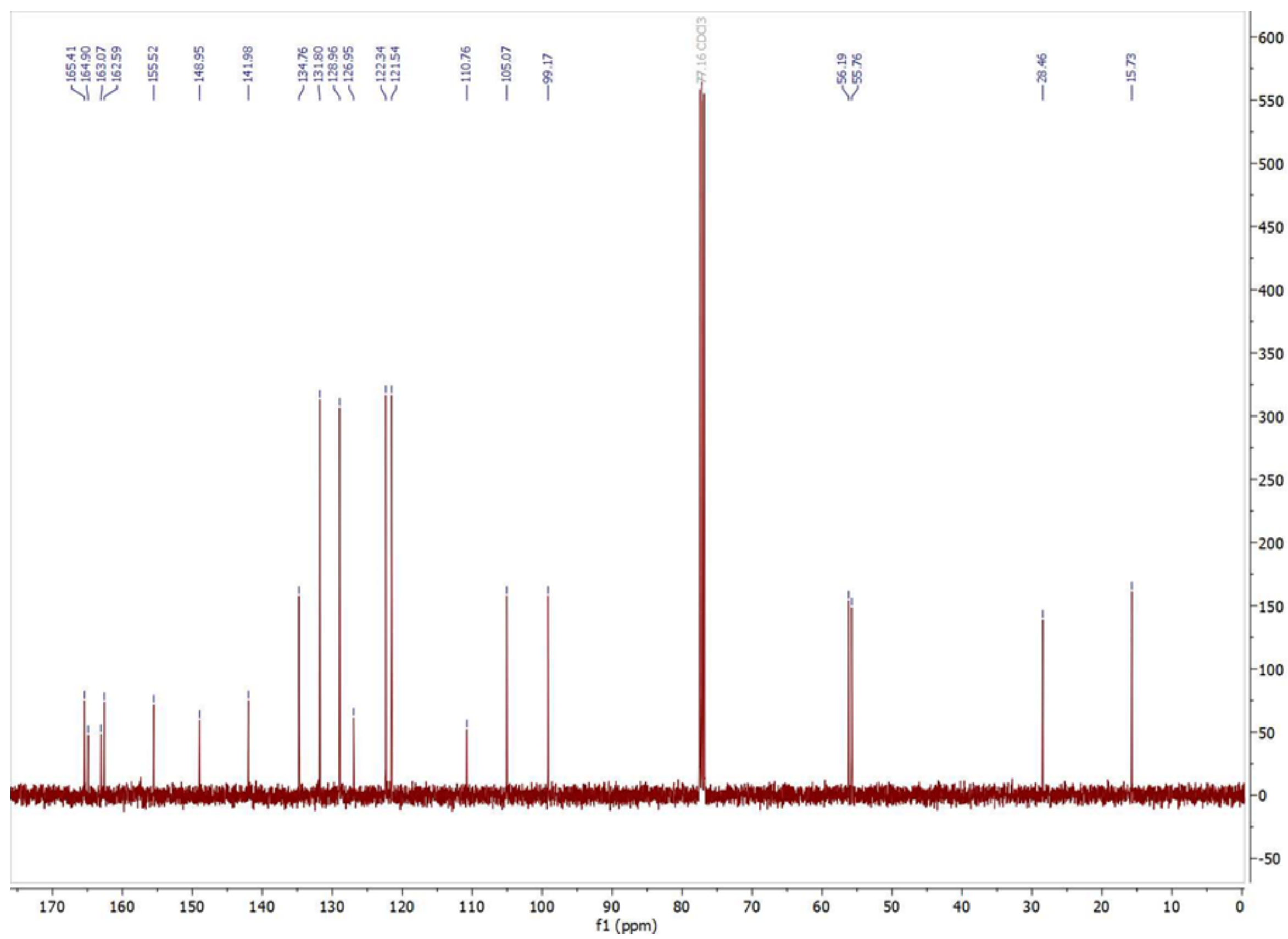

**Figure SI-70:** $^{13}C\{^{1}H\}$ NMR for 4-((4-Ethylphenoxy)carbonyl)phenyl-2,4-dimethoxybenzoate (COB30 IDX 1014)

### 7.8.2. 4-((4-formylphenyloxy)carbonyl)phenyl 2,4-dimethoxybenzoate (COB27 IDX 1407)

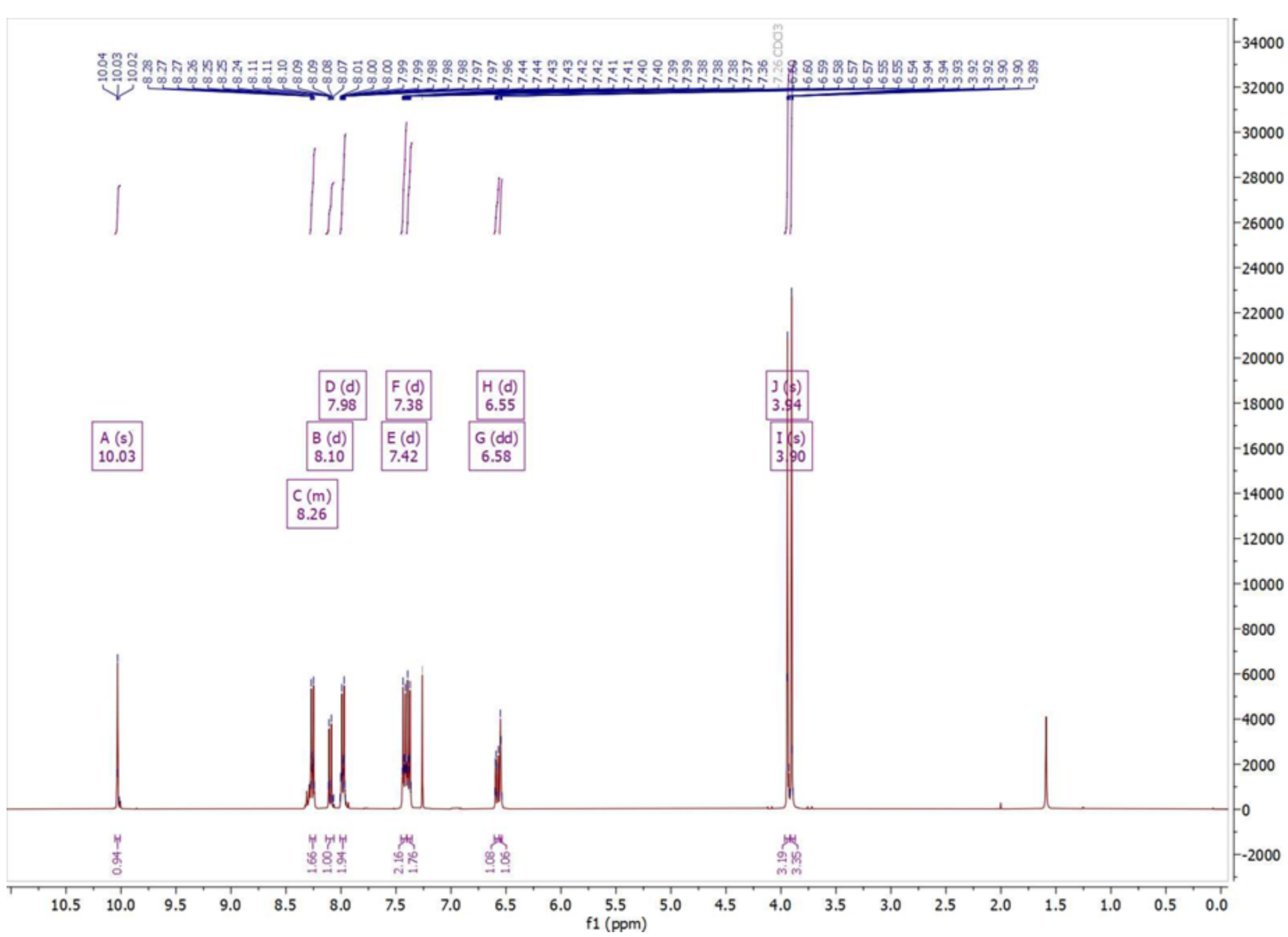

**Figure SI-71:** $^{1}$H NMR for IDX 4-((4-formylphenyloxy)carbonyl)phenyl 2,4-dimethoxybenzoate

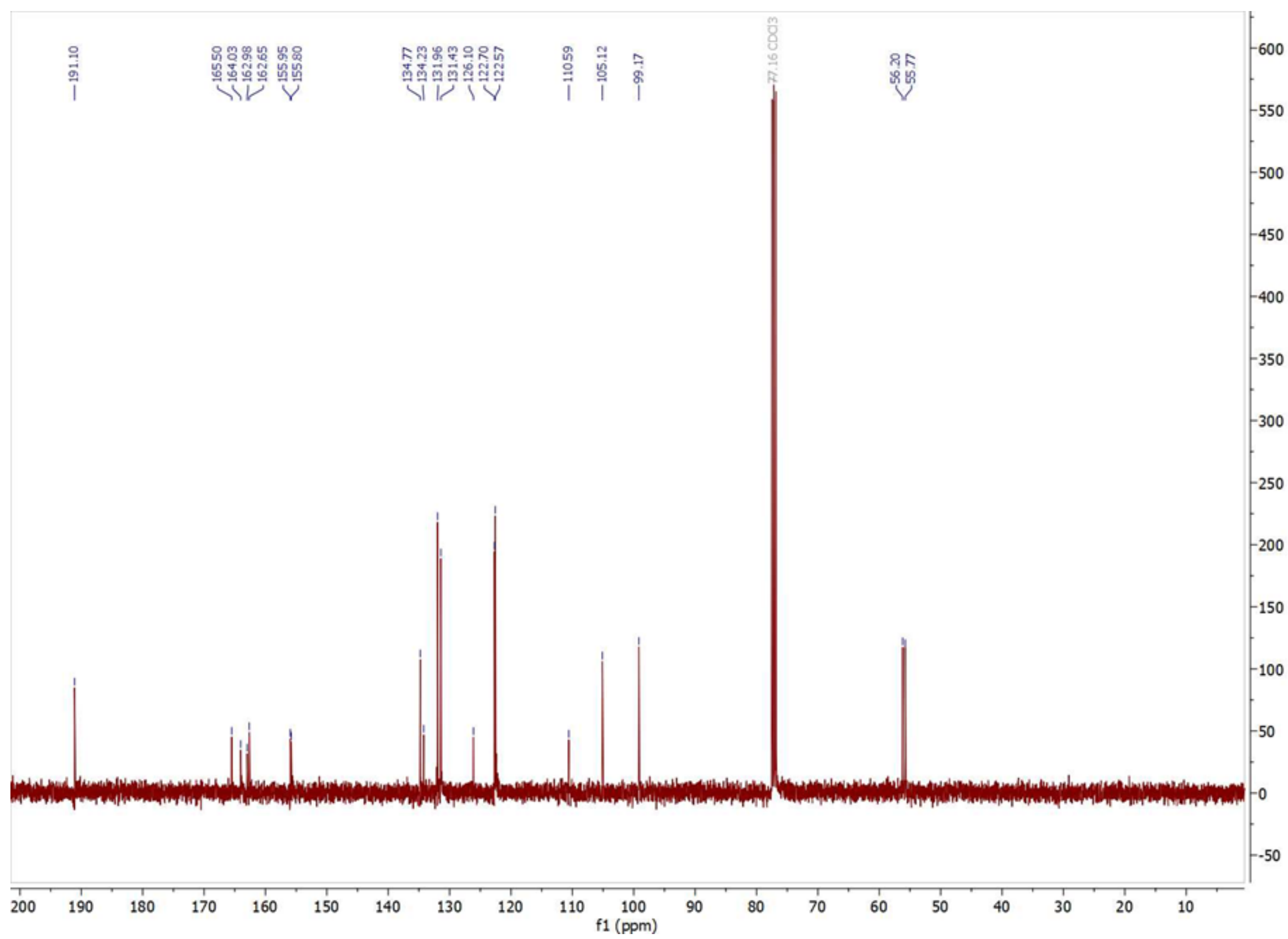

**Figure SI-72:** $^{13}C\{^{1}H\}$ NMR for 4-((4-formylphenyloxy)carbonyl)phenyl 2,4-dimethoxybenzoate